\documentclass[12pt]{article}
\usepackage{latexsym,cite,subeqn}
\input amssym.def
\input amssym.tex

\makeatletter  
\renewcommand{\theequation}{\thesection.\arabic{equation}}
\@addtoreset{equation}{section}
\makeatother

\font\msym=msbm10

\def\Real{{\mathop{\hbox{\msym \char  '122}}}} 
 
\def\Z{{\mathop{\hbox{\msym\char '132}}}}

\def\Y{{\rm Y}}
\def\tX{\tilde{{\rm X}}}

\def\tx{\tilde{{\rm x}}}
\def\x{{\rm x}} 
\def\y{{\rm y}} 
\def\ta{\tilde{{\rm a}}}
 
\def\da{\dot{\alpha}}

\def\bt{\bar{\theta}}

\begin{document}
\begin{titlepage}
\title{\vskip -60pt
{\small
\begin{flushright} 
DAMTP 98-88 \\
hep-th/9807186
\end{flushright}}
\vskip 15pt
Superconformal Symmetry in Six-dimensions \\and Its Reduction to
Four\\ ~~~ }
\vspace{2.0cm}
\author{Jeong-Hyuck Park\thanks{E-mail address:\,J.H.Park@damtp.cam.ac.uk}}
\date{}
\maketitle
\vspace{-1.0cm}
\begin{center}
\textit{Department of Applied Mathematics and Theoretical Physics}\\
\textit{University of Cambridge}\\
\textit{Silver Street, Cambridge, CB3 9EW, England}
\end{center}
\vspace{2.7cm}
\begin{abstract}
Superconformal  
symmetry in six-dimensions is analyzed  in terms of 
coordinate transformations on superspace. A superconformal Killing 
equation is derived and its  solutions are  
identified in terms of   supertranslations, dilations, Lorentz
transformations, $R$-symmetry transformations and special superconformal 
transformations. The full superconformal symmetry, which is shown to
form the group $\mbox{OSp}(2,6|N)$, is possible only if the 
supersymmetry algebra has  $N$ spinorial generators of the same  
chirality, corresponding to $(N,0)$ supersymmetry. The  
$R$-symmetry group is then  $\mbox{Sp}(N)$ and the corresponding superspace
is $\Real^{6|8N}$. We define superinversion as a map to the associated
superspace of opposite chirality.  
General  formulae for two-point  and three-point  
correlation functions of quasi-primary superfields are exhibited. 
The superconformal group in six-dimensions is reduced to 
 a corresponding  extended superconformal group in four-dimensions. 
Superconformally covariant differential operators are also discussed.
\end{abstract}
\thispagestyle{empty}
\end{titlepage}
\newpage

\section{Introduction}
Theories with 
extended supersymmetry in four-dimensions may  have a  close relation
to simpler theories in higher
dimensions, as  dimensional reduction of the latter often leads to
the former. In particular, reductions of the six and ten dimensional
$N=1$ 
supersymmetric theories  yield $N=2$ and $N=4$ supersymmetric theories
  in four-dimensions. Six-dimensional supersymmetric theories have
been studied some time ago by various authors, 
e.g.~\cite{NPB174335,NPB221331,NPB222319,CQG2815,PLB19857}. 
More recently there is
evidence from M/string theory for the presence of  renormalization
group fixed points  
in  six-dimensions and hence the existence of 
non-trivial six-dimensional theories
with superconformal invariance; 
see \cite{9609161,9702038} and references
given there. 
It  remains an open problem  to construct  explicitly    
 six-dimensional superconformal  theories in a  superspace
formalism, although  some actions have been proposed  in terms
 of component fields - scalar, spinor and two-form tensor 
fields~\cite{9710127,9711161} such that the three-form  
formed by its exterior derivative is self-dual.  Note that  
in six-dimensions  the self-duality condition is a real constraint 
with Minkowskian signature and  
theories with  a two-form field may be conformal, since, in general, 
  under conformal transformations   the exterior derivative of $p$-form
  transforms  homogeneously only in $d=2(p+1)$ dimensions.\newline
It is   well known that   conformal transformations may be defined 
for spacetimes of any dimension. 
However, not  all spacetime dimensions allow the corresponding 
supersymmetry algebra to be extended to a superconformal algebra. 
All possible superconformal algebras  were analyzed by   
Nahm~\cite{nahm} based on the
classification of real simple Lie superalgebras~\cite{kac}. 
According to \cite{nahm} the standard supersymmetry algebra 
admits an extension to a superconformal algebra only if $d\leq 6$. 
In particular, the highest dimension admitting superconformal algebra 
is six and in $d=4,6$ dimensions, 
which are of our interest in this paper, 
the bosonic part of the superconformal algebra is 
\begin{equation}
\begin{array}{ll}
d=4;~~&\mbox{o}(4,2)\oplus\mbox{u}(N),~N\neq 4~~\mbox{or}~~
\mbox{o}(4,2)\oplus\mbox{su}(4)\\
{}&{}\\
d=6;~~&\mbox{o}(6,2)\oplus\mbox{sp}(N)
\end{array}
\label{bosonic46}
\end{equation}
On six-dimensional Minkowskian spacetime it is possible to define 
Weyl spinors of opposite chiralities   
and so the six-dimensional supersymmetry may be 
 denoted by two numbers, $(N,\tilde{N})$, 
where $N$ and $\tilde{N}$ are the numbers of chiral and
anti-chiral supercharges. The $R$-symmetry group is then 
$\mbox{Sp}(N)\times\mbox{Sp}(\tilde{N})$. The analysis of 
Nahm  shows that to admit a superconformal  algebra
either $N$ or $\tilde{N}$ should be zero. Although both 
 $(1,1)$ and $(2,0)$ supersymmetry give rise  $N=4$ 
four-dimensional supersymmetry after dimensional reduction, only $(2,0)$
supersymmetry theories can be superconformal~\cite{seiberg16}.\newline
The above analysis  is essentially based on the 
classification of real simple Lie superalgebras and identification  of
the  bosonic part with  the  usual spacetime conformal symmetry. 
This approach does not rely on  any definition of 
superconformal transformation in superspace. In fact, by analyzing  
possible forms of 
the commutators between conformal symmetry and supersymmetry and
ensuring compatibility with the  Jacobi identity,  the 
six-dimensional superconformal algebra may be
constructed~\cite{9711161}. \newline
In this paper, we analyze   six-dimensional superconformal
symmetry directly in terms of  
  coordinate transformations on  superspace. The   
superconformal group is independently derived  and we also  obtain 
 concise formulae for correlation functions of superfields which
transform simply under superconformal transformations. 
This approach has previously been applied in 
four-dimensions;  for the analysis of superconformally invariant correlation
functions in four-dimensions see our earlier
work~\cite{paper1} which follows analogous discussions in the non
supersymmetric case~\cite{hughpetkou,hughjohanna}. 
General discussions of superconformal invariance
in four spacetime dimensions 
are contained in  
\cite{buchbinder,westscf} which have further references.\newline 
The organization of this paper is as follows.  In section 2,   
we review  six-dimensional $(N,\tilde{N})$ 
supersymmetry. For six-dimensional Weyl spinors  a pseudo-Majorana 
notation~\cite{kugotownsend} is  introduced, which simplifies the
formalism significantly.   In section 3, 
 we  first define the six-dimensional superconformal group in terms of 
superspace coordinate transformations as  a generalization
of the definition of ordinary conformal transformations. We then
derive a sufficient and necessary condition for a supercoordinate 
transformation to be superconformal.  
The general solutions are  identified in terms of  
supertranslations, dilations, Lorentz
transformations, $R$-symmetry transformations and special superconformal 
transformations for  $(N,0)$ supersymmetry. 
On the other hand, if $N,\tilde{N}>0$, 
 we show that although scale transformations may be introduced,   
there exist no special superconformal
transformations as expected from  Nahm's result. 
We also define   superinversion
as a map from $(N,0)$ superspace to $(0,N)$ superspace.   
The $(N,0)$ superconformal
group is identified with a 
supermatrix group, $\mbox{OSp}(2,6|N)$ having  dimensions 
$(28+N(2N+1)|16N)$. In section 4, we obtain an explicit formula for
the finite superconformal transformations of the supercoordinates. 
We also discuss about functions depending on  two or   three points in
superspace which transform 
covariantly under superconformal transformations. These functions are
crucial to obtain two and three points correlation functions later. In
section 5, the $(N,0)$ superconformal group in six-dimensions is reduced to
 a corresponding  superconformal group in four-dimensions. 
In particular, the
$R$-symmetry group 
is reduced from $\mbox{Sp}(N)$ to $\mbox{U}(N)$ if $N\neq
4$  and $\mbox{SU}(4)$ if $N=4$,  as in eq.(\ref{bosonic46}). 
In section 6, we discuss the conditions of  
superconformal invariance for correlation
functions for  quasi-primary superfields. General forms of   
two-point and three-point  functions are exhibited and   
explicit formulae for two-point  functions of
 superfields in simple cases are given as well.  
In an appendix, 
some useful equations  and $(N,0)$ superconformal algebra
are given. Superconformally  
covariant differential operators are also discussed by considering the
conditions for
superfields  formed by the action of  
spinor derivatives on quasi-primary superfields to remain
quasi-primary.

\section{Supersymmetry in Six-dimensions}
With the six-dimensional  Minkowskian metric $\eta^{AB} =
\mbox{diag}(+1,-1,-1\cdots -1)$, the  $8\times 8$ gamma matrices,
$\Gamma^{A},~A=0,1,\cdots 5$,   
satisfy the Clifford algebra
\begin{equation}
\Gamma^{A}\Gamma^{B}+\Gamma^{B}\Gamma^{A}=2\eta^{AB}
\label{GC}
\end{equation}
The gamma matrices for even dimensions can be chosen in general
  to have the form
\begin{equation}
\Gamma^{A}=\left(\begin{array}{cc}
                 0 & \gamma^{A}\\
                 \tilde{\gamma}^{A} & 0 \end{array}\right)
\label{rep}
\end{equation} 
where the generalization of $\gamma_{5}$ is diagonal
\begin{equation}
\Gamma^{7}=\Gamma^{0}\Gamma^{1}\Gamma^{2}\Gamma^{3}\Gamma^{4}\Gamma^{5}=
\left(\begin{array}{cc}
                           1&0\\
                           0&-1 \end{array}\right)
\label{g5}
\end{equation}
We also assume the hermiticity condition
\begin{equation}
\Gamma^{0}\Gamma^{A\dagger}\Gamma^{0}=\Gamma^{A}
\label{Gdagger}
\end{equation}
The $4\times 4$ matrices, $\gamma^{A},\tilde{\gamma}^{A}$
satisfy from eq.(\ref{GC}) 
\begin{equation}
\gamma^{A}\tilde{\gamma}^{B}+\gamma^{B}\tilde{\gamma}^{A}=2\eta^{AB}
\label{gamma}
\end{equation}
and
\begin{equation}
\mbox{tr}(\gamma^{A}\tilde{\gamma}^{B})=4\eta^{AB}
\label{Tr2}
\end{equation}
From eq.(\ref{Gdagger}) we have
\begin{equation}
\begin{array}{cc}
\gamma^{0}\gamma^{A}{}^{\dagger}\gamma^{0}
=\gamma^{A}~~~~&~~~~\gamma^{A}{}^{\dagger}=\tilde{\gamma}_{A}
\label{gaga}
\end{array}
\end{equation}
In six-dimensions, $\gamma^{A},\,\tilde{\gamma}^{A}$ 
may also be taken to be anti-symmetric
\begin{equation}
\begin{array}{cc}
(\gamma^{A})_{\alpha\beta}=-(\gamma^{A})_{\beta\alpha}~~~~ &~~~~ 
(\tilde{\gamma}^{A})^{\alpha\beta}=-(\tilde{\gamma}^{A})^{\beta\alpha}
\end{array}
\label{anti-sym}
\end{equation}
$\{\gamma^{A}\}$ and $\{\tilde{\gamma}^{A}\}$ are  separately bases 
of
$4 \times 4$ anti-symmetric matrices so that
\begin{equation}
(\gamma^{A})_{\alpha\beta}(\tilde{\gamma}_{A})^{\gamma\delta}=
2(\delta^{~\delta}_{\alpha}\delta^{~\gamma}_{\beta}-
\delta^{~\gamma}_{\alpha}\delta^{~\delta}_{\beta}) 
\label{contract}
\end{equation}
with the coefficient determined by eq.(\ref{Tr2}). \newline
$\{\gamma^{[A}\tilde{\gamma}^{B]}, 1\}$ also forms  
a basis of general $4\times 4$ matrices with the completeness relation
\begin{equation}
-\textstyle{\frac{1}{8}}(\gamma^{[A}\tilde{\gamma}^{B]})_{\alpha}^{~\beta}
(\gamma_{[A}\tilde{\gamma}_{B]})_{\gamma}^{~\delta}+
\textstyle{\frac{1}{4}}\delta_{\alpha}^{~\beta}\delta_{\gamma}^{~\delta}=
\delta_{\alpha}^{~\delta}\delta_{\gamma}^{~\beta}
\label{complete}
\end{equation}
With this choice of gamma matrices we get\footnote{We put 
$\epsilon_{1234}=\epsilon^{1234}=1$. See Appendix A for further discussion.}
\begin{subeqnarray}
(\gamma^{A})_{\alpha\beta}=\textstyle{\frac{1}{2}}
\epsilon_{\alpha\beta\gamma\delta}(\tilde{\gamma}^{A})^{\gamma\delta}~~~~
\label{geg1}\\
{}\nonumber\\
(\tilde{\gamma}^{A})^{\alpha\beta}=\textstyle{\frac{1}{2}}
\epsilon^{\alpha\beta\gamma\delta}(\gamma^{A})_{\gamma\delta}~~~~
\label{geg2}
\end{subeqnarray}
The supersymmetry algebra has the standard form
\begin{equation}
\{ {\cal Q}^{a}, \bar{{\cal Q}}_{b}\}
=2\delta^{a}_{~b}\Gamma^{A}P_{A} 
\label{6Dsusy}
\end{equation}
The eight-component spinorial supercharge, ${\cal Q}^{a}$ and Grassman 
variable, $\vartheta^{a}$ ($a=1,2,\cdots N$) can be written 
 in the basis given by eqs.(\ref{rep},\,\ref{g5})
\begin{equation}
\begin{array}{cc}
{\cal Q}^{a}=\left(\begin{array}{c}
        {\cal Q}^{a}_{+}\\
        {\cal Q}^{a}_{-}
       \end{array}\right)        ~~~~ &~~~~  
\bar{{\cal Q}}_{a}={{\cal Q}^{a}}^{\dagger}\Gamma^{0}=
(\bar{{\cal Q}}_{a}^{-
},
        \bar{{\cal Q}}_{a}^{+}  )
\end{array}
\label{decomQ}
\end{equation}
\begin{equation}
\begin{array}{cc}
\vartheta^{a}=\left(\begin{array}{c}
        \vartheta^{a}_{+}\\
        \vartheta^{a}_{-}
       \end{array}\right)          ~~~~&~~~~  
\bar{\vartheta}_{a}
={\vartheta^{a}}^{\dagger}\Gamma^{0}=(\bar{\vartheta}_{a}^{-},
        \bar{\vartheta}_{a}^{+}  )
\end{array}
\label{decomT}
\end{equation}
where  the superscripts/subscripts $\pm$ denote Weyl spinors.\newline 
With the decomposition (\ref{decomQ}) and (\ref{decomT}) we 
define for $1\leq i\leq 2N$
\begin{equation}
\begin{array}{cc}
Q^{i}=\left(\begin{array}{c}
              {\cal Q}^{a}_{+}\\
              (\bar{{\cal Q}}^{+}_{a})^{t}
             \end{array}\right)           
~~~~~&~~~~~\tilde{Q}^{i}=\left(\begin{array}{c}
              {\cal Q}^{a}_{-}\\
              (\bar{{\cal Q}}^{-}_{a})^{t}
             \end{array}\right)  
\end{array}
\end{equation}
\begin{equation}
\begin{array}{cc}
\theta^{i}=\left(\begin{array}{c}
              \vartheta^{a}_{-}\\
              (\bar{\vartheta}^{-}_{a})^{t}
             \end{array}\right)
~~~~~&~~~~~\tilde{\theta}^{i}=\left(\begin{array}{c}
              \vartheta^{a}_{+}\\
              (\bar{\vartheta}^{+}_{a})^{t}
             \end{array}\right)
\end{array}
\end{equation}
With these definitions the supersymmetry  algebra~(\ref{6Dsusy}) is
equivalent in six-dimensions  to
\begin{subequations}
\begin{equation}
\{Q^{i},(Q^{j})^{t}\}=2{\cal E}^{ij}\gamma^{A}P_{A}
\label{N,0alge}
\end{equation}
\begin{equation}
\{\tilde{Q}^{i},(\tilde{Q}^{j})^{t}\}=2{\cal E}^{ij}\tilde{\gamma}^{A}P_{A}
\label{0,Nalge}
\end{equation}
\end{subequations}
where ${\cal E}^{ij}$ is 
\begin{equation}
{\cal E}^{ij}=\left(\begin{array}{cc}
                       0  & \delta^{a}_{~b}\\
                    -\delta^{~b}_{a} & 0
                    \end{array}\right)
\end{equation}
and the inverse, $\bar{\cal E}_{ij}$, is defined by 
${\cal E}^{ij}\bar{\cal E}_{jk}=\delta^{i}_{~k}$
\begin{equation}
\bar{\cal E}_{jk}=\left(\begin{array}{cc}
                     0& -\delta_{b}^{~c}\\
                 \delta^{b}_{~c}&0
                    \end{array}\right)
\end{equation}
$Q^{i},\theta^{i},\tilde{Q}^{i},\tilde{\theta}^{i}$ satisfy 
the pseudo-Majorana conditions
\begin{equation}
\begin{array}{cc}
\bar{Q}_{i}=Q^{i\dagger}\gamma^{0}=(Q^{j})^{t}\bar{\cal E}_{ji}
~~~~&~~~~\bar{\tilde{Q}}_{i}=\tilde{Q}^{i\dagger}\tilde{\gamma}^{0}
=(\tilde{Q}^{j})^{t}\bar{\cal E}_{ji}
\end{array}
\end{equation}
\begin{equation}
\begin{array}{cc}
\bar{\theta}_{i}=\theta^{i\dagger}\tilde{\gamma}^{0}
=(\theta^{j})^{t}\bar{\cal E}_{ji}~~~~&~~~~
\bar{\tilde{\theta}}_{i}=\tilde{\theta}^{i\dagger}\gamma^{0}
=(\tilde{\theta}^{j})^{t}\bar{\cal E}_{ji}
\end{array}
\label{pseudoM}
\end{equation}
which give
\begin{equation}
\bar{\vartheta}_{a}{\cal Q}^{a}+\bar{{\cal Q}}_{a}\vartheta^{a}=
\bar{Q}_{i}\theta^{i}+\bar{\tilde{Q}}_{i}\tilde{\theta}^{i}
\end{equation}
\begin{equation}
\begin{array}{cc}
\bar{Q}_{i}\theta^{i}=\bar{\theta}_{i}Q^{i}~~~~~&~~~~~
\bar{\tilde{Q}}_{i}\tilde{\theta}^{i}
=\bar{\tilde{\theta}}_{i}\tilde{Q}^{i}
\end{array}
\end{equation}
The six-dimensional supersymmetry algebra obtained here 
can be generalized by replacing 
eq.(\ref{0,Nalge}) by
\begin{equation}
\begin{array}{cc}
\{\tilde{Q}^{\tilde{\imath}},(\tilde{Q}^{\tilde{\jmath}})^{t}\}=
2\tilde{{\cal 
E}}{}^{\tilde{\imath}\tilde{\jmath}}\tilde{\gamma}^{A}P_{A}~~~&~~~
\tilde{\imath},\tilde{\jmath}=1,2,\cdots,2\tilde{N}
\end{array}
\end{equation}
with $\tilde{{\cal E}}$, the appropriate anti-symmetric matrix, 
so that with eq.(\ref{N,0alge}) they form the six-dimensional 
$(N,\tilde{N})$  
supersymmetry algebra.  
The above discussion therefore demonstrates that the supersymmetry
algebra~(\ref{6Dsusy}) corresponds to $(N,N)$ supersymmetry.\newline
The exterior derivative,~${\rm d}$, on superspace with coordinates 
$z^{{\cal 
A}}
 = (x^{A},\theta^{i\alpha},\tilde{\theta}^{\tilde{\imath}}_{\alpha})$, is 
defined as
\begin{equation}
{\rm d}
\equiv {\rm d}z^{{\cal A}}\frac{\partial~}{\partial z^{{\cal A}}}
=e^{{\cal A}}D_{{\cal A}}=e^{A}\partial_{A}+{\rm 
d}\theta^{i\alpha}D_{i\alpha}
+{\rm d}
\tilde{\theta}^{\tilde{\imath}}_{\alpha}
{\tilde{D}}^{\alpha}_{\tilde{\imath}}
\end{equation}
where $e^{{\cal A}}=(e^{A},
{\rm d}\theta^{i\alpha},{\rm
d}\tilde{\theta}^{\tilde{\imath}}_{\alpha})$ 
are
supertranslation  invariant  one forms
\begin{equation}
e^{A}(z)={\rm d}x^{A}-i\bar{\theta}_{i}\gamma^{A}{\rm d}\theta^{i}-
i\bar{\tilde{\theta}}_{\tilde{\imath}}\tilde{\gamma}^{A}{\rm 
d}\tilde{\theta}^{\tilde{\imath}}
\label{infint}
\end{equation}
and   $D_{{\cal 
A}}=(\partial_{A},D_{i\alpha},\tilde{D}_{\tilde{\imath}}^{\alpha})$ are 
covariant  derivatives  
\begin{equation}
\begin{array}{ccc}
\displaystyle{\partial_{A}=\frac{\partial~}{\partial x^{A}}}~~~&~~~
\displaystyle{
D_{i\alpha}=\frac{\partial~}{\partial\theta^{i\alpha}}-i(\bar{\theta}_{i}
\gamma^{A})_{\alpha}\frac{\partial~}{\partial x^{A}}}~~~&~~~\displaystyle{
\tilde{D}_{\tilde{\imath}}^{\alpha}=
\frac{\partial~}{\partial\tilde{\theta}^{\tilde{\imath}}_{\alpha}}-
i(\bar{\tilde{\theta}}_{\tilde{\imath}}
\tilde{\gamma}^{A})^{\alpha}\frac{\partial~}{\partial x^{A}}}
\end{array}
\end{equation}
satisfying the anti-commutator relations
\begin{equation}
\begin{array}{cc}
\displaystyle{
\{D_{i\alpha},D_{j\beta}\}=
-2i\bar{\cal E}_{ij}\gamma^{A}_{\alpha\beta}\partial_{A}}~~~&~~~
\displaystyle{
\{\tilde{D}_{\tilde{\imath}}^{\alpha},\tilde{D}_{\tilde{\jmath}}^{\beta}\}
=-
2i\bar{\tilde{\cal E}}_{\tilde{\imath}\tilde{\jmath}}
\tilde{\gamma}^{A\alpha\beta}\partial_{A}}
\end{array}
\label{anticomD}
\end{equation}
Under an arbitrary superspace coordinate transformation,   
$z \rightarrow z^{\prime}$,   
$e^{{\cal A}}$ and $D_{{\cal A}}$  transform as
\begin{equation}
\begin{array}{cc}
e^{{\cal A}}(z^{\prime})
=e^{{\cal B}}(z){\cal R}_{{\cal B}}{}^{{\cal A}}(z)~~~~&~~~~
D^{\prime}_{{\cal A}}={\cal R}^{-1}{}_{{\cal A}}{}^{{\cal B}}(z)D_{{\cal
B}}
\end{array}
\label{ReD}
\end{equation}
so that the exterior derivative is left invariant
\begin{equation}
e^{{\cal A}}(z)D_{{\cal A}}=e^{{\cal A}}(z^{\prime})D^{\prime}_{{\cal A}}
\end{equation}

\section{Superconformal Symmetry in Six-dimensions}
The superconformal group is  defined here as a group of 
superspace coordinate
transformations, $z \stackrel{g}{\longrightarrow} z^{\prime}$ that 
preserve the 
infinitesimal supersymmetric interval length, $e^{2}=\eta_{AB}e^{A}e^{B}$
  upto a local scale factor, so that
\begin{equation}
{e^{2}}(z)~\rightarrow~e^{2}(z^{\prime})=\Omega^{2}(z;g)e^{2}(z) 
\label{scdef}
\end{equation}
where $\Omega(z;g)$ is a local scale factor.\newline
This requires 
\begin{equation}
e^{A}(z^{\prime})=e^{B}(z)R^{~A}_{B}(z;g)
\label{ehomogeneous}
\end{equation}
and
\begin{equation}
R^{~C}_{A}(z;g)R^{~D}_{B}(z;g)\eta_{CD}=\Omega^{2}(z;g)\eta_{AB}
\label{R2}
\end{equation}
giving
\begin{equation}
\det R(z;g)=\Omega^{6}(z;g)
\end{equation}
To ensure eq.(\ref{ehomogeneous}), 
${\cal R}_{{\cal A}}{}^{{\cal B}}(z;g)$ in eq.(\ref{ReD}) 
must be  of the form 
\begin{equation}
{\cal R}_{{\cal A}}{}^{{\cal B}}(z;g)=
\left(\begin{array}{cll}
R^{~B}_{A}(z;g)&\partial_{A}\theta^{\prime j\beta}&
\partial_{A}\tilde{\theta}^{\prime \tilde{\jmath}}_{\beta}\\
0&D_{i\alpha}\theta^{\prime j\beta}&
D_{i\alpha}\tilde{\theta}^{\prime \tilde{\jmath}}_{\beta}\\
0&\tilde{D}_{\tilde{\imath}}^{\alpha}\theta^{\prime j\beta}&
\tilde{D}_{\tilde{\imath}}^{\alpha}\tilde{\theta}^{\prime 
\tilde{\jmath}}_{\beta}
\end{array}\right)
\label{calR}
\end{equation}
To achieve the form~(\ref{calR}) we must impose
\begin{subeqnarray}
\label{BB=0}
&D_{i\alpha}x^{\prime A}+i\bar{\theta}^{\prime}_{j}
\gamma^{A}D_{i\alpha}\theta^{\prime j}
+i\bar{\tilde{\theta}}{}^{\prime}_{\tilde{\jmath}}\tilde{\gamma}^{A}
D_{i\alpha}\tilde{\theta}^{\prime \tilde{\jmath}}=0\\
&{}\nonumber\\
&\tilde{D}_{\tilde{\imath}}^{\alpha}x^{\prime A}
+i\bar{\theta}^{\prime}_{j}
\gamma^{A}\tilde{D}_{\tilde{\imath}}^{\alpha}\theta^{\prime j}
+i\bar{\tilde{\theta}}{}^{\prime}_{\tilde{\jmath}}\tilde{\gamma}^{A}
\tilde{D}_{\tilde{\imath}}^{\alpha}\tilde{\theta}^{\prime \tilde{\jmath}}=0
\end{subeqnarray}
while $R_{A}^{~B}$ is given by  
\begin{equation}
R^{~B}_{A}(z;g)=\frac{\partial 
x^{\prime B}}{\partial x^{A}}-i\bar{\theta}^{\prime}_{i}\gamma^{B}
\frac{\partial\theta^{\prime i}}{\partial 
x^{A}}-i\bar{\tilde{\theta}}{}^{\prime}_{\tilde{\imath}}\tilde{\gamma}^{B}
\frac{\partial 
\tilde{\theta}^{\prime \tilde{\imath}}}{\partial x^{A}} 
\label{Rh}
\end{equation}
and $R(z;g)$ must also satisfy eq.(\ref{R2}).  $R(z;g)$ forms  
a representation of the superconformal group, since  under  
the successive superconformal transformations, 
$z\stackrel{g}{\longrightarrow}z^{\prime}
\stackrel{g^{\prime}}{\longrightarrow}z^{ \prime\prime}$ giving 
$z\stackrel{g^{\prime\prime}}{\longrightarrow}z^{\prime\prime}$, we have
\begin{equation}
R(z;g)R(z^{\prime};g^{\prime})=R(z;g^{\prime\prime})
\label{Rrep}
\end{equation}
Infinitesimally   
$z^{\prime}\simeq z+\delta z$, eq.(\ref{BB=0})  gives, if $N> 0$,
\begin{equation}
D_{i\alpha}h^{A}=-2i(\bar{\lambda}_{i}\gamma^{A})_{\alpha}
\label{enough1}
\end{equation}
and for $\tilde{N}>0$
\begin{equation}
\tilde{D}^{\alpha}_{\tilde{\imath}}h^{A}
=-2i(\bar{\tilde{\lambda}}_{\tilde{\imath}}\tilde{\gamma}^{A})^{\alpha}
\label{enough2}
\end{equation}
where we define
\begin{equation}
\begin{array}{cc}
\lambda^{i}=\delta\theta^{i}~~~&~~~
\tilde{\lambda}^{\tilde{\imath}}=\delta\tilde{\theta}^{\tilde{\imath}}
\end{array}
\end{equation}
\begin{equation}
h^{A}=\delta
x^{A}-i\bar{\theta}_{i}\gamma^{A}\delta\theta^{i}-
i\bar{\tilde{\theta}}_{\tilde{\imath}}\tilde{\gamma}^{A}
\delta\tilde{\theta}^{\tilde{\imath}}
\end{equation}
Infinitesimally  $R_{A}^{~B}$ from eq.(\ref{Rh}) is of the form
\begin{equation} 
R_{A}^{~B}\simeq\delta_{A}^{~B}+\partial_{A}h^{B}
\label{infR}
\end{equation}
so that the condition~(\ref{R2}) reduces to the ordinary conformal
Killing equation
\begin{equation}
\partial_{A}h_{B}+\partial_{B}h_{A}\propto \eta_{AB}
\label{ordiKi}
\end{equation}
This result  follows from eq.(\ref{enough1}), since  
using the anti-commutator relation for 
$D_{i\alpha}$~(\ref{anticomD}) we get from eq.(\ref{enough1})
\begin{equation}
\bar{\cal E}_{ij}\partial_{B}h^{A}
=\textstyle{\frac{1}{4}}\left(
D_{i\alpha}(\bar{\lambda}_{j}\gamma^{A}\tilde{\gamma}_{B})^{\alpha}
-D_{j\alpha}(\bar{\lambda}_{i}\gamma^{A}\tilde{\gamma}_{B})^{\alpha}\right)
\end{equation}
and hence
\begin{equation}
\bar{\cal E}_{ij}(\partial_{A}h_{B}+\partial_{B}h_{A})
=\textstyle{\frac{1}{2}}(D_{i\alpha}\bar{\lambda}^{\alpha}_{j}
-D_{j\alpha}\bar{\lambda}^{\alpha}_{i})\eta_{AB}
\label{ordinaryKilling}
\end{equation}
which implies
\begin{subeqnarray}
\partial_{A}h_{B}+\partial_{B}h_{A}=2\rho{}\eta_{AB}\\
{}\nonumber\\
D_{i\alpha}\bar{\lambda}^{\alpha}_{j}
-D_{j\alpha}\bar{\lambda}^{\alpha}_{i}=4\rho{}\bar{\cal E}_{ij}
\end{subeqnarray}
A similar equation may also be obtained from
eq.(\ref{enough2}).\newline  
Hence eqs.(\ref{enough1},\,\ref{enough2}) are sufficient and necessary
conditions for $(N,\tilde{N})$ superconformal symmetry, although if
$\tilde{N}=0$ eq.(\ref{enough1}) alone is sufficient.\footnote{While 
this work was being completed,  
similar equations to (\ref{enough1},\,\ref{ordiKi}) were
discussed by Grojean and Mourad~\cite{9807055}.}\newline
In six-dimensions there is a unique correspondence between a  general six
vector,~$v^{A}$, and  an anti-symmetric matrix,~${\rm
v}_{\alpha\beta}$ or $\tilde{{\rm v}}^{\alpha\beta}$ 
through
\begin{subeqnarray}
{\rm v}_{\alpha\beta}=v^{A}\gamma_{A}{}_{\alpha\beta}~~~~&~~~
v^{A}=\textstyle{\frac{1}{4}}\mbox{tr}(\tilde{\gamma}^{A}{\rm v})\\
{}\nonumber\\
\tilde{{\rm v}}^{\alpha\beta}=v^{A}\tilde{\gamma}_{A}^{\alpha\beta}
~~~~&~~~~
v^{A}=\textstyle{\frac{1}{4}}\mbox{tr}(\gamma^{A}\tilde{{\rm v}})
\end{subeqnarray}
where
\begin{equation}
\begin{array}{cc}
{\rm v}_{\alpha\beta}=\textstyle{\frac{1}{2}}
\epsilon_{\alpha\beta\gamma\delta}\tilde{{\rm v}}^{\gamma\delta}~~~~&~~~~
\tilde{{\rm v}}^{\alpha\beta}=\textstyle{\frac{1}{2}}
\epsilon^{\alpha\beta\gamma\delta}{\rm v}_{\gamma\delta}
\end{array}
\end{equation}
With this notation and using eq.(\ref{contract}), 
eq.(\ref{enough1}) is equivalent to
\begin{equation}
D_{i\alpha}\tilde{{\rm h}}^{\beta\gamma}=\textstyle{\frac{1}{3}}(
\delta_{\alpha}^{~\beta}D_{i\delta}\tilde{{\rm h}}^{\delta\gamma}-
\delta_{\alpha}^{~\gamma}D_{i\delta}\tilde{{\rm h}}^{\delta\beta})
\label{master}
\end{equation}
and $\bar{\lambda}^{\alpha}_{i}$ is given by
\begin{equation}
\bar{\lambda}^{\alpha}_{i}=i\textstyle{\frac{1}{12}}D_{i\beta}\tilde{{\rm 
h}}^{\beta\alpha}
\label{lambdah}
\end{equation} 
Eq.(\ref{master}) may therefore be regarded as the fundamental 
$(N,0)$ superconformal
Killing equation.\newline
To solve these equations we first write a general solution of
eq.(\ref{ordiKi}) as
\begin{equation}
h^{A}(z)=a^{A}(\theta)+\lambda(\theta)x^{A}+w^{A}_{~B}(\theta)x^{B}
+2x{\cdot b}(\theta)x^{A}-x^{2}b^{A}(\theta)
\label{hA}
\end{equation}
where $w_{AB}(\theta)+w_{BA}(\theta)=0.$ 
It is convenient to  introduce the variables
\begin{equation}
\tx_{\pm}=\tx\pm2i\theta^{i}\bar{\theta}_{i}
\label{tildeX}
\end{equation}
where 
\begin{equation}
\begin{array}{cc}
\tx^{t}_{\pm}=-\tx_{\mp} ~~~~~~~&~~~~~~
\tx_{\pm}=\tilde{\gamma}^{0}\tx^{\dagger}_{\mp}\tilde{\gamma}^{0}
\end{array}
\label{Xprop}
\end{equation}
Then eq.(\ref{hA}) can be written in terms of  
$\tilde{{\rm h}}=h^{A}\tilde{\gamma}_{A}$   as
\begin{equation}
\tilde{{\rm h}}(z)=\tx_{-}{\rm b}(\theta)\tx_{+}+W(\theta)\tx_{+}
+\tx_{-}W(\theta)^{t}+\tilde{{\rm A}}(\theta)
\label{tildeh}
\end{equation}
where
\begin{equation}
W(\theta)={\textstyle\frac{1}{4}}w_{AB}(\theta)\tilde{\gamma}^{A}\gamma^{B}
+\textstyle{\frac{1}{2}}\lambda(\theta) +2i\theta^{i}\bar{\theta}_{i}{\rm
b}(\theta)
\end{equation}
\begin{equation}
\tilde{{\rm A}}(\theta)=\tilde{{\rm a}}(\theta)
+i{\textstyle\frac{1}{2}}w_{AB}(\theta)
(\theta^{i}\bar{\theta}_{i}\gamma^{B}\tilde{\gamma}^{A}-
\tilde{\gamma}^{A}\gamma^{B}\theta^{i}\bar{\theta}_{i})+4(\bar{\theta}_{i} 
{\rm b}(\theta)\theta^{j})\theta^{i}\bar{\theta}_{j}
=-(\tilde{{\rm A}}(\theta))^{t}
\end{equation}
Essentially we may regard   $W(\theta)$ as  an 
arbitrary $4\times 4$  matrix and $\tilde{{\rm A}}(\theta)$ as an arbitrary
$4\times 4$ anti-symmetric matrix.\newline
The variables,~$\tx_{\pm}$ defined in eq.(\ref{tildeX}), satisfy
\begin{equation}
\begin{array}{cc}
D_{i\alpha}\tx^{\beta\gamma}_{+}=
4i\delta_{\alpha}^{~\gamma}\bar{\theta}_{i}^{\beta}~~~~&~~~~
D_{i\alpha}\tx^{\beta\gamma}_{-}=-
4i\delta_{\alpha}^{~\beta}\bar{\theta}_{i}^{\gamma}
\end{array}
\label{DX}
\end{equation}
which ensures that substituting  eq.(\ref{tildeh})  into
eq.(\ref{master}) leads  independent   equations 
for ${\rm b}(\theta),~W(\theta),~\tilde{{\rm A}}(\theta)$. Here,
we outline the method of solution.\footnote{The details of deriving 
eqs.(\ref{btheta},\,\ref{Wtheta1},\,\ref{Wtheta2},\,\ref{atheta1},\,\ref{atheta2}) 
can be found in  
Appendix B.}\newline
Considering the $x^{2}$-terms one can show 
\begin{equation}
D_{i\alpha}{\rm b}(\theta)=0
\label{btheta}
\end{equation}
and so $b^{A}(\theta)$ is independent of $\theta$ and hence
 eq.(\ref{DX}) shows  that 
$\tx_{-}{\rm b}\tx_{+}$ is a solution of
eq.(\ref{master}).  Now focusing on the  linear terms in $x$, one can show 
\begin{equation}
D_{i\alpha}W^{\beta}_{~\gamma}(\theta)
=\delta_{\alpha}^{~\beta}D_{i\gamma}W^{\delta}_{~\delta}(\theta)
\label{Wtheta1}
\end{equation}
\begin{equation}
D_{i\alpha}D_{j\beta}W^{\gamma}_{~\delta}(\theta)=0
\label{Wtheta2}
\end{equation}
These two relations determine $W(\theta)$ 
\begin{equation}
\begin{array}{cc}
W(\theta)=-4\theta^{i}\bar{\rho}_{i}+\omega
+{\textstyle\frac{1}{2}\lambda}~~~~&~~~
W(\theta)^{t}=-4\rho^{i}\bar{\theta}_{i}-\tilde{\omega}
+{\textstyle\frac{1}{2}\lambda}
\end{array}
\end{equation}
where for $\omega_{AB}=-\omega_{BA}$ 
\begin{equation}
\begin{array}{cc}
\omega={\textstyle
\frac{1}{4}}\omega_{AB}\tilde{\gamma}^{A}\gamma^{B}~~~~~~&
~~~~~~
\tilde{\omega}
={\textstyle \frac{1}{4}}\omega_{AB}\gamma^{A}\tilde{\gamma}^{B}
\end{array}
\label{omega2}
\end{equation}
and $\rho^{i}_{\alpha},~\bar{\rho}_{i\alpha}$ satisfy
\begin{equation}
\bar{\rho}_{i}=(\rho^{j})^{t}\bar{\cal E}_{ji}
\label{pseudorho}
\end{equation}
$W(\theta)\tx_{+}
+\tx_{-}W(\theta)^{t}$ is a solution of eq.(\ref{master}) as well. 
Therefore 
the remaining
terms are
\begin{equation}
D_{i\alpha}\tilde{{\rm A}}^{\beta\gamma}(\theta)=\textstyle{\frac{1}{3}}
\left(\delta_{\alpha}^{~\beta}
D_{i\delta}\tilde{{\rm A}}^{\delta\gamma}(\theta)-
\delta_{\alpha}^{~\gamma}
D_{i\delta}\tilde{{\rm A}}^{\delta\beta}(\theta)\right)
\label{remain}
\end{equation}
from which one can derive
\begin{equation}
D_{i\alpha}D_{j\beta}\tilde{{\rm A}}^{\gamma\delta}(\theta)=
\textstyle{\frac{1}{12}}(\delta_{\alpha}^{~\gamma}\delta_{\beta}^{~\delta}
-\delta_{\alpha}^{~\delta}\delta_{\beta}^{~\gamma})
D_{i\epsilon}D_{j\eta}\tilde{{\rm A}}^{\epsilon\eta}(\theta)
\label{atheta1}
\end{equation}
\begin{equation}
D_{i\alpha}D_{j\beta}D_{k\gamma}\tilde{{\rm A}}(\theta)=0
\label{atheta2}
\end{equation}
and so  $\tilde{{\rm A}}(\theta)$ is at most quadratic in
$\theta$. By virtue of these two equations, it is straightforward to
get the following  form of $\tilde{{\rm A}}(\theta)$ 
\begin{equation}
\tilde{{\rm A}}(\theta)=-4i\theta^{i}T_{i}^{~j}\bar{\theta}_{j}
+4i(\varepsilon^{i}\bar{\theta}_{i}-\theta^{i}\bar{\varepsilon}_{i})+
\tilde{{\rm a}} 
\label{Atheta}
\end{equation}
where
$T_{i}^{~j},~\varepsilon^{i\alpha},~\bar{\varepsilon}^{\alpha}_{i}$ 
satisfy
\begin{equation}
\begin{array}{c}
T^{t}{\cal E}+{\cal E}T=0\\
\nonumber\\
\bar{\varepsilon}_{i}=(\varepsilon^{j})^{t}\bar{\cal E}_{ji}
\end{array}
\label{pseudovar}
\end{equation}
Each term appearing in eq.(\ref{Atheta}) is an independent solution of
eq.(\ref{master}). \newline
All together, we now 
have the following general solution of eq.(\ref{master})
\begin{equation}
\tilde{{\rm h}}=\tx_{-}{\rm b}\tx_{+}
+(\omega+{\textstyle\frac{1}{2}}\lambda-4\theta^{i}\bar{\rho}_{i})\tx_{+}
-\tx_{-}(\tilde{\omega}
-{\textstyle\frac{1}{2}}\lambda +4\rho^{i}\bar{\theta}_{i})
-4i\theta^{i}T_{i}^{~j}\bar{\theta}_{j}
+4i(\varepsilon^{i}\bar{\theta}_{i}-\theta^{i}\bar{\varepsilon}_{i})
+\tilde{{\rm a}}
\label{solmaster}
\end{equation}
which induces 
\begin{equation}
\lambda^{i}(z)=\varepsilon^{i}+{\textstyle\frac{1}{2}}\lambda\theta^{i}
+\omega\theta^{i}-\theta^{j}T_{j}^{~i}+\tx_{-}{\rm b}\theta^{i}
+i\tx_{-}\rho^{i}
-4(\bar{\rho}_{j}\theta^{i})\theta^{j}
\label{dlambda}
\end{equation}
We may also rewrite the result~(\ref{solmaster}) as
\begin{subeqnarray}
\delta \tx_{+}=\tx_{+}{\rm b}\tx_{+}
-4\tx_{+}\rho^{i}\bar{\theta}_{i}
+\lambda \tx_{+}+\omega \tx_{+}-\tx_{+}\tilde{\omega}
+4i\varepsilon^{i}\bar{\theta}_{i}+\tilde{{\rm a}}\label{dX+}\\
{}\nonumber\\
\delta \tx_{-}=\tx_{-}{\rm b}\tx_{-}
-4\theta^{i}\bar{\rho}_{i}\tx_{-}+\lambda \tx_{-}+
\omega \tx_{-}-\tx_{-}\tilde{\omega}
-4i\theta^{i}\bar{\varepsilon}_{i}+\tilde{{\rm a}}\label{dX-}
\end{subeqnarray}
Furthermore, imposing the reality condition for $x^{A}$ and 
$\bar{\theta}_{i}=\theta^{i\dagger}\tilde{\gamma}^{0}$ gives the
 following extra conditions
\begin{equation}
\begin{array}{c}
a^{A*}=a^{A}~~~~b^{A*}=b^{A}~~~~\omega_{AB}{}^{*}=\omega_{AB}~~~~
\lambda^{*}=\lambda~~~~T^{\dagger}=-T\\
{}\\
\bar{\rho}_{i}=\rho^{i\dagger}\gamma^{0}~~~~~~~
\bar{\varepsilon}_{i}
=\varepsilon^{i\dagger}\tilde{\gamma}^{0}
\end{array}
\end{equation}
The solutions we obtained here read the generators of $(N,0)$ superconformal
transformations.

\subsection{$(N,0)$ Superconformal Transformations} 
In summary the generators of  $(N,0)$ superconformal transformations in
six-dimensions acting  on the  six-dimensional chiral  superspace, 
$\Real^{6|8N}_{+}$, with coordinates, $z^{{\cal
A}}=(x^{A},\theta^{i})$ can be reduced to 
\begin{enumerate}
\item  Supertranslations
\begin{equation}
\begin{array}{cc}
\delta x^{A}=a^{A}-i\bar{\theta}_{i}\gamma^{A}\varepsilon^{i}
~~~&~~~
\delta\theta^{i}=\varepsilon^{i}
\end{array}
\label{trans}
\end{equation}
\item Dilations
\begin{equation}
\begin{array}{cc}
\delta x^{A}=\lambda x^{A}~~~&~~~
\delta\theta^{i}=\textstyle{\frac{1}{2}}\lambda\theta^{i}
\end{array}
\end{equation}
\item Lorentz transformations, with $\omega$ defined in eq.(\ref{omega2}),
\begin{equation}
\begin{array}{cc}
\delta x^{A}=\omega^{A}_{~B}x^{B} ~~~&~~~ 
\displaystyle{\delta\theta^{i}=\omega \theta^{i}}
\end{array}
\end{equation}
\item $R$-symmetry transformations belonging to  
 $\mbox{Sp}(N)$, of dimension $N(2N+1)$,
\begin{equation}
\begin{array}{cc}
\delta x^{A}=0~~~&~~~
\delta\theta^{i}=-\theta^{j}T_{j}^{~i}
\end{array}
\end{equation}
where the $2N\times 2N$ matrix,~$T$, is a $\mbox{Sp}(N)$ generator, i.e. 
$T^{\dagger}=-T,~T^{t}{\cal E}+{\cal E}T=0$.
\item Special superconformal transformations
\begin{equation}
\begin{array}{l}
\delta x^{A}=2x{\cdot b}\,
x^{A}-x^{2}b^{A}+\bar{\theta}_{i}{\rm b}\theta^{j}\,\bar{\theta}_{j}
\gamma^{A}\theta^{i}+
\bar{\theta}_{i}\gamma^{A}\tx_{+}\rho^{i}\\
{}\\
\delta\theta^{i}=\tx_{-}{\rm b}\theta^{i}
+i\tx_{-}\rho^{i}
-4(\bar{\rho}_{j}\theta^{i})\theta^{j}
\end{array}
\label{special}
\end{equation}
\end{enumerate}

\subsection{$(N,\tilde{N}),~  N,\tilde{N} \neq 0$ Case}
To get the  generators of  $(N,\tilde{N})$ possible superconformal
transformations, where both  $N,\tilde{N} \neq 0$, 
we need to require $h^{A}$ to satisfy eq.(\ref{enough2}) as
well. \newline
$\{D_{i\alpha},\tilde{D}_{\tilde{\jmath}}^{\beta}\}=0$ implies
\begin{equation}
\tilde{D}_{\tilde{\jmath}}^{\beta}(\bar{\lambda}_{i}\gamma^{A})_{\alpha}=
-
D_{i\alpha}(\bar{\tilde{\lambda}}_{\tilde{\jmath}}
\tilde{\gamma}^{A})^{\beta}
\end{equation}
Contracting this with $\tilde{\gamma}_{A}^{\gamma\delta}$,  
using eqs.(\ref{contract},\,\ref{tildetilde}), leads to
\begin{equation}
2\tilde{D}^{\beta}_{\tilde{\jmath}}\bar{\lambda}_{i}^{[\gamma}
\delta_{\alpha}^{\delta]}
=D_{i\alpha}\bar{\tilde{\lambda}}_{\tilde{\jmath}\eta}
\epsilon^{\eta\beta\gamma\delta}
\end{equation}
Furthermore considering the contraction with $\delta_{\beta\gamma}$, 
it is easy to show $D_{i\alpha}\bar{\tilde{\lambda}}_{\tilde{\jmath}\beta}=
\tilde{D}_{\tilde{\imath}}^{\alpha}\bar{\lambda}_{j}^{\beta}=0$ and so
\begin{equation}
D_{i\alpha}\tilde{D}_{\tilde{\jmath}}^{\beta}h^{A}=0
\end{equation}
Consequently
\begin{equation}
\begin{array}{c}
D_{i\alpha}\partial_{B}h^{A}=\tilde{D}_{\tilde{\imath}}^{\alpha}
\partial_{B}h^{A}=0\\
{}\\
\partial_{B}\partial_{C}h^{A}=0
\end{array}
\end{equation}
which imply that if  $N,\tilde{N}\neq 0$ then  there are
 no special superconformal
transformations as in eq.(\ref{special}). \newline
The infinitesimal  transformations  satisfying eq.(\ref{scdef}) are just
\begin{enumerate}
\item  Supertranslations
\begin{equation}
\begin{array}{ccc}
\delta x^{A}=a^{A}-i\bar{\theta}_{i}\gamma^{A}\varepsilon^{i}
-i\bar{\tilde{\theta}}_{\tilde{\imath}}\tilde{\gamma}^{A}
\tilde{\varepsilon}^{\tilde{\imath}} ~~~&~~~
\delta\theta^{i}=\varepsilon^{i}~~~ &~~~
\delta\tilde{\theta}^{\tilde{\imath}}=\tilde{\varepsilon}^{\tilde{\imath}}
\end{array}
\label{strNN}
\end{equation}
\item Dilations
\begin{equation}
\begin{array}{ccc}
\delta x^{A}=\lambda x^{A}~~~&~~~
\delta\theta^{i}=\textstyle{\frac{1}{2}}\lambda\theta^{i}~~~&~~~
\delta\tilde{\theta}^{\tilde{\imath}}
=\textstyle{\frac{1}{2}}\lambda\tilde{\theta}^{\tilde{\imath}}
\end{array}
\end{equation}
\item Lorentz transformations
\begin{equation}
\begin{array}{ccc}
\delta x^{A}=\omega^{A}_{~B}x^{B} ~~~&~~~ 
\displaystyle{\delta\theta^{i}=\omega\theta^{i}}~~~&~~~
\displaystyle{
\delta\tilde{\theta}^{\tilde{\imath}}=\tilde{\omega}
\tilde{\theta}^{\tilde{\imath}
}}
\end{array}
\end{equation}
\item $R$-symmetry transformations, 
$\mbox{Sp}(N)\times \mbox{Sp}(\tilde{N})$
\begin{equation}
\begin{array}{ccc}
\delta x^{A}=0~~~&~~~
\delta\theta^{i}=-\theta^{j}T_{j}^{~i} ~~~&~~~
\delta\tilde{\theta}^{\tilde{\imath}}
=-
\tilde{\theta}^{\tilde{\jmath}}\tilde{T}_{\tilde{\jmath}}^{~\tilde{\imath}}
\end{array}
\end{equation}
where $T,\tilde{T}$ are $\mbox{Sp}(N),\mbox{Sp}(\tilde{N})$ 
generators respectively.
\end{enumerate}
The supersymmetric interval in this case, which is invariant under 
supertranslations\,(\ref{strNN}),  is given by
\begin{equation}
z^{{\cal A}}_{12}=
(x_{12}^{A},\theta^{i}_{12},\tilde{\theta}^{\tilde{\imath}}_{12})
=-z_{21}^{{\cal A}}
\end{equation}
where
\begin{equation}
\begin{array}{ccc}
x_{12}^{A}=x_{1}^{A}-x_{2}^{A}-i\bar{\theta}_{2i}\gamma^{A}\theta_{1}^{i}-
i\bar{\tilde{\theta}}_{2\tilde{\imath}}\tilde{\gamma}^{A}
\tilde{\theta}_{1}^{\tilde{\imath}}~~~~&~~~
\theta_{12}^{i}=\theta^{i}_{1}-\theta_{2}^{i}
~~~~&~~~\tilde{\theta}_{12}^{\tilde{\imath}}
=\tilde{\theta}_{1}^{\tilde{\imath}}-\tilde{\theta}_{2}^{\tilde{\imath}}
\end{array}
\label{susyint}
\end{equation}

\subsection{Superinversion and Spatial Reflection}
In this subsection we show how superinversion may be defined as a map
from  $\Real_{+}^{6|8N}$ superspace   to $\Real_{-}^{6|8N}$
superspace, where  $\Real_{-}^{6|8N}$ has coordinates
$\tilde{z}^{\tilde{{\cal A}}}=(y^{A},\phi^{i})$   
with $\phi^{i}_{\alpha}$ having the opposite chirality to  
$\theta^{i\alpha}$. \newline
For any $z^{{\cal A}}=(x^{A},\theta^{i})\in \Real^{6|8N}_{+}$   if 
we define  
$\y_{\pm},\,\phi^{i},\,\bar{\phi}_{i}$ by
\begin{equation}
\begin{array}{ccc}
\y_{\pm}=-\tx_{\pm}^{-1}~~~~&~~~
\phi^{i}=-i\tx_{-}^{-1}\theta^{i}~~~~&~~~
\bar{\phi}_{i}=i\bar{\theta}_{i}\tx_{+}^{-1}
\end{array}
\label{inversion}
\end{equation}
then from eqs.(\ref{pseudoM},\,\ref{Xprop}) these satisfy
\begin{equation}
\begin{array}{c}
\y_{+}-\y_{-}=\tx_{-}^{-1}(\tx_{+}-\tx_{-})\tx^{-1}_{+}
=4i\phi^{i}\bar{\phi}_{i}\\
{}\\
\phi{}^{i\dagger}\gamma^{0}
=\phi{}^{jt}\bar{\cal E}_{ji}
=\bar{\phi}_{i}
\label{consistency}
\end{array}
\end{equation}
Hence
\begin{equation}
{\rm y}=\textstyle{\frac{1}{2}}(\y_{+}+\y_{-})=-{\rm
y}^{t}=y^{A}\gamma_{A}
\end{equation}
and in consequence 
$\tilde{z}^{\tilde{{\cal A}}}=(y^{A},\phi^{i})\in
\Real^{6|8N}_{-}$.  We may therefore use
eq.(\ref{inversion}) to define superinversion   $z^{{\cal A}}
\stackrel{i}{\longrightarrow}
\tilde{z}^{\tilde{{\cal A}}}$  as a map, ~   
$\Real_{+}^{6|8N}\rightarrow\Real_{-}^{6|8N}$. \newline
From eq.(\ref{inversion}) the
inverse $i^{-1}$ is easily seen to be given by 
\begin{equation}
\begin{array}{ccc}
\tx_{\pm}=-\y_{\pm}^{-1}~~~~&~~~
\theta^{i}=-i\y_{-}^{-1}\phi^{i}~~~~&~~~
\bar{\theta}_{i}=i\bar{\phi}_{i}\,\y_{+}^{-1}
\end{array}
\label{inverseofi}
\end{equation}
From eq.(\ref{infint}) the infinitesimal supersymmetric
intervals,~$e^{A}(z),\,f^{A}(\tilde{z})$ for the  $\Real^{6|8N}_{+}$  and 
$\Real^{6|8N}_{-}$ 
superspaces   are respectively 
\begin{equation}
\begin{array}{cc}
e^{A}(z)={\rm d}x^{A}-i\bar{\theta}_{i}\gamma^{A}{\rm d}\theta^{i}~~~~&~~~
f^{A}(\tilde{z})
={\rm d}y^{A}-i\bar{\phi}_{i}\tilde{\gamma}^{A}{\rm 
d}\phi^{i}
\end{array}
\label{defef}
\end{equation}
Under superinversion, $e^{A}$   transforms as
\begin{equation}
{\rm f}(\tilde{z})=\tx^{-1}_{-}
\tilde{{\rm e}}(z)\tx^{-1}_{+}
\end{equation}
or equivalently  
\begin{equation}
\begin{array}{cc}
f^{A}(\tilde{z})=e^{B}(z)R_{B}^{~A}(z;i)~~~~&~~~~
R_{B}^{~A}(z;i)=\textstyle{\frac{1}{4}}\mbox{tr}(\tx^{-1}_{-}
\tilde{\gamma}_{B}\tx^{-1}_{+}\tilde{\gamma}^{A})
\end{array}
\label{Rsuperinversion}
\end{equation}
If we consider a transformation,$~\tilde{z}\,
\stackrel{i^{-1}}{\longrightarrow}\,z\,\stackrel{g}{\longrightarrow}\,
z^{\prime}\,\stackrel{i}{\longrightarrow}\,\tilde{z}^{\prime}$,  where
$g$ is a $(N,0)$ superconformal transformation, then  this becomes a
$(0,N)$ superconformal transformation $~\tilde{z}\,
\stackrel{\tilde{g}}{\longrightarrow}\,\tilde{z}^{\prime}$. To
demonstrate we consider the variations 
$\delta \y_{+},\,\delta \phi^{i}$ arising from the 
infinitesimal $(N,0)$ 
superconformal transformation~(\ref{dlambda},\,\ref{dX+}) 
\begin{equation}
\begin{array}{l}
\delta \y_{+}=\tx^{-1}_{+}\delta\tx_{+}\tx_{+}^{-1}=
\y_{+}\tilde{{\rm a}}\y_{+}
-4\y_{+}\varepsilon^{i}\bar{\phi}_{i}
-\lambda \y_{+}+\tilde{\omega} \y_{+}-\y_{+}\omega
+4i\rho^{i}\bar{\phi}_{i}+{\rm b}\\
{}\\
\delta \phi{}^{i}=
\rho^{i}-{\textstyle\frac{1}{2}}\lambda\phi^{i}
+\tilde{\omega}\phi^{i}-
\phi^{j}T_{j}^{~i}+\y_{-}\tilde{{\rm a}}
\phi^{i}+i\y_{-}\varepsilon^{i}
-4(\bar{\varepsilon}_{j}\phi^{i})
\phi^{j}
\end{array}
\end{equation}
Hence, under  superinversion, the superconformal transformations
are related by 
\begin{equation}
\begin{array}{ccc}
\left(\begin{array}{c}
a\\
b\\
\varepsilon^{i}\\
\rho^{i}\\
\lambda\\
\omega_{AB}\\
T_{i}^{~j}
\end{array}\right)~~~~& \longrightarrow &~~~~
\left(\begin{array}{c}
b\\
a\\
\rho^{i}\\
\varepsilon^{i}\\
-\lambda\\
\omega_{AB}\\
T_{i}^{~j}
\end{array}\right)
\end{array}
\label{resulti}
\end{equation}
which manifestly shows a duality between $(N,0)$ and $(0,N)$
superconformal transformations. In particular, $(N,0)$ special 
superconformal
transformations~(\ref{special}) can be obtained by 
$z\stackrel{i}{\longrightarrow}
\tilde{z}\stackrel{(b,\rho)}{-\!\!\!-\!\!\!-\!\!\!\longrightarrow}
\tilde{z}^{\prime}\stackrel{i^{-1}}{\longrightarrow}z^{\prime}$ where
$(b,\rho)$ is a supertranslation on $\Real^{6|8N}_{-}$ superspace.\newline
We may also define a reflection map, 
$r_{B}~\,B\neq 0$, from $\Real^{6|8N}_{+}$
superspace to  $\Real^{6|8N}_{-}$ superspace so that $z^{{\cal A}}
\stackrel{r_{B}}{\longrightarrow}
\tilde{z}^{\tilde{{\cal A}}}$ by 
\begin{equation}
\begin{array}{ccc}
y^{A}=\left\{\begin{array}{cl}
            x^{A}~&~A\neq B\\
            -x^{A}~&~A=B
              \end{array}\right.~~&\mbox{or}~~&
{\rm y}=\gamma^{B}\tilde{{\rm x}}\gamma^{B}
\end{array}
\label{rB1}
\end{equation}
and, introducing in addition a $\mbox{Sp}(N)$ transformation, $R$,
\begin{equation}
\begin{array}{cc}
\phi^{i}=i\gamma^{B}\theta^{j}R_{j}{}^{i}~~~&~~~
\bar{\phi}_{i}=-i(R^{-1})_{i}{}^{j}\bar{\theta}_{j}\gamma^{B}
\end{array}
\label{rB2}
\end{equation}
This gives with the definitions~(\ref{defef})
\begin{equation}
{\rm f}(\tilde{z})=\gamma^{B}\tilde{{\rm e}}(z)\gamma^{B}
\end{equation}
We may now define a map $z
\stackrel{i}{\longrightarrow}
\tilde{z}\stackrel{r^{-1}_{B}}{-\!\!\!\longrightarrow}z^{\prime}$ or 
$z\stackrel{i_{B}}{-\!\!\!\longrightarrow}z^{\prime}$ which leads
\begin{equation}
\begin{array}{ccc}
\tx_{\pm}^{\prime}=-\tilde{\gamma}^{B}\tx_{\pm}^{-1}
\tilde{\gamma}^{B}~~~~&~~~
\theta^{\prime}{}^{i}=\tilde{\gamma}^{B}\tx_{-}^{-1}\theta^{j}
(R^{-1})_{j}{}^{i}~~~~&~~~
\bar{\theta}_{i}^{\prime}=R_{i}{}^{j}\bar{\theta}_{j}
\tx_{+}^{-1}\tilde{\gamma}^{B}
\end{array}
\label{iB}
\end{equation}
It is easy to see since
$\tx_{\pm}^{\prime}{}^{-1}=-\gamma^{B}\tx_{\pm}\gamma^{B}$ that
\begin{equation}
(x^{A},\theta^{i})\stackrel{i^{2}_{B}}{\longrightarrow}
(x^{A},-\theta^{j}(R^{2})^{-1}{}_{j}{}^{i})
\end{equation}
If we choose $R$ so that
\begin{equation}
R^{2}=-1
\label{Rsquare}
\end{equation}
then
\begin{equation}
i_{B}^{2}=1
\end{equation}
\subsection{$(N,0)$ Superconformal Algebra}
The generator of  infinitesimal $(N,0)$ 
superconformal transformations,~${\cal L}$,
is given by
\begin{equation}
{\cal L}
=h^{A}\partial_{A}+\lambda^{i\alpha}D_{i\alpha}
\label{calL}
\end{equation}
If we write the commutator of two generators,~${\cal L}_{1},{\cal
L}_{2}$  as
\begin{equation}
[{\cal L}_{2},{\cal L}_{1}]={\cal L}_{3}
=h^{A}_{3}\partial_{A}+\lambda^{i\alpha}_{3}D_{i\alpha}
\label{hcommutator}
\end{equation}
then $h_{3}^{A},~\lambda^{i\alpha}_{3}$ are given by 
\begin{equation}
\begin{array}{l}
h^{A}_{3}=h^{B}_{2}\partial_{B}h^{A}_{1}
-i\bar{\lambda}_{2i}\gamma^{A}\lambda^{i}_{1}-
(1\leftrightarrow 2)\\
{}\\
\lambda^{i\alpha}_{3}=h^{A}_{2}\partial_{A}\lambda^{i\alpha}_{1}
+\lambda^{j\beta}_{2}D_{j\beta}\lambda^{i\alpha}_{1}-
(1\leftrightarrow 2)
\end{array}
\end{equation}
and $h^{A}_{3},~\lambda^{i\alpha}_{3}$ satisfy eq.(\ref{enough1})
verifying the closure of the Lie algebra.\newline
Explicitly we get\footnote{From eq.(\ref{MMcom}) we can read off the
six-dimensional $(N,0)$ superconformal algebra. The explicit
form of it is given in  Appendix C.}
\begin{equation}
\begin{array}{l}
\displaystyle{a^{A}_{3}={\omega^{A}_{1}}_{B}a^{B}_{2}+\lambda_{1}a^{A}_{2}+
i\bar{\varepsilon}_{1i}\gamma^{A}\varepsilon_{2}^{i} -
(1\leftrightarrow 2)}\\
{}\\
\varepsilon^{i}_{3}=
\omega_{1}\varepsilon^{i}_{2}+
\textstyle{\frac{1}{2}}\lambda_{1}\varepsilon^{i}_{2}+
i\tilde{\rm a}_{2}\rho^{i}_{1}-\varepsilon^{j}_{2}{T_{1j}}^{i}-
(1\leftrightarrow 2)\\
{}\\
\displaystyle{\lambda_{3}=2a_{2}{\cdot b_{1}}
+2\bar{\rho}_{1i}\varepsilon^{i}_{2}-
(1\leftrightarrow 2)}\\
{}\\
\displaystyle{\omega^{AB}_{3}=\omega^{A}_{1C}\omega^{CB}_{2}
+2(a_{2}^{A}b_{1}^{B}-a_{2}^{B}b_{1}^{A})+
2\bar{\varepsilon}_{2i}\gamma^{[A}\tilde{\gamma}^{B]}
\rho^{i}_{1}-(1\leftrightarrow 2)}\\
{}\\
b^{A}_{3}={\omega^{A}_{1}}_{B}b^{B}_{2}-\lambda_{1}b^{A}_{2}+
i\bar{\rho}_{1i}\tilde{\gamma}^{A}\rho_{2}^{i} -
(1\leftrightarrow 2)\\
{}\\
\rho^{i}_{3}=\tilde{\omega}_{1}\rho^{i}_{2}-
\frac{1}{2}\lambda_{1}\rho^{i}_{2}+i{\rm b}_{2}\varepsilon^{i}_{1}-
\rho^{j}_{2}{T_{1j}}^{i}-(1\leftrightarrow 2)\\
{}\\
\displaystyle{T_{3}{}_{i}{}^{j}=(T_{1}T_{2})_{i}^{~j}
+4(\bar{\varepsilon}_{1i}\rho^{j}_{2}+\bar{\rho}_{1i}\varepsilon^{j}_{2})
-(1\leftrightarrow 2)}
\end{array}
\label{MMcom}
\end{equation}
Now, if we define a $(8+2N)\times (8+2N)$ matrix,~$M$ as 
\begin{equation}
\displaystyle{
M=\left(\begin{array}{ccc}
        \omega +\frac{1}{2}\lambda & 
        -i\tilde{{\rm a}}&2\varepsilon^{j}\\
       -i{\rm b} & 
       \tilde{\omega}-\frac{1}{2}\lambda &
         2\rho^{j}\\
       2\bar{\rho}_{i}&2\bar{\varepsilon}_{i}&T^{~j}_{i}
        \end{array}\right)}
\label{Mform}
\end{equation}
then the relation above~(\ref{MMcom}) agrees with the matrix commutator  
\begin{equation}
[M_{1},M_{2}]=M_{3}
\label{Mcommutator}
\end{equation}
This can be verified from eqs.(\ref{contract},\,\ref{complete}) using
\begin{equation}
\begin{array}{c}
\bar{\varepsilon}_{1i}\gamma^{A}\varepsilon^{i}_{2}\tilde{\gamma}_{A}=
2(\varepsilon_{1}^{i}\bar{\varepsilon}_{2i}-\varepsilon_{2}^{i}
\bar{\varepsilon}_{1i})\\
{}\\
\rho^{i}_{1}\bar{\varepsilon}_{2i}=\textstyle{\frac{1}{8}}
\bar{\varepsilon}_{2i}\gamma^{[A}\tilde{\gamma}^{B]}\rho_{1}^{i}
\gamma_{[A}\tilde{\gamma}_{B]}-\textstyle{\frac{1}{4}}
\bar{\varepsilon}_{2i}\rho^{i}_{1}1
\end{array}
\end{equation} 
In general, $M$ can be defined as a $(8,2N)$ supermatrix subject to the two
conditions\footnote{We define the transpose of a supermatrix
as $\left(\begin{array}{cc}
a&b\\
c& d\end{array}\right)^{t}=\left(\begin{array}{cc}
a^{t}&c^{t}\\
-b^{t}& d^{t}\end{array}\right)$.}
\begin{equation}
BMB^{-1}=-M^{\dagger}
\label{Mdagger}
\end{equation}
\begin{equation}
CMC^{-1}=-M^{t}
\label{Mt}
\end{equation}
where
\begin{equation}
\begin{array}{cc}
B=B^{-1}=\left(\begin{array}{ccc}
0&\tilde{\gamma}^{0}&0\\
\gamma^{0}&0&0\\
0&0&-1\end{array}\right),~~~~&~~~~
C=\left(\begin{array}{ccc}
0&1&0\\
1&0&0\\
0&0&-{\cal E}\end{array}\right)
\end{array}
\end{equation}
The $8\times 8$ matrix appearing in $M$
\begin{equation}
\left(\begin{array}{cc}
\omega+\textstyle{\frac{1}{2}}\lambda &-i\ta\\
-i{\rm b} & \tilde{\omega}-\textstyle{\frac{1}{2}}\lambda
\end{array}\right)
\end{equation}
corresponds to a generator of $\mbox{O}(2,6)$ as demonstrated in 
 Appendix D. Thus, 
the $(N,0)$ superconformal group in six-dimensions may be identified
with the matrix group generated by matrices of the form $M$, which is
$\mbox{OSp}(2,6|N)\equiv\mbox{G}_{s}$ having dimensions\newline
$(28+N(2N+1)|16N)$.

\section{Coset Realization of  Transformations}
To obtain an explicit  formula for the finite superconformal
transformations, we first identify the superspace, $\Real^{6|8N}_{+}$,
as a coset, 
$\mbox{G}_{s}/\mbox{G}_{0}$,
where $\mbox{G}_{0}\subset \mbox{G}_{s}$ is the subgroup generated by
matrices $M_{0}$
of the form~(\ref{Mform}) with $a^{A}=0,\,\varepsilon^{i}=0$ and
depending on 
parameters $b^{A},\,\rho^{i},\,
\lambda,\,\omega_{AB},\,T_{i}{}^{j}$.\newline
The group of supertranslations, $\mbox{G}_{T}$,  
is parameterized by coordinates, 
$z^{{\cal A}}\in\Real^{6|8N}_{+}$, and  is  defined in terms of
supermatrices\footnote{The subscript, $T$, denotes supertranslations.}
\begin{equation}
G_{T}(z)=\mbox{exp}\left(\begin{array}{ccc}
               0&-i\tilde{{\rm x}}&2\theta^{j}\\
               0&0&0\\
               0&2\bar{\theta}_{i}&0
                \end{array}\right)=
\left(\begin{array}{ccc}
      1&  -i\tx_{+}&2\theta^{j}\\
      0&  1&0\\
      0&  2\bar{\theta}_{i}&\delta_{i}^{~j}
        \end{array}\right)
\end{equation}
where $\tx_{+}$ is given in eq.(\ref{tildeX}).\newline
We may write
\begin{equation}
\begin{array}{cc}
G_{T}(z_{2})^{-1}G_{T}(z_{1})=G_{T}(z_{12})~~~~&~~~~G_{T}(z)^{-1}=G_{T}(-z)
\end{array}
\end{equation}
where 
\begin{equation}
\begin{array}{cc}
\multicolumn{2}{c}{z^{{\cal A}}_{12}=(x^{A}_{12},\theta^{i}_{12})=
-z^{{\cal A}}_{21}}\\
{}&{}\\
x_{12}^{A}=x_{1}^{A}-x_{2}^{A}-i\bar{\theta}_{2i}\gamma^{A}\theta^{i}_{1}
~~~~&~~~~\theta^{i}_{12}=\theta^{i}_{1}-\theta^{i}_{2}
\end{array}
\label{susyinterval+}
\end{equation}
which defines supersymmetric interval  for $\Real^{6|8N}_{+}$
superspace, as in eq.(\ref{susyint}).\newline
In general an element of $\mbox{G}_{s}$  can be uniquely decomposed  as
$G_{T}G_{0}$. Thus for any  element $G(g)\in\mbox{G}_{s}$ we may define a
superconformal transformation, $z \stackrel{g}{\longrightarrow}
z^{\prime}$, and an associated element $G_{0}(z;g)\in\mbox{G}_{0}$ by
\begin{equation}
G(g)^{-1}G_{T}(z)G_{0}(z;g)=G_{T}(z^{\prime})
\label{GTFINITE}
\end{equation}
If $G(g)\in\mbox{G}_{T}$ then clearly $G_{0}(z;g)=1$.\newline
Infinitesimally eq.(\ref{GTFINITE}) becomes
\begin{equation}
\delta G_{T}(z)=MG_{T}(z)-G_{T}(z)\hat{M}_{0}(z) 
\label{MGL}
\end{equation}
If  $M$ is given by eq.(\ref{Mform}) then letting  
$\delta G_{T}(z)={\cal L}G_{T}(z)$ we may verify that ${\cal
L}$ is identical with the form given by eq.(\ref{calL}) with  
eqs.(\ref{solmaster},\,\ref{dlambda}) and further  
$\hat{M}_{0}(z)$, the generator of $\mbox{G}_{0}$,  has the form 
\begin{equation}
\displaystyle{
\hat{M}_{0}(z)=\left(\begin{array}{ccc}
        \hat{\omega}(z) +\frac{1}{2}\hat{\lambda}(z) & 
        0&0\\
       -i{\rm b} & 
       \tilde{\hat{\omega}}(z)-\frac{1}{2}\hat{\lambda}(z) &
         2\hat{\rho}^{j}(z)\\
       2\bar{\hat{\rho}}_{i}(z)&0&\hat{T}^{~j}_{i}(z)
        \end{array}\right)}
\end{equation}
where the elements of $\hat{M}_{0}(z)$ depending on $z$ are given by 
\begin{equation}
\begin{array}{l}
\hat{\omega}(z)=\omega-4\theta^{i}\bar{\rho}_{i}+\tx_{-}{\rm b}
+\textstyle{\frac{1}{4}}
\mbox{tr}(4\theta^{i}\bar{\rho}_{i}-\tx_{-}{\rm b})\,1 \\
{}\\
\tilde{\hat{\omega}}(z)
=\tilde{\omega}+4\rho^{i}\bar{\theta}_{i}-{\rm b}\tx_{+}
-\textstyle{\frac{1}{4}}
\mbox{tr}(4\rho^{i}\bar{\theta}_{i}-{\rm b}\tx_{+})\,1 
=-\hat{\omega}(z)^{t}        \\
{}\\
\hat{\lambda}(z)=\lambda+2b{\cdot x}+2\bar{\theta}_{i}\rho^{i}
=\frac{1}{6}\partial_{A}h^{A}(z)\\
{}\\
\hat{T}^{~j}_{i}(z)=T^{~j}_{i}+4i\bar{\theta}_{i}{\rm b}\theta^{j}+4
(\bar{\rho}_{i}\theta^{j}-\bar{\theta}_{i}\rho^{j})\\
{}\\
\hat{\rho}^{j}(z)=\rho^{j}-i{\rm b}\theta^{j}=-i\textstyle{\frac{1}{6}}
\gamma^{A}\partial_{A}\lambda^{j}(z)
\end{array}
\label{hats}
\end{equation}
$\hat{\omega}(z),\,\tilde{\hat{\omega}}(z)$ can be also written as 
$\hat{\omega}(z)
=\textstyle{\frac{1}{4}}
\hat{\omega}_{AB}(z)\tilde{\gamma}^{A}\gamma^{B},\,
\tilde{\hat{\omega}}(z)
=\textstyle{\frac{1}{4}}
\hat{\omega}_{AB}(z)\gamma^{A}\tilde{\gamma}^{B}$ where
\begin{equation}
\hat{\omega}_{AB}(z)=\omega_{AB}+4x_{[A}b_{B]}+\bar{\theta}_{i}
\gamma_{[A}\tilde{\gamma}_{B]}(2\rho^{i}-i{\rm b}\theta^{i})
=-\partial_{[A}h_{B]}(z)
\end{equation}
The definitions\,(\ref{hats}) may be summarized by 
\begin{equation}
D_{i\alpha}\lambda^{j\beta}(z)=
\delta_{i}^{~j}
(\textstyle{\frac{1}{2}}\hat{\lambda}(z)\delta^{~\beta}_{\alpha}
-\tilde{\hat{\omega}}_{\alpha}{}^{\beta}(z))-
\hat{T}^{~j}_{i}(z)\delta^{~\beta}_{\alpha}
\end{equation}
and they  give
\begin{equation}
[D_{i\alpha},{\cal L}]=\textstyle{\frac{1}{2}}\hat{\lambda}(z)D_{i\alpha}
-\tilde{\hat{\omega}}_{\alpha}{}^{\beta}(z)
D_{i\beta}-\hat{T}_{i}^{~j}(z)D_{j\alpha}
\label{comDL}
\end{equation}
For later use we note 
\begin{equation}
\begin{array}{l}
D_{i\alpha}\tilde{\hat{\omega}}_{\beta}{}^{\gamma}(z)
=4\delta_{\alpha}^{~\gamma}
\bar{\hat{\rho}}_{i\beta}(z)-\delta_{\beta}^{~\gamma}
\bar{\hat{\rho}}_{i\alpha}(z)\\
{}\\
D_{i\alpha}\hat{\lambda}(z)=-2\bar{\hat{\rho}}_{i\alpha}(z)\\
{}\\
D_{i\alpha}\hat{T}_{j}^{~k}(z)=-4\delta_{i}^{~k}
\bar{\hat{\rho}}_{j\alpha}(z)-4\bar{\cal E}_{ij}\hat{\rho}{}^{k}_{\alpha}(z)
\end{array}
\label{Ddelta}
\end{equation}
The above analysis  can be simplified by reducing $G_{0}(z;g)$. To
achieve this we let 
\begin{equation}
Z_{0}=\left(\begin{array}{cc}
              0&0\\
              1&0\\
              0&1
            \end{array}\right)
\end{equation}
and then
\begin{equation}
\begin{array}{cc}
M_{0}Z_{0}=Z_{0}H_{0}~~~~&~~~~
H_{0}=
\left(\begin{array}{cc}
           \tilde{\omega}-\textstyle{\frac{1}{2}}\lambda & 2\rho^{j}\\
                             0&T_{i}^{~j}                
                          \end{array}\right)
\end{array}
\label{M0Z0}
\end{equation}
Now if we define
\begin{equation}
Z(z)\equiv G_{T}(z)Z_{0}=
\left(\begin{array}{cc}
        -i\tx_{+}&2\theta^{j}\\
        1&0\\
        2\bar{\theta}_{i}&\delta_{i}^{~j}
        \end{array}\right)
\label{defZ}
\end{equation}
then $Z(z)$ transforms  
under infinitesimal  superconformal transformations as
\begin{equation}
\delta Z(z)={\cal L}Z(z)=MZ(z)-Z(z)H(z) 
\label{infZ}
\end{equation}
where $H(z)$ is given by  
\begin{equation}
\begin{array}{cc}
\hat{M}_{0}(z)Z_{0}=Z_{0}H(z)~~~~&~~~~
H(z)=
\left(\begin{array}{cc}
\tilde{\hat{\omega}}(z)-\frac{1}{2}\hat{\lambda}(z)&2\hat{\rho}^{j}(z)\\
0&\hat{T}_{i}^{~j}(z)
\end{array}\right)
\end{array}
\label{defH}
\end{equation}
From eqs.(\ref{hcommutator},\,\ref{Mcommutator}) we have
\begin{equation}
H_{3}(z)={\cal L}_{2}H_{1}(z)-{\cal L}_{1}H_{2}(z)+[H_{1}(z),H_{2}(z)]
\label{Kcom}
\end{equation}
which gives separate equations for
$\tilde{\hat{\omega}},\hat{\lambda},\hat{\rho}^{i}$ 
and $\hat{T}_{i}^{~j}$, thus 
$\hat{\lambda}_{3}={\cal L}_{2}\hat{\lambda}_{1}-{\cal
L}_{1}\hat{\lambda}_{2}$, etc.\newline
As a conjugate of  $Z(z)$ we   define  $\bar{Z}(z)$ by
\begin{equation}
\bar{Z}(z)=\left(\begin{array}{cc}
               \tilde{\gamma}^{0}&0\\
                0&-1 \end{array}\right)Z(z){}^{\dagger}B=
              \left(\begin{array}{cc}
               1&0\\
               0&-\bar{\cal E} \end{array}\right) Z(z){}^{t}C
=\left(\begin{array}{ccc}
                 1&i\tx_{-}&-2\theta^{j}\\
                 0&-2\bar{\theta}_{i}&\delta_{i}^{~j}\end{array}\right)
\label{defbarZ}
\end{equation}
This satisfies
\begin{equation}
\bar{Z}(z)=\bar{Z}(0)G_{T}(z)^{-1}
\end{equation}
and corresponding to eq.(\ref{infZ}) we have
\begin{equation}
\delta \bar{Z}(z)={\cal L}\bar{Z}(z)
=\bar{H}(z)\bar{Z}(z)-\bar{Z}(z)M 
\label{infbarZ}
\end{equation}
where
\begin{equation}
\bar{H}(z)=\left(\begin{array}{cc}
\hat{\omega}(z)+\textstyle{\frac{1}{2}}\hat{\lambda}(z)&0\\
2\bar{\hat{\rho}}_{i}(z)&\hat{T}_{i}^{~j}(z)
\end{array}\right)
\end{equation}
\subsection{Functions of Two Points}
To consider functions of two points, $z_{1},z_{2}$, 
which transform covariantly 
under superconformal transformations, we first define   
$F(z)$ for $z\in\Real^{6|8N}_{+}$ by
\begin{equation}
F(z)=\left(\begin{array}{cc}
              -i\tx_{+}&2\theta^{j}\\
           2\bar{\theta}_{i}& \delta_{i}^{~j}\end{array}\right)
\label{Fform}
\end{equation}
This   satisfies the conditions
\begin{equation}
\begin{array}{ll}
F(-z)&=\left(\begin{array}{cc}
\tilde{\gamma}^{0}&0\\
0&-1
\end{array}\right)F(z)^{\dagger}\left(\begin{array}{cc}
\tilde{\gamma}^{0}&0\\
0&-1
\end{array}\right)=
\left(\begin{array}{cc}
1&0\\
0&-\bar{{\cal E}}
\end{array}\right)F(z)^{t}\left(\begin{array}{cc}
1&0\\
0&-{\cal E}
\end{array}\right)\\
{}&{}\\
{}&=\left(\begin{array}{cc}
              i\tx_{-}&-2\theta^{j}\\
           -2\bar{\theta}_{i}& \delta_{i}^{~j}\end{array}\right)
\label{-Fform}
\end{array}
\end{equation}
and the superdeterminant of $F(z)$ is given by
\begin{equation}
\begin{array}{cl}
\mbox{sdet}\:F(z)&=\det(-i\tx_{+}-
4\theta^{i}\bar{\theta}_{i})=\det(-i\tx_{-})\\
{}&{}\\
{}&=\det\tx_{+}=\det\tx_{-}
\end{array}
\label{sdetZZ}
\end{equation}
If we consider 
\begin{equation}
\left(\begin{array}{cc}
               1&0\\
                -2i\bar{\theta}_{i}\tx_{+}^{-1}&1
\end{array}\right)F(z)
\left(\begin{array}{cc}
               1&-2i\tx_{+}^{-1}\theta^{j}\\
               0&1
\end{array}\right)=
\left(\begin{array}{cc}
               -i\tx_{+}&0\\
               0&V_{i}{}^{j}(-z)
\end{array}\right)
\label{XV}
\end{equation}
then this defines $V_{i}{}^{j}(z)$ as 
\begin{equation}
V_{i}{}^{j}(z)=\delta_{i}^{~j}+4i\bar{\theta}_{i}
\tx^{-1}_{-}\theta^{j}
\label{Vzz}
\end{equation}
From eqs.(\ref{sdetZZ},\,\ref{XV}) it is evident that
\begin{equation}
\det V(z)=1
\end{equation}
and from eq.(\ref{Xprop}) $V_{i}{}^{j}(z)$ also satisfies 
eq.(\ref{Uprop}) 
and hence $V_{i}{}^{j}(z)\in \mbox{Sp}(N)$. Furthermore we have
\begin{equation}
V_{i}{}^{j}(-z)=V^{-1}{}_{i}{}^{j}(z)
=V^{\dagger}{}_{i}{}^{j}(z)=\delta_{i}^{~j}-
4i\bar{\theta}_{i}\tx_{+}^{-1}\theta^{j}
\end{equation}
We may also show from eq.(\ref{XV})
\begin{equation}
F(z)^{-1}=\left(\begin{array}{cc}
i\tx_{-}^{-1}&-2i\tx_{-}^{-1}\theta^{j}\\
-2i\bar{\theta}_{i}\tx_{-}^{-1}&V_{i}^{~j}(z)
\end{array}\right)
\label{Finverse}
\end{equation}
Using the definition of $F(z)$~(\ref{Fform}), 
we may now write from eqs.(\ref{defZ},\,\ref{defbarZ})
\begin{equation}
\displaystyle{\bar{Z}(z_{2})Z(z_{1})=F(z_{12})=
              \left(\begin{array}{cc}
              i{\cal X}_{21}&-2\theta_{21}^{j}\\
                -2\bar{\theta}_{21i}& \delta_{i}^{~j}\end{array}\right)}
\label{ZZ}
\end{equation}
where from eq.(\ref{tildeX})
\begin{equation}
\begin{array}{ll}
{\cal X}_{21}&=\tx_{2-}-\tx_{1+}
+4i\theta^{i}_{2}\bar{\theta}_{1i}\\
{}&{}\\
{}&=\tx_{21}-2i\theta^{i}_{21}\bar{\theta}_{21i}=(\tx_{21})_{-}
=-(\tx_{12})_{+}
\label{X1221}
\end{array}
\end{equation}
From eq.(\ref{Xprop}) we also have
\begin{equation}
\begin{array}{cc}
{\cal X}_{21}^{t}={\cal X}_{12}~~~~&~~~~
{\cal X}_{21}^{\dagger}=-\gamma^{0}{\cal X}_{12}\gamma^{0}
\end{array}
\label{X12dagger}
\end{equation}
where
\begin{equation}
{\cal X}_{12}=\tx_{12}-2i\theta^{i}_{12}\bar{\theta}_{12i}=-{\cal X}_{21}
-4i\theta_{21}^{i}\bar{\theta}_{21i}
\end{equation}
Infinitesimally,   from eqs.(\ref{infZ},\,\ref{infbarZ}), $F(z_{12})$
transforms as
\begin{equation}
\delta F(z_{12})=\bar{H}(z_{2})F(z_{12})-F(z_{12})H(z_{1})
\label{infF}
\end{equation}
For consistency with the form~(\ref{ZZ}) it is necessary that
\begin{equation}
\hat{T}_{i}^{~j}(z_{2})-\hat{T}_{i}^{~j}(z_{1})
-4\bar{\hat{\rho}}_{i}(z_{2})\theta^{j}_{21}+
4\bar{\theta}_{21i}\hat{\rho}^{j}(z_{1})=0
\label{cons1}
\end{equation}
In particular, we have from eq.(\ref{infF})
\begin{equation}
\delta{\cal X}_{21}=\hat{\omega}(z_{2}){\cal X}_{21}
-{\cal X}_{21}\tilde{\hat{\omega}}(z_{1})+
\textstyle{\frac{1}{2}}(\hat{\lambda}(z_{1})+\hat{\lambda}(z_{2}))
{\cal X}_{21}
\end{equation}
and hence 
\begin{equation}
\mbox{sdet}\:F(z_{12})=\det{\cal X}_{12}=\det{\cal X}_{21}
\end{equation}
which is  a  scalar depending on $z_{1},z_{2}$  transforming homogeneously 
\begin{equation}
\delta\det{\cal X}_{12}=2(\hat{\lambda}(z_{1})+\hat{\lambda}(z_{2}))
\det{\cal X}_{12}
\end{equation}
From eq.(\ref{Finverse}) we get
\begin{equation}
F(z_{12})^{-1}=\left(\begin{array}{cc}
i{\cal X}_{12}^{-1}&-2i{\cal X}_{12}^{-1}\theta_{12}^{j}\\
-2i\bar{\theta}_{12i}{\cal X}_{12}^{-1}&V_{i}^{~j}(z_{12})
\end{array}\right)
\label{Finv12}
\end{equation}
From eq.(\ref{infF}) $F(z_{12})^{-1}$ transforms as 
\begin{equation}
\delta
F(z_{12})^{-1}=H(z_{1})F(z_{12})^{-1}-F(z_{12})^{-1}\bar{H}(z_{2})
\end{equation}
Consistency with the form~(\ref{Finv12}) requires
\begin{equation}
\tilde{\hat{\omega}}(z_{1})-\textstyle{\frac{1}{2}}\hat{\lambda}(z_{1})-
4\hat{\rho}^{i}(z_{1})\bar{\theta}_{12i}+{\rm b}{\cal X}_{12}=
\tilde{\hat{\omega}}(z_{2})-\textstyle{\frac{1}{2}}\hat{\lambda}(z_{2}) 
\label{cons2}
\end{equation}
$V_{i}{}^{j}(z_{21})$ transforms infinitesimally as
\begin{equation}
\delta V(z_{21})=\hat{T}(z_{2})V(z_{21})-V(z_{21})\hat{T}(z_{1})
\label{infV}
\end{equation}

\subsection{Finite Transformations}
Finite superconformal transformations are  obtained by exponentiation
of infinitesimal transformations. To obtain a superconformal
transformation $z\stackrel{g}{\longrightarrow}z^{\prime}$, we
therefore solve the differential equation
\begin{equation}
\begin{array}{ccc}
\displaystyle{
\frac{{\rm d}~}{{\rm d}s}z^{{\cal A}}_{s}={\cal L}^{{\cal
A}}(z_{s})}~~~~&~~~~z_{0}=z~~~&~~~z_{1}=z^{\prime}
\end{array}
\end{equation}
where, with ${\cal L}$ given in eq.(\ref{calL}), ${\cal L}^{{\cal
A}}(z)$ is defined by 
\begin{equation}
{\cal L}={\cal L}^{{\cal A}}(z)\partial_{{\cal A}}
\end{equation}
From eq.(\ref{infZ}) we get
\begin{equation}
\displaystyle{
\frac{{\rm d}~}{{\rm d}s}Z(z_{s})=MZ(z_{s})-Z(z_{s})H(z_{s}) }
\end{equation}
which integrates to
\begin{equation}
Z(z_{s})=e^{sM}Z(z)K(z,s)
\label{finites}
\end{equation}
where $K(z,s)$ satisfies
\begin{equation}
\begin{array}{cc}
\displaystyle{
\frac{{\rm d}~}{{\rm d}s}K(z,s)=-K(z,s)H(z_{s})}~~~~&~~~~
K(z,0)=\left(\begin{array}{cc}
               1&0\\
               0&1
             \end{array}\right)
\end{array}
\end{equation}
Hence for $s=1$ with $K(z,1)\equiv K(z;g)$  the superconformal
transformation,~$z\stackrel{g}{\longrightarrow}z^{\prime}$, from 
eq.(\ref{finites}) becomes
\begin{equation}
\begin{array}{cc}
Z(z^{\prime})=G(g)^{-1}Z(z)K(z;g)~~~~&~~~~G(g)^{-1}=e^{M}
\end{array}
\label{Ztr}
\end{equation}
$K(z;g)$ is a representation of the superconformal group, since under  
the successive superconformal transformations, 
$z\stackrel{g}{\longrightarrow}z^{\prime}
\stackrel{g^{\prime}}{\longrightarrow}z^{ \prime\prime}$ giving 
$z\stackrel{g^{\prime\prime}}{\longrightarrow}z^{\prime\prime}$, we have
\begin{equation}
K(z;g)K(z^{\prime};g^{\prime})=K(z;g^{\prime\prime})
\label{Krep}
\end{equation}
In general  $K(z;g)$ is of the form
\begin{equation}
K(z;g)=\left(\begin{array}{cc}
                L(z;g)&2\Sigma^{j}(z;g)\\
                 0&U_{i}^{~j}(z;g)
                \end{array}\right)
\label{Kform}
\end{equation}
Since $\frac{\rm{d}~}{\rm{d}s}U(z,s)=-U(z,s)\hat{T}(z_{s})$ and
$\hat{T}(z)$ satisfies the conditions
\begin{equation}
\begin{array}{ccc}
\hat{T}^{t}{\cal E}+{\cal
E}\hat{T}=0~~~~&~~~~\hat{T}^{\dagger}=-\hat{T}
~~~~&~~~~\mbox{tr}(\hat{T})=0
\end{array}
\end{equation}
$U\in \mbox{Sp}(N)$, i.e. 
\begin{equation}
\begin{array}{ccc}
U^{t}{\cal E} U={\cal E} ~~~~&~~~U^{\dagger}=U^{-1}~~~~&~~~ \det U=1
\end{array}
\label{Uprop}
\end{equation}
Similarly $L$ satisfies
\begin{equation}
L^{\dagger}=\gamma^{0}L^{t}\tilde{\gamma}^{0}
\label{Lreality}
\end{equation}
$G_{0}(z;g)$ in eq.(\ref{GTFINITE}) is related to 
$K(z;g)$ from eq.(\ref{Ztr})  by
\begin{equation}
G_{0}(z;g)Z_{0}=Z_{0}K(z;g)
\end{equation}
From eq.(\ref{Ztr}) $\bar{Z}$ transforms as
\begin{equation}
\bar{Z}(z^{\prime})=\bar{K}(z;g)\bar{Z}(z)G(g)
\label{barZtr}
\end{equation}
where 
\begin{equation}
\begin{array}{cl}
\bar{K}(z;g)&=\left(\begin{array}{cc}
               \tilde{\gamma}^{0}&0\\
                0&-1 \end{array}\right)K(z;g){}^{\dagger}
\left(\begin{array}{cc}
               \gamma^{0}&0\\
                0&-1 \end{array}\right)
=\left(\begin{array}{cc}
               1&0\\
                0&-\bar{\cal E} \end{array}\right)K(z;g){}^{t}
\left(\begin{array}{cc}
               1&0\\
                0&-{\cal E} \end{array}\right)\\
{}&{}\\
{}&=\left(\begin{array}{cc}
               L(z;g)^{t}&0\\
                -2\bar{\Sigma}(z;g)&U^{-1}(z;g) \end{array}\right)
\end{array}
\label{Kbarform}
\end{equation}
From eqs.(\ref{Ztr},\,\ref{barZtr})  $F(z_{12})$ transforms as
\begin{equation}
F(z^{\prime}_{12})
=\bar{K}(z_{2};g)F(z_{12})K(z_{1};g)
\label{zz}
\end{equation}
In particular,  with eqs.(\ref{Kform},\,\ref{Kbarform}), this  gives the
transformation rule 
\begin{equation}
{\cal X}^{\prime}_{21}=L(z_{2};g)^{t}{\cal X}_{21}L(z_{1};g)~~~~~
\label{green2}
\end{equation}
\begin{equation}
\det{\cal X}_{12}^{\prime}=
\Omega^{2}(z_{1};g)\Omega^{2}(z_{2};g)\det{\cal X}_{12}~~~~~
\label{green4}
\end{equation}
where $\Omega(z;g)$  is given by  
\begin{equation}
\Omega(z;g)^{2}=\det L(z;g)=
\mbox{sdet}\:K(z;g)
\label{LOmega2}
\end{equation} 
If we  write 
\begin{equation}
\hat{L}(z;g)=L(z;g)/\Omega(z;g)^{1/2}
\end{equation}
then $\hat{L}\in \mbox{G}_{L}$, the group of $4\times 4$ matrices
satisfying the reality condition eq.(\ref{Lreality}) and
$\det \hat{L}=1$.\newline
Corresponding to eq.(\ref{infV}) $V_{i}{}^{j}(z_{21})$ transforms as
\begin{equation}
V(z^{\prime}_{21})=U^{-1}(z_{2};g)V(z_{21})U(z_{1};g)
\label{green3}
\end{equation}


\subsection{Transformation of Vectors}
If
\begin{equation}
\gamma^{B}R^{~A}_{B}(z;g)=L(z;g)\gamma^{A}L(z;g)^{t}
\label{green1}
\end{equation}
then $R^{~B}_{A}(z;g)$ is identical to the
definition~(\ref{ehomogeneous}), since infinitesimally 
\begin{equation}
\hat{\lambda}(z)\gamma^{A}-\tilde{\hat{\omega}}(z)\gamma^{A}
-\gamma^{A}\tilde{\hat{\omega}}(z)^{t}=\hat{\lambda}(z)\gamma^{A}
-\gamma^{B}\hat{\omega}(z)_{B}^{~A}=\gamma^{B}\partial_{B}h^{A}(z)
\end{equation}
as a consequence of eq.(\ref{hats}), which agrees with
eq.(\ref{infR}). Furthermore eq.(\ref{green1}) shows that the
definition\,(\ref{R2}) of $\Omega(z;g)$  
coincides with eq.(\ref{LOmega2}).\newline
From eq.(\ref{green1}) 
\begin{equation}
\gamma^{B}\hat{R}^{~A}_{B}(z;g)=\hat{L}(z;g)\gamma^{A}\hat{L}(z;g)^{t}
\label{hatR}
\end{equation}
where
 $\hat{R}=R/\Omega\in \mbox{SO}(1,5)\simeq\mbox{G}_{L}/\Z_{2} $.\newline
The matrix, ${\cal R}_{{\cal A}}{}^{{\cal B}}(z;g)$, given in
eq.(\ref{calR}) for the $(N,0)$ case becomes
\begin{equation}
{\cal R}_{{\cal A}}{}^{{\cal B}}(z;g)=\left(
\begin{array}{cl}
R_{A}^{~B}(z;g)&
i(\Sigma^{j}(z;g)\tilde{\gamma}_{A}L(z;g))^{\beta}\\
0&L_{\alpha}^{~\beta}(z;g)U_{i}^{~j}(z;g)
\end{array}\right)
\end{equation}
and so  $D_{{\cal A}}$ transforms  under  
the superconformal transformations as
\begin{equation} 
 D_{{\cal A}}={\cal R}_{{\cal A}}{}^{{\cal B}}(z;g)D_{{\cal B}}^{\prime}
\label{Dtrans}
\end{equation}
From eqs.(\ref{green2},\,\ref{green1}) 
$\mbox{tr}(\gamma^{A}{\cal X}_{12}\gamma^{B}{\cal X}_{21})$ 
transforms covariantly as
\begin{equation}
\mbox{tr}(\gamma^{A}{\cal X}^{\prime}_{12}\gamma^{B}{\cal X}^{\prime}_{21})=
\mbox{tr}(\gamma^{C}{\cal X}_{12}\gamma^{D}{\cal X}_{21})
R_{C}{}^{A}(z_{1};g)
R_{D}{}^{B}(z_{2};g)
\label{green5}
\end{equation}

\subsection{Relation to Superinversion}
If we define
\begin{equation}
\begin{array}{l}
I(z)=\left(\begin{array}{cc}
   i\tx_{+}^{-1}&-2i\tx_{-}^{-1}\theta^{j}\\
                 0&V_{i}{}^{j}(z)
                \end{array}\right)\\
{}\\
\bar{I}(z)=
\left(\begin{array}{cc}
   -i\tx_{-}^{-1}&0\\
-2i\bar{\theta}_{i}\tx_{+}^{-1}
                 &V^{-1}{}_{i}{}^{j}(z)
                \end{array}\right)=\left(\begin{array}{cc}
   \gamma^{0}&0\\
     0&-1          
     \end{array}\right)I(z)^{\dagger}\left(\begin{array}{cc}
   \gamma^{0}&0\\
     0&-1          
     \end{array}\right)
\end{array}
\end{equation}
where $V_{i}{}^{j}(z)$ is given by eq.(\ref{Vzz}), then using
$\theta^{i}V_{i}{}^{j}(z)=\tx_{+}\tx_{-}^{-1}\theta^{j}$ we
get 
\begin{equation}
\begin{array}{l}
\tilde{Z}(\tilde{z})\equiv Z(z)I(z)=\left(\begin{array}{cc}
1&0\\
-i\y_{+}&2\phi^{j}\\
2\bar{\phi}_{i}&\delta_{i}^{~j}
\end{array}\right)\\
{}\\
\bar{\tilde{Z}}(\tilde{z})\equiv \bar{I}(z)\bar{Z}(z)
=\left(\begin{array}{ccc}
i\y_{-}&1&-2\phi^{j}\\
-2\bar{\phi}_{i}&0&\delta_{i}^{~j}
\end{array}\right)
\end{array}
\label{tildeZZ}
\end{equation}
where
$\tilde{z}^{\tilde{{\cal A}}}=(y^{A},\phi^{i})\in
\Real_{-}^{6|8N}$ is 
defined in eq.(\ref{inversion}).\newline
We may also write 
\begin{equation}
G_{T}(z)G_{I}(z)=G_{\tilde{T}}(\tilde{z})=\left(\begin{array}{ccc}
1&0&0\\
-i\y_{+}&1&2\phi^{j}\\
2\bar{\phi}_{i}&0&\delta_{i}^{~j}
\end{array}\right)
\end{equation}
where 
\begin{equation}
G_{I}(z)=\left(\begin{array}{ccc}
0&i\tx_{-}&0\\
i\tx^{-1}_{+}&1&-2i\tx^{-1}_{-}\theta^{j}\\
0&-2\bar{\theta}_{i}&V_{i}{}^{j}(z)
\end{array}\right)
\end{equation}
Note 
\begin{equation}
\begin{array}{cc}
\tilde{Z}(\tilde{z})=G_{\tilde{T}}(\tilde{z})\tilde{Z}(0)~~~~&~~~~
\bar{\tilde{Z}}(\tilde{z})=\bar{\tilde{Z}}(0)G_{\tilde{T}}(\tilde{z})^{-1}
\end{array}
\end{equation}
and
\begin{equation}
G_{I}(z)\tilde{Z}(0)=Z(0)I(z)
\end{equation}
$G_{\tilde{T}}(\tilde{z})$ forms the  group of supertranslations
acting on $\tilde{z}\in \Real_{-}^{6|8N}$.\newline
$\tilde{Z}(\tilde{z}),\bar{\tilde{Z}}(\tilde{z})$ 
transform under the superconformal 
transformation,\,$\tilde{z}\stackrel{i^{-1}}{\longrightarrow}
z\stackrel{g}{\longrightarrow}z^{\prime}\stackrel{i}{\longrightarrow}
\tilde{z}^{\prime} $, 
from eqs.(\ref{Ztr},\,\ref{barZtr})
\begin{equation}
\begin{array}{l}
\tilde{Z}(\tilde{z}^{\prime})
=G(g)^{-1}\tilde{Z}(\tilde{z})\tilde{K}(\tilde{z};g)
~~~\\
{}\\
\bar{\tilde{Z}}(\tilde{z}^{\prime})=\bar{\tilde{K}}(\tilde{z};g)
\bar{\tilde{Z}}(\tilde{z})G(g)
\end{array}
\end{equation}
where
\begin{equation}
\begin{array}{l}
\tilde{K}(\tilde{z};g)=I(z)^{-1}K(z;g)I(z^{\prime})~~~~~~~~~~\\
{}\\
\bar{\tilde{K}}(\tilde{z};g)=\bar{I}(z^{\prime})\bar{K}(z;g)\bar{I}(z)^{-1}
=
\left(\begin{array}{cc}
   \gamma^{0}&0\\
     0&-1          
     \end{array}\right)\tilde{K}(\tilde{z};g)^{\dagger}
\left(\begin{array}{cc}
   \tilde{\gamma}^{0}&0\\
     0&-1          
     \end{array}\right)
\end{array}
\end{equation}
With $z_{21}\stackrel{i}{\longrightarrow}\widetilde{z_{21}}$ we may
write from eq.(\ref{tildeZZ}) 
\begin{equation}
\begin{array}{cc}
\tilde{Z}(\widetilde{z_{21}})=\left(\begin{array}{cc}
1&0\\
-i{\cal X}_{12}^{-1}& -2i{\cal X}_{21}^{-1}\theta_{21}^{j}\\
-2i\bar{\theta}_{21i}{\cal X}^{-1}_{12}&\delta_{i}^{~j}
\end{array}\right)~~~~&~~~~\bar{\tilde{Z}}(\widetilde{z_{21}})=
\left(\begin{array}{ccc}
-i{\cal X}_{21}^{-1}&1& 2i{\cal X}_{21}^{-1}\theta_{21}^{j}\\
2i\bar{\theta}_{21i}{\cal X}^{-1}_{12}&0&\delta_{i}^{~j}
\end{array}\right)
\end{array}
\end{equation}
which transform, using eqs.(\ref{cons1},\,\ref{cons2}), according to 
\begin{equation}
\begin{array}{l}
\delta\tilde{Z}(\widetilde{z_{21}})=
\hat{M}_{0}(z_{1})\tilde{Z}(\widetilde{z_{21}})-
 \tilde{Z}(\widetilde{z_{21}})\left(\begin{array}{cc}   
\hat{\omega}(z_{1})+\textstyle{\frac{1}{2}}\hat{\lambda}(z_{1})&0\\
0&\hat{T}(z_{1})
\end{array}\right)\\
{}\\
\delta\bar{\tilde{Z}}(\widetilde{z_{21}})
=\left(\begin{array}{cc}   
\tilde{\hat{\omega}}(z_{1})-\textstyle{\frac{1}{2}}\hat{\lambda}(z_{1})&0\\
0&\hat{T}(z_{1})
\end{array}\right)\bar{\tilde{Z}}(\widetilde{z_{21}})
-\bar{\tilde{Z}}(\widetilde{z_{21}})\hat{M}_{0}(z_{1})
\end{array}
\label{deltatildeZ}
\end{equation}
Taking
\begin{equation}
\begin{array}{cc}
G(i_{B})^{-1}=\left(\begin{array}{ccc}
            0&i\tilde{\gamma}^{B}&0\\
                 -i\gamma^{B}&0&0\\
            0&0&R_{i}^{~j}
                \end{array}\right)=e^{\frac{\pi}{2}M_{B}}~~~~&~~~~
M_{B}=\left(\begin{array}{ccc}
            0&i\tilde{\gamma}^{B}&0\\
           -i\gamma^{B}&0&0\\
            0&0&\frac{2}{\pi}\mbox{ln}R
            \end{array}\right)
\end{array}
\label{GiB}
\end{equation}
and
\begin{equation}
K(z;i_{B})=I(z)\left(\begin{array}{cc}
   -i\tilde{\gamma}^{B}&0\\
                     0&R^{-1}{}_{i}^{~j}        \end{array}\right)
\label{KiB}
\end{equation}
gives the realization of the transformation,\,
$z\stackrel{i_{B}}{\longrightarrow}z^{\prime}$,  
given in eq.(\ref{iB}) 
\begin{equation}
Z(z^{\prime})=G(i_{B})^{-1}Z(z)K(z;i_{B})
\label{Zp}
\end{equation}
From eq.(\ref{Xprop}) 
$L(z;i_{B})\equiv\tx_{+}^{-1}\tilde{\gamma}^{B}$ satisfies 
eq.(\ref{Lreality}).\newline
Eq.(\ref{Zp})  induces
\begin{equation}
\bar{Z}(z^{\prime})=\bar{K}(z;i_{B})\bar{Z}(z)G(i_{B})
\end{equation}
where
\begin{equation}
\bar{K}(z;i_{B})=
\left(\begin{array}{cc}
   -i\tilde{\gamma}^{B}&0\\
                     0&R        \end{array}\right)\bar{I}(z)
\end{equation}
Note that $G(i_{B})^{2}=-1$ and 
$K(z;i_{B})K(z^{\prime};i_{B})=-1$.

\subsection{Functions of Three Points}
In a similar fashion to eq.(\ref{Fform}), a corresponding function   
$\tilde{F}(\tilde{z})$ for 
$\tilde{z}^{\tilde{{\cal A}}}=
(y^{A},\phi^{i})\in\Real^{6|8N}_{-}$ is given by
\begin{equation}
\tilde{F}(\tilde{z})=\left(\begin{array}{cc}
              -i\y_{+}&2\phi^{j}\\
           2\bar{\phi}_{i}& \delta_{i}^{~j}\end{array}\right)
\label{tildeFform}
\end{equation}
With this definition  we may write
\begin{equation}             
\bar{\tilde{Z}}(\widetilde{z_{31}})\tilde{Z}(\widetilde{z_{21}})
=\tilde{F}(\widetilde{{\Z}_{1}})=
\left(\begin{array}{cc}    
-i\Y_{1+}&2\Phi^{j}_{1}\\         
2\bar{\Phi}_{1i}&\delta_{i}^{~j}     
\end{array}\right)  
\label{tildeFZ1}
\end{equation}
where
\begin{equation}
\begin{array}{c}
\Y_{1+}={\cal X}_{31}^{-1}{\cal X}_{32}
{\cal X}_{12}^{-1}\\
{}\\
\Phi^{i}_{1}=i({\cal X}_{31}^{-1}\theta^{i}_{31}-
{\cal X}_{21}^{-1}\theta^{i}_{21})
\end{array}
\label{W1}
\end{equation}
with 
\begin{equation}
\bar{\Phi}_{1i}= 
\Phi_{1}^{i}{}^{\dagger}\gamma^{0}
=\Phi_{1}^{j}{\cal E}_{ji}=i(\bar{\theta}_{31i}{\cal X}_{13}^{-1}
-\bar{\theta}_{21i}{\cal X}_{12}^{-1})
\end{equation}
Using
\begin{equation}
{\cal X}_{12}+{\cal X}_{23}={\cal X}_{13}
+4i\theta^{i}_{12}\bar{\theta}_{23i}
\end{equation}
one can show 
\begin{equation}
\begin{array}{c}
\Y_{1-}=\Y_{1+}-4i\Phi^{i}_{1}\bar{\Phi}_{1i}
=-{\cal X}^{-1}_{21}{\cal X}_{23}{\cal X}_{13}^{-1}
=-\Y_{1+}^{t}\\
{}\\
\Y_{1}=\textstyle{\frac{1}{2}}(\Y_{1+}+\Y_{1-})=-\Y_{1}^{t}
=Y_{1}^{A}\gamma_{A}
\end{array}
\label{Ypro}
\end{equation}
Hence we may define  
$\widetilde{{\Z}_{1}}{}^{\tilde{{\cal A}}}
=(Y^{A}_{1},\Phi^{i}_{1})
\in\Real^{6|8N}_{-}$.\newline
It is evident from eqs.(\ref{W1},\,\ref{Ypro}) that under 
$z_{2}\leftrightarrow z_{3}$,  $\widetilde{\Z_{1}}\rightarrow -
\widetilde{\Z_{1}}$.\newline
From eq.(\ref{deltatildeZ}), $\tilde{F}(\widetilde{{\Z}_{1}})$ transforms
infinitesimally as
\begin{equation}
\delta \tilde{F}(\widetilde{{\Z}_{1}})=\left(\begin{array}{cc}   
\tilde{\hat{\omega}}(z_{1})-\textstyle{\frac{1}{2}}\hat{\lambda}(z_{1})&0\\
0&\hat{T}(z_{1})
\end{array}\right)
\tilde{F}(\widetilde{{\Z}_{1}})-\tilde{F}(\widetilde{{\Z}_{1}})
\left(\begin{array}{cc}   
\hat{\omega}(z_{1})+\textstyle{\frac{1}{2}}\hat{\lambda}(z_{1})&0\\
0&\hat{T}(z_{1})
\end{array}\right)
\label{inftildeF}
\end{equation}
and hence for finite transformations
\begin{equation}
\tilde{F}(\widetilde{\Z_{1}}{}^{\prime})=\left(\begin{array}{cc}
L(z_{1};g)^{-1}&0\\
0&U(z_{1};g)^{-1}
\end{array}\right) \tilde{F}(\widetilde{{\Z}_{1}})
\left(\begin{array}{cc}    
L(z_{1};g)^{-1}{}^{t}&0\\
0&U(z_{1};g)
\end{array}\right)
\label{FFtrans}
\end{equation}
Thus $\widetilde{{\Z}_{1}}$ transforms homogeneously at $z_{1}$. 
Explicitly we have from eq.(\ref{FFtrans})  
\begin{equation}
\begin{array}{cc}
\multicolumn{2}{c}{\,\,\,\,\,\Y^{\prime}_{1+}
=L(z_{1};g)^{-1}\Y_{1+}L(z_{1};g)^{-1}{}^{t}}\\
{}&{}\\
\Phi^{\prime i}_{1}
=L(z_{1};g)^{-1}\Phi^{j}_{1}U_{j}^{~i}(z_{1};g)~~~~
&~~~~\bar{\Phi}{}^{\prime}_{1i}=
U^{-1}{}_{i}^{~j}(z_{1};g)\bar{\Phi}_{1j}L(z_{1};g)^{-1}{}^{t}
\end{array}
\label{expX}
\end{equation}
$\Y_{1-}$ also transforms in the same way as  $\Y_{1+}$ 
in eq.(\ref{expX}).\newline  
In a similar fashion to eq.(\ref{sdetZZ}) we get 
\begin{equation}
\mbox{sdet}\:\tilde{F}(\widetilde{{\Z}_{1}})=\det \Y_{1+}
=\det \Y_{1-}
=\displaystyle{
\frac{\det {\cal X}_{32}}{\det {\cal X}_{31}\,
\det {\cal X}_{12}}}
\end{equation}
It is useful to consider
 $\widetilde{{\Z}_{1}}=(Y^{A}_{1},\Phi_{1}^{i})
\stackrel{i^{-1}}{\longrightarrow}\Z_{1}=(X^{A}_{1},\Theta^{i}_{1})$,
where $i^{-1}$ is the inverse of superinversion given in 
eq.(\ref{inverseofi})
\begin{equation}
\begin{array}{cc}
\tX_{1\pm}=-\Y_{1\pm}^{-1}~~~~&~~~~~
\Theta^{i}_{1}=-i\Y_{1-}^{-1}\Phi_{1}^{i}
\end{array}
\end{equation}
By taking cyclic permutations of $z_{1},z_{2},z_{3}$ in eq.(\ref{W1})
we may define $\widetilde{\Z_{2}},\widetilde{\Z_{3}}$ and hence 
$\Z_{2},\Z_{3}$.    
We find $\Z_{2},\Z_{3}$ are related to $\Z_{1}$ in a simple form
\begin{equation}
\begin{array}{l}
F(\Z_{2})=\left(\begin{array}{cc}
                 i{\cal X}_{21}&0\\
                 0&V(z_{21})
            \end{array}\right)\tilde{F}(-\widetilde{\Z_{1}})
\left(\begin{array}{cc}
                 i{\cal X}_{12}&0\\
                 0&V(z_{21})^{-1}
            \end{array}\right)\\
{}\\
\tilde{F}(\widetilde{\Z_{3}})=\left(\begin{array}{cc}
                 -i{\cal X}_{13}^{-1}&0\\
                 0&V(z_{31})
            \end{array}\right) F(-\Z_{1})
\left(\begin{array}{cc}
                 -i{\cal X}_{31}^{-1}&0\\
                 0&V(z_{13})
            \end{array}\right)
\end{array}
\end{equation}
where $F$ is given in eq.(\ref{Fform}).\newline
Explicitly we have
\begin{subeqnarray}
\label{Z2Z3}
&\tX_{2+}={\cal X}_{21}\Y_{1-}{\cal X}_{12}~~~~~&~~~~
\Theta^{i}_{2}=-i{\cal X}_{21}\Phi_{1}^{j}
V_{j}^{~i}(z_{12})\\
{}&{}\nonumber\\
&\Y_{3+}={\cal X}_{13}^{-1}\tX_{1-}{\cal X}_{31}^{-1}~~~&~~~~
\Phi^{i}_{3}=i{\cal X}_{13}^{-1}\Theta_{1}^{j}
V_{j}^{~i}(z_{13})
\end{subeqnarray}
If we define a function $\tilde{V}(\tilde{z})\in\mbox{Sp}(N)$ for 
$\tilde{z}\in\Real^{6|8N}_{-}$, analogous to $V(z)$ in eq.(\ref{Vzz}), by 
\begin{equation}
\tilde{V}_{i}^{~j}(\tilde{z})=\delta_{i}^{~j}
+4i\bar{\phi}_{i}\Y_{-}^{-1}\phi^{j}
\end{equation}
then one can show
\begin{equation}
\tilde{V}(\widetilde{{\Z}_{1}})=V(z_{13})V(z_{32})V(z_{21})
\label{tVdecom}
\end{equation} 
$\tilde{V}(\widetilde{{\Z}_{1}})$ transforms as 
\begin{equation}
\tilde{V}(\widetilde{{\Z}_{1}}{}^{\prime})=
U^{-1}(z_{1};g)V(\widetilde{{\Z}_{1}})U(z_{1};g)
\end{equation}
Note also that with $z\stackrel{i}{\longrightarrow}\tilde{z}$
\begin{equation}
\tilde{V}(\tilde{z})=V(z)^{-1}
\end{equation}


\section{Reduction to Four Dimensions}
The reduction of the  conformal group  to four-dimensions  may be 
defined as the subgroup such that 
\begin{equation}
\begin{array}{cccc}
a^{A}=0~~~&~~~b^{A}=0~~~&~~~\omega^{\mu A}=0~~~&~~~A=4,5
\end{array}
\end{equation}
where  $\mu=0,1,2,3$.\newline
In the superconformal case,  from eq.(\ref{MMcom}), it is necessary
therefore to require 
\begin{equation}
\begin{array}{cccc}
\bar{\varepsilon}_{i}\gamma^{A}\varepsilon^{\prime i}=0~~~&~~~
\bar{\rho}_{i}\tilde{\gamma}^{A}\rho^{\prime i}=0~~~&~~~
\bar{\varepsilon}_{i}\gamma^{[A}\tilde{\gamma}^{\mu]}\rho^{i}=0~~~&~~~
A=4,5
\end{array}
\label{SpNinv}
\end{equation}
To solve  this, we first decompose $\varepsilon^{i}=\varepsilon_{+}^{i}
+\varepsilon_{-}^{i}$ where
\begin{equation}
i\tilde{\gamma}_{4}\gamma_{5}\varepsilon^{i}_{\pm}=\pm\varepsilon_{\pm}^{i}
\label{45eigen}
\end{equation}
Defining $\bar{\varepsilon}_{\pm i}=(\varepsilon^{i}_{\pm})^{\dagger}
\tilde{\gamma}^{0}$ as in eq.(\ref{pseudoM}) we have
\begin{equation}
i\tilde{\gamma}_{4}\gamma_{5}\bar{\varepsilon}^{t}_{\pm i}=\mp
\bar{\varepsilon}^{t}_{\pm i}
\label{ccv}
\end{equation}
Since $\tilde{\gamma}^{0}\gamma^{1}\tilde{\gamma}^{2}\gamma^{3}
\tilde{\gamma}^{4}\gamma^{5}=-1$, $\varepsilon^{i}_{\pm}$ are chiral
spinors in four-dimensions. We may also decompose 
 $\rho^{i}$ in a similar fashion as 
$\rho^{i}=\rho_{+}^{i}+\rho_{-}^{i}$ where now  $
i\gamma_{4}\tilde{\gamma}_{5}\rho^{i}_{\pm}=\pm\rho_{\pm}^{i}$.\newline
Since $i\tilde{\gamma}_{4}\gamma_{5}$ satisfies
\begin{equation}
\begin{array}{ll}
\gamma^{A}(i\tilde{\gamma}_{4}\gamma_{5})=\pm
(i\tilde{\gamma}_{4}\gamma_{5})^{t}\gamma^{A}~~~~&~~~~
\left\{\begin{array}{l}
                                      +:~A=4,5\\
                                      -:~A=\mu=0,1,2,3
                                    \end{array}\right.
\end{array}
\label{fourgg}
\end{equation}
eq.(\ref{SpNinv}) becomes  
\begin{equation}
\begin{array}{ll}
\multicolumn{2}{l}{
\bar{\varepsilon}_{+i}\gamma^{A}\varepsilon^{\prime i}_{-}+
\bar{\varepsilon}_{-i}\gamma^{A}\varepsilon^{\prime i}_{+}=
\bar{\rho}_{+i}\tilde{\gamma}^{A}\rho^{\prime i}_{-}+
\bar{\rho}_{-i}\tilde{\gamma}^{A}\rho^{\prime i}_{+}}=0\\
{}&{}\\
\bar{\varepsilon}_{+i}\gamma^{[A}\tilde{\gamma}^{\mu]}\rho^{i}_{-}+
\bar{\varepsilon}_{-i}\gamma^{[A}\tilde{\gamma}^{\mu]}\rho^{i}_{+}=0~~~&~~~
A=4,5
\end{array}
\label{4con}
\end{equation}
To solve the conditions~(\ref{4con}) we require 
$\bar{\varepsilon}_{+i},\bar{\rho}_{+i}$ to
be orthogonal to   $\varepsilon^{i}_{-},\rho^{i}_{-}$ when contracting
 the indices~$i$.  The solutions for $\varepsilon^{i}_{\pm}$ 
define  spaces $V_{\pm}$ where
$\varepsilon^{i}_{\pm}\in V_{\pm}$ and since, by virtue of
eq.(\ref{ccv}),  complex conjugation interchanges $V_{+}$ and
$V_{-}$, we must have  $\mbox{dim}V_{+}=\mbox{dim}V_{-}$.  
The maximal solution of eq.(\ref{4con}) occurs when we reduce  half of
the degrees of freedom for the spinors so that 
$\mbox{dim}V_{+}=\mbox{dim}V_{-}=N$. Using a basis where 
\begin{equation}
i\tilde{\gamma}_{4}\gamma_{5},\,-i\gamma_{4}\tilde{\gamma}_{5}
                             =\left(\begin{array}{cc}
                                        1&0\\
                                        0&-1
                                    \end{array}\right)
\end{equation}
the  solution can be written, using the freedom of
$\mbox{Sp}(N)$ transformations,  in the form 
\begin{equation}
\begin{array}{ccc}
\varepsilon^{i}=\left(\begin{array}{cc}
             \bar{\varepsilon}_{a} &0\\
               0&\varepsilon^{a}{}^{t}
              \end{array}\right)~~~~&~~~~
\rho^{i}=\left(\begin{array}{cc}
             0&\bar{\rho}^{a}{}^{t} \\
             \rho_{a}&0
              \end{array}\right)~~~~&~~~~a=1,2,\cdots ,N
\end{array}
\label{vrforms}
\end{equation}
and so following eqs.(\ref{pseudorho},\,\ref{pseudovar})
\begin{equation}
\begin{array}{cc}
\bar{\varepsilon}_{i}=\left(\begin{array}{cc}
                0&\varepsilon^{a} \\
               -\bar{\varepsilon}_{a}^{t}&0
              \end{array}\right)~~~~&~~~~
\bar{\rho}_{i}=\left(\begin{array}{cc}
             \bar{\rho}^{a}&0 \\
             0&-\rho_{a}^{t}
              \end{array}\right)
\end{array}
\label{4vr}
\end{equation}
The representation\footnote{See Appendix A for a particular
choice of  representation for $\gamma^{\mu},\tilde{\gamma}^{\mu}$
compatible with eq.(\ref{4vr}).}
 of $\gamma^{\mu},\tilde{\gamma}^{\mu}$ may be
chosen such that  $\bar{\varepsilon}_{a}^{\dot{\alpha}}
=\varepsilon^{a\alpha}{}^{\dagger},\,
\bar{\rho}^{a}_{\dot{\alpha}}=\rho_{a\alpha}^{\dagger}
~\alpha,\dot{\alpha}=1,2$.\newline
Since, from eq.(\ref{MMcom}),  
$\varepsilon^{j}T_{j}^{~i},\rho^{j}T_{j}^{~i}$ should remain of  
the form~(\ref{vrforms}), it is necessary to restrict the
$\mbox{Sp}(N)$  generators,~$T_{i}^{~j}$ to the form 
\begin{equation}
T_{j}^{~i}=\left(\begin{array}{cc}
                t^{a}_{~b}+i\phi\delta^{a}_{~b}&0\\
               0&-(t^{t})_{a}^{~b}-i\phi\delta_{a}^{~b}
              \end{array}\right)
\end{equation}
where $t\in \mbox{su}(N),\,\phi\in S^{1}$. With this form
$\mbox{Sp}(N)\rightarrow \mbox{SU}(N)\times \mbox{U}(1)$\newline
After reduction to four-dimensions, the matrix representation of
superconformal algebra,~$M$, given in eq.(\ref{Mform}) decomposes  as
\begin{equation}
\begin{array}{cc}
M=M_{+}+M_{-}~~~~&~~~~M_{-}=-C^{-1}M^{t}_{+}C
\end{array}
\label{M+-}
\end{equation}
where, with $\omega_{45}=-\psi,\,w=\textstyle{\frac{1}{4}}
\omega_{\mu\nu}\tilde{\sigma}^{\mu}\sigma^{\nu},\,\tilde{w}=
\textstyle{\frac{1}{4}}
\omega_{\mu\nu}\sigma^{\mu}\tilde{\sigma}^{\nu}$,
\begin{equation}
\displaystyle{
M_{+}=\left(\begin{array}{ccc}
       (w+\textstyle{\frac{1}{2}}\lambda+
     i\textstyle{\frac{1}{2}}\psi)\,p_{+} &
         -ia{\cdot\tilde{\sigma}}\,\tau_{+}&2\bar{\varepsilon}_{b}\,p_{+}\\
       -ib{\cdot\sigma}\,\tau_{-} &(\tilde{w}
        -\textstyle{\frac{1}{2}}\lambda +i 
     \textstyle{\frac{1}{2}}\psi)\,p_{-} &2\rho_{b}\,\tau_{-}\\
       2\bar{\rho}^{a}\,p_{+}&2\varepsilon^{a}\,\tau_{+}
     &(t^{a}_{~b}+i\phi\,\delta^{a}_{~b})\,p_{+}
        \end{array}\right)}
\label{M+}
\end{equation}
and we define 
\begin{equation}
\begin{array}{cccc}
p_{+}=\left(\begin{array}{cc}
              1&0\\
              0&0
              \end{array}\right)~~~&~~~
p_{-}=\left(\begin{array}{cc}
              0&0\\
              0&1
              \end{array}\right)~~~&~~~
\tau_{+}=\left(\begin{array}{cc}
              0&1\\
              0&0
              \end{array}\right)~~~&~~~
\tau_{-}=\left(\begin{array}{cc}
              0&0\\
              1&0
              \end{array}\right)
\end{array} 
\end{equation}
Since
$p^{2}_{+}=\tau_{+}\tau_{-}=p_{+},\,p^{2}_{-}=\tau_{-}\tau_{+}=p_{-},\,
p_{+}\tau_{+}=\tau_{+}p_{-}=\tau_{+},\,p_{-}\tau_{-}=\tau_{-}p_{+}
=\tau_{-}$,
the form of $M_{+}$ is closed under multiplication and also 
$[M_{+},M^{\prime}_{-}]=0$. Hence  $M_{+}$ alone defines  the 
four-dimensional superconformal algebra, and it  can be naturally
reduced to the $(4+N)\times (4+N)$ matrix,~${\cal M}$
\begin{equation}
\displaystyle{
{\cal M}=\left(\begin{array}{ccc}
       w+ \textstyle{\frac{1}{2}}\lambda +i 
     \textstyle{\frac{1}{2}}\psi
&         -ia{\cdot\tilde{\sigma}}&2\bar{\varepsilon}_{b}\\
       -ib{\cdot\sigma} &\tilde{w} -\textstyle{\frac{1}{2}}\lambda +i 
     \textstyle{\frac{1}{2}}\psi  &2\rho_{b}\\
       2\bar{\rho}^{a}&2\varepsilon^{a}&t^{a}_{~b}+i\frac{2}{N}
\psi\delta^{a}_{~b}
        \end{array}\right)}
\label{calM}
\end{equation}
where we impose $\mbox{str}\,{\cal
M}=\mbox{str}\,M_{+}=i(2\psi-N\phi)=0$,  removing a $\mbox{U}(1)$
factor, which is consistent with  
$\mbox{str}\,[{\cal M},{\cal M}^{\prime}]=0$.  For  general $N$ the
$R$-symmetry  group is $\mbox{SU}(N)\times \mbox{U}(1)$. When 
$N=4$,  however, it is evident from eq.(\ref{calM}) that  ${\cal M}={\cal
M}_{0}+i\textstyle{\frac{1}{2}}\psi\,1$ where ${\cal M}_{0}$ is of the
form~(\ref{calM}) with $\psi=0$, so that $\psi$ parameterizes
a $\mbox{U}(1)$ invariant subalgebra. Hence in this case we may set
$\psi=0$ and reduce ${\cal M}$ to ${\cal M}_{0}$. The associated
$\mbox{U}(1)$ 
symmetry is an ideal of four-dimensional $N=4$ superconformal
symmetry, so that the $R$-symmetry group is just $\mbox{SU}(4)$.\newline
From eq.(\ref{Mdagger}) the hermiticity condition becomes 
\begin{equation}
\begin{array}{cc}
{\cal B}{\cal M}{\cal B}^{-1}=-{\cal M}^{\dagger}~~~~&~~~~
{\cal B}=\left(\begin{array}{ccc}
                 0&1&0\\ 
                 1&0&0\\
                 0&0&-1
              \end{array}\right)
\end{array}
\end{equation}
\subsection{Reduction of Superspace}
The reduction of  $\Real_{+}^{6|8N}$ superspace to four-dimensions 
is defined for $z^{{\cal A}}=(x^{A},\theta^{i})$ by setting  
$x^{4}=x^{5}=0$ and, as in eqs.(\ref{vrforms},\,\ref{4vr}),
restricting $\theta^{i}$ to the form
\begin{equation}
\begin{array}{ccc}
\theta^{i}=\left(\begin{array}{cc}
             \bar{\theta}_{a} &0\\
               0&\theta^{a}{}^{t}
              \end{array}\right)~~~&\Longrightarrow&~~~
\bar{\theta}_{i}=\left(\begin{array}{cc}
                0&\theta^{a} \\
               -\bar{\theta}_{a}^{t}&0
              \end{array}\right)
\end{array}
\label{spinor1}
\end{equation}
Hence $z^{{\cal A}}\rightarrow z^{{\cal N}}=
(x^{\mu},\theta^{a\alpha},\bar{\theta}^{\dot{\alpha}}_{a})\in
\Real^{4|4N}$ and   the six-dimensional infinitesimal supersymmetric
interval is furthermore projected  into  four-dimensions, 
$e^{{\cal A}}(z)\rightarrow e^{{\cal N}}(z)$, where
\begin{equation}
\begin{array}{cc}
e^{{\cal N}}(z)=(e^{\mu}(z),{\rm d}\theta^{a\alpha},
{\rm d}\bar{\theta}^{\dot{\alpha}}_{a})~~~~&~~~~
e^{\mu}(z)={\rm d}x^{\mu}+i{\rm d}\theta^{a}\sigma^{\mu}\bar{\theta}_{a}
-i\theta^{a}\sigma^{\mu}{\rm d}\bar{\theta}_{a}
\end{array}
\end{equation}
$\tx_{\pm}$ defined in eq.(\ref{tildeX}) is of the form
\begin{equation}
\begin{array}{cc}
\tx_{\pm}=\left(\begin{array}{cc}
                  0&x_{\pm}{\cdot\tilde{\sigma}}\\
               -(x_{\mp}{\cdot\tilde{\sigma}})^{t}&0
               \end{array}\right)~~~~&~~~~
x_{\pm}^{\mu}=x^{\mu}\mp
i\theta^{a}\sigma^{\mu}\bar{\theta}_{a}
\end{array}
\label{4Xpm}
\end{equation}
Note that 
\begin{equation}
x_{\pm}{\cdot\tilde{\sigma}}=x{\cdot\tilde{\sigma}}
\pm 2i\bar{\theta}_{a}\theta^{a}
\end{equation}
The four-dimensional superconformal transformations are now realized
by restricting the six-dimensional $(N,0)$ superconformal
transformations as in $M\rightarrow{\cal M}$ with the group of
dimensions $(15+N^{2}|8N)$ if $N\neq 4$. We then 
get $N$-extended superconformal
transformations in four-dimensions from eqs.(\ref{dlambda},\,\ref{dX+})
\begin{equation}
\begin{array}{l}
\delta\theta^{a}=\varepsilon^{a}+\textstyle{\frac{1}{2}}(\lambda+i\Omega)
\theta^{a}-\theta^{a}\tilde{w}
+t^{a}_{~b}\theta^{b}+\theta^{a}
b{\cdot\sigma}\,
x_{+}{\cdot\tilde{\sigma}}-i\bar{\rho}^{a}x_{+}{\cdot\tilde{\sigma}}-
4(\theta^{a}\rho_{b})\,\theta^{b}\\
{}\\
\delta x_{+}{\cdot\tilde{\sigma}}=
x_{+}{\cdot\tilde{\sigma}}\,b{\cdot\sigma}\,x_{+}{\cdot\tilde{\sigma}}-4
x_{+}{\cdot\tilde{\sigma}}\,\rho_{a}\theta^{a}+\lambda
x_{+}{\cdot\tilde{\sigma}}+w\,x_{+}{\cdot\tilde{\sigma}}
-x_{+}{\cdot\tilde{\sigma}}\,\tilde{w}
+4i\bar{\varepsilon}_{a}\theta^{a}+a{\cdot\tilde{\sigma}}
\end{array}
\label{4dimsupercon}
\end{equation}
where $\Omega=(\textstyle{\frac{4}{N}}-1)\psi$. When 
$N=4$, we get $\Omega=0$ and the transformations are independent of
$\psi$ which is consistent with the reduction of the superconformal
algebra in this case.\newline 
$Z(z)$ and $\bar{Z}(z)$ defined by eqs.(\ref{defZ},\,\ref{defbarZ})
can also be decomposed in a similar fashion to
eqs.(\ref{M+-},\,\ref{M+})   by writing 
\begin{equation}
\begin{array}{cc}
Z(z)=Z_{+}(z)+Z_{-}(z)~~~~&~~~~
\bar{Z}(z)=\bar{Z}_{+}(z)+\bar{Z}_{-}(z)
\end{array}
\end{equation}
where
\begin{equation}
\begin{array}{cc}
Z_{+}(z)=\left(\begin{array}{cc}
-ix_{+}{\cdot\tilde{\sigma}}\,\tau_{+}&2\bar{\theta}_{b}\,p_{+}\\
p_{-}&0\\
2\theta^{a}\,\tau_{+}&\delta^{a}_{~b}\,p_{+}
\end{array}\right)~~~&~~~
\bar{Z}_{+}(z)=\left(\begin{array}{ccc}
p_{+}&ix_{-}{\cdot\tilde{\sigma}}\,\tau_{+}&-2\bar{\theta}_{b}\,p_{+}\\
0&-2\theta^{a}\,\tau_{+}&\delta^{a}_{~b}\,p_{+}
\end{array}\right)
\end{array}
\label{Z+}
\end{equation}
and $Z_{-}(z),\bar{Z}_{-}(z)$ are given by
\begin{equation}
\begin{array}{cc}
Z_{-}(z)=C^{-1}\bar{Z}^{t}_{+}(z)\left(\begin{array}{cc}
            1&0\\
      0&\bar{\cal E} \end{array}\right)
~~&~~
\bar{Z}_{-}(z)=\left(\begin{array}{cc}
            1&0\\
      0&-\bar{\cal E} \end{array}\right) Z_{+}(z){}^{t}C 
\end{array}
\end{equation}
Just as  $M_{+}\rightarrow {\cal M}$,  we can  similarly have
\begin{equation}
\begin{array}{l}
Z_{+}(z)\rightarrow {\cal Z}(z)=\left(\begin{array}{cc}
-ix_{+}{\cdot\tilde{\sigma}}&2\bar{\theta}_{b}\\
1&0\\
2\theta^{a}&\delta^{a}_{~b}
\end{array}\right)\\
{}\\
\bar{Z}_{+}(z)\rightarrow \bar{{\cal Z}}(z)=\left(\begin{array}{ccc}
1&ix_{-}{\cdot\tilde{\sigma}}&-2\bar{\theta}_{b}\\
0&-2\theta^{a}&\delta^{a}_{~b}
\end{array}\right)=\left(\begin{array}{cc}
            1&0\\
            0&-1\end{array}\right){\cal Z}(z)^{\dagger}{\cal B}
\end{array}
\label{calZ}
\end{equation}
Infinitesimally ${\cal Z}(z),\bar{{\cal Z}}(z)$ transform,   
from eqs.(\ref{Ztr},\,\ref{barZtr}), as
\begin{equation}
\begin{array}{l}
\delta {\cal Z}(z)={\cal M}{\cal Z}(z)-{\cal Z}(z){\cal H}(z)\\
{}\\
\delta \bar{{\cal Z}}(z)=\bar{{\cal H}}(z)\bar{{\cal Z}}(z)-
\bar{{\cal Z}}(z){\cal M}
\end{array}
\label{calZtr}
\end{equation}
where
\begin{equation}
\begin{array}{c}
{\cal H}(z)=\left(\begin{array}{cc}
\tilde{\hat{w}}(z)-\textstyle{\frac{1}{2}}\hat{\lambda}(z)+
i\textstyle{\frac{1}{2}}\hat{\psi}(z)
&2\hat{\rho}_{b}(z)\\
0&\hat{t}^{a}_{~b}(z)+
i\textstyle{\frac{2}{N}}\hat{\psi}(z)\delta^{a}_{~b}
\end{array}\right)\\
{}\\
\bar{{\cal H}}(z)=\left(\begin{array}{cc}
\hat{w}(z)+\textstyle{\frac{1}{2}}\hat{\lambda}(z)+
i\textstyle{\frac{1}{2}}\hat{\psi}(z)
&0\\
2\bar{\hat{\rho}}{}^{a}(z)~~~~~&\hat{t}^{a}_{~b}(z)+
i\textstyle{\frac{2}{N}}\hat{\psi}(z)\delta^{a}_{~b}
\end{array}\right)
\end{array}
\end{equation}
with
\begin{equation}
\begin{array}{ll}
\multicolumn{2}{l}{ 
\tilde{\hat{w}}(z)-\textstyle{\frac{1}{2}}\hat{\lambda}(z)
+i\textstyle{\frac{1}{2}}\hat{\psi}(z)=
\tilde{w}+4\rho_{a}\theta^{a}-b{\cdot\sigma}x_{+}{\cdot\tilde{\sigma}}
-\textstyle{\frac{1}{2}}\lambda +i\textstyle{\frac{1}{2}}\psi}\\
{}&{}\\
\multicolumn{2}{l}{
\hat{w}(z)+\textstyle{\frac{1}{2}}\hat{\lambda}(z)+
i\textstyle{\frac{1}{2}}\hat{\psi}(z)=w
-4\bar{\theta}_{a}\bar{\rho}^{a}+x_{-}{\cdot\tilde{\sigma}}b{\cdot\sigma}
+\textstyle{\frac{1}{2}}\lambda+i\textstyle{\frac{1}{2}}\psi
=-(\tilde{\hat{w}}(z)-\textstyle{\frac{1}{2}}\hat{\lambda}(z)+
i\textstyle{\frac{1}{2}}\hat{\psi}(z))^{\dagger}}\\
{}&{}\\
\multicolumn{2}{l}{
\hat{t}^{a}_{~b}(z)+
i\textstyle{\frac{2}{N}}\hat{\psi}(z)\delta^{a}_{~b}=t^{a}_{~b}+4i\theta^{a}
b{\cdot\sigma}\bar{\theta}_{b}+4(\bar{\rho}^{a}\bar{\theta}_{b}-
\theta^{a}\rho_{b})+i\textstyle{\frac{2}{N}}\psi\delta^{a}_{~b}}\\
{}&{}\\
\hat{\rho}_{a}(z)=\rho_{a}-ib{\cdot\sigma}\bar{\theta}_{a}~~~~&~~~~
\bar{\hat{\rho}}{}^{a}(z)=\bar{\rho}{}^{a}+i\theta^{a}b{\cdot\sigma}
=\hat{\rho}_{a}(z)^{\dagger}
\end{array}
\end{equation}
In general, with $z^{{\cal A}}\rightarrow z^{{\cal
N}}\in\Real^{4|4N}$,  
$F(z)$ defined in eq.(\ref{Fform}) can also be decomposed as 
\begin{equation}
F(z)=F_{+}(z)+F_{-}(z)
\label{F+-1}
\end{equation}
where 
\begin{equation}
\begin{array}{l}
F_{+}(z)=\left(\begin{array}{cc}
-ix_{+}{\cdot\tilde{\sigma}}\,\tau_{+}&2\bar{\theta}_{b}\,p_{+}\\
2\theta^{a}\,\tau_{+}&\delta^{a}_{~b}\,p_{+} 
\end{array}\right)\\
{}\\
F_{-}(z)=\left(\begin{array}{cc}
1&0\\
0&-\bar{{\cal E}}
\end{array}\right)F_{+}(-z)^{t}\left(\begin{array}{cc}
1&0\\
0&-{\cal E}
\end{array}\right)
\end{array}
\label{F+-2}
\end{equation}
Analogous to eqs.(\ref{Fform},\,\ref{XV},\,\ref{Vzz}) we define for any 
$z^{{\cal N}}=(x^{\mu},\theta^{a\alpha},\bar{\theta}^{\da}_{a})
\in\Real^{4|4N}$ 
\begin{equation}
{\cal F}(z)=\left(\begin{array}{cc}
-ix_{+}{\cdot\tilde{\sigma}}&2\bar{\theta}_{b}\\
2\theta^{a}&\delta^{a}_{~b}
\end{array}\right)
\label{calF}
\end{equation}
so that $F_{+}(z)\rightarrow{\cal F}(z)$. Furthermore 
\begin{equation}
\left(\begin{array}{cc}
1&0\\
-2i\theta^{a}
(x_{+}{\cdot\tilde{\sigma}})^{-1}&1
\end{array}\right){\cal F}(z)
\left(\begin{array}{cc}
1&-2i(x_{+}{\cdot\tilde{\sigma}})^{-1}
\bar{\theta}_{b}\\
0&1
\end{array}\right)=\left(\begin{array}{cc}
-ix_{+}{\cdot\tilde{\sigma}}&0\\
0&v^{a}_{~b}(-z)
\end{array}\right)
\label{fv4}
\end{equation}
defines
\begin{equation}
v^{a}_{~b}(z)=\displaystyle{\delta^{a}_{~b}+4i
\frac{1}{x_{-}^{2}}\theta^{a}x_{-}{\cdot\sigma}
\bar{\theta}_{b}}
\end{equation}
which satisfies
\begin{equation}
v^{a}_{~b}(-z)=
v^{-1}{}^{a}_{~b}(z)=v^{\dagger}{}^{a}_{~b}(z)
\end{equation} 
The superdeterminant of ${\cal F}(z)$ is given by
\begin{equation}
\mbox{sdet}\:{\cal F}(z)=-x_{-}^{2}
\label{sdetf}
\end{equation}
and also from eq.(\ref{fv4})
\begin{equation}
\det v(z)=\displaystyle{\frac{x^{2}_{-}}{x^{2}_{+}}}
\label{detv}
\end{equation}
In the above, we considered the reduction $\Real^{6|8N}_{+}\rightarrow
 \Real^{4|4N}$.  Alternatively we may consider the reduction of 
$\Real^{6|8N}_{-}$ superspace.   
For  $\tilde{z}^{\tilde{{\cal A}}}=(y^{A},\phi^{i})\in
\Real^{6|8N}_{-}$ we set $y^{4}=y^{5}=0$ and, analogous to 
eq.(\ref{spinor1}),  
\begin{equation}
\begin{array}{ccc}
\phi^{i}=\left(\begin{array}{cc}
               0&\bar{\phi}{}^{at}\\
               \phi_{a}&0
              \end{array}\right)~~~&\Longrightarrow &~~~
\bar{\phi}_{i}=\left(\begin{array}{cc}
                \bar{\phi}^{a}&0 \\
                0&-\phi_{a}^{t}
              \end{array}\right)
\end{array}
\end{equation}
Hence  $\tilde{z}^{\tilde{{\cal A}}}\rightarrow
\tilde{z}^{\tilde{{\cal
N}}}=(y^{\mu},\phi_{a},\bar{\phi}^{a})$. \newline
After reducing  to four-dimensions, 
$\Real^{6|8N}_{\pm}\rightarrow\Real^{4|4N}$,  the
superspaces are no longer inequivalent. This is apparent by using the 
reflection  transformation $r_{5}$ defined by  
eqs.(\ref{rB1},\,\ref{rB2}) with the $\mbox{Sp}(N)$ transformation   
$R$  given by 
\begin{equation}
\begin{array}{cccc}
R_{i}^{~j}=\left(\begin{array}{cc}
                   0& -\zeta^{ab}\\
                 \bar{\zeta}_{ab}&0
                    \end{array}\right)~~~~&~~~~~
\zeta^{ab}\bar{\zeta}_{bc}=\delta^{a}_{~c}~~&~~\zeta^{ab}=\zeta^{ba}~~&~~
\bar{\zeta}_{ab}=(\zeta^{ab})^{\ast}
\end{array}
\label{Rzeta}
\end{equation}
which satisfies eq.(\ref{Rsquare}).   Letting now  
$\Real^{6|8N}_{-}\stackrel{r_{5}^{-1}}{-\!\!\!-\!\!\!\longrightarrow}
\Real^{6|8N}_{+}\longrightarrow\Real^{4|4N}$,   we get       
$\tilde{z}^{\tilde{{\cal A}}}\rightarrow 
z^{{\cal N}}=(x^{\mu},\theta^{a},\bar{\theta}_{a})$ where
$x^{\mu}=y^{\mu}$ and 
\begin{equation}
\begin{array}{cc}
\theta^{a}=-\zeta^{ab}\phi^{t}_{b}\epsilon^{-1}~~~~~&~~~~
\bar{\theta}_{a}=\bar{\epsilon}^{-1}\bar{\phi}^{bt}\bar{\zeta}_{ba}
\end{array}
\label{xtbt}
\end{equation}
In a similar fashion to eq.(\ref{4Xpm}),  
${\rm y}_{\pm}={\rm y}\pm
2i\phi{}^{i}\bar{\phi}_{i}$ reduces to
\begin{equation}
\begin{array}{cc}
{\rm y}_{\pm}=\left(\begin{array}{cc}
0&-(y_{\pm}{\cdot\sigma})^{t}\\
y_{\mp}{\cdot\sigma}&0
\end{array}\right)~~~~&~~~~
y^{\mu}_{\pm}=y^{\mu}
\pm i\bar{\phi}{}^{a}\tilde{\sigma}^{\mu}\phi_{a}=x^{\mu}_{\pm}
\end{array}
\end{equation}
where we may also write 
\begin{equation}
y_{\pm}{\cdot\sigma}=y{\cdot\sigma}\mp 
2i\phi_{a}\bar{\phi}{}^{a}
\end{equation}
In a similar fashion to eqs.(\ref{F+-1},\,\ref{F+-2}), 
$\tilde{F}(\tilde{z})$ defined  in eq.(\ref{tildeFform}) 
can  be decomposed as 
\begin{equation}
\tilde{F}(\tilde{z})=\tilde{F}_{+}(\tilde{z})+\tilde{F}_{-}(\tilde{z})
\label{tildeF+-1}
\end{equation}
where 
\begin{equation}
\begin{array}{l}
\tilde{F}_{+}(\tilde{z})=\left(\begin{array}{cc}
-iy_{-}{\cdot\sigma}\,\tau_{-}&2\phi_{b}\,\tau_{-}\\
2\bar{\phi}{}^{a}\,p_{+}&\delta^{a}_{~b}\,p_{+} 
\end{array}\right)\\
{}\\
\tilde{F}_{-}(\tilde{z})=\left(\begin{array}{cc}
1&0\\
0&-\bar{{\cal E}}
\end{array}\right)\tilde{F}_{+}(-\tilde{z})^{t}\left(\begin{array}{cc}
1&0\\
0&-{\cal E}
\end{array}\right)
\end{array}
\label{tildeF+-2}
\end{equation}
Hence 
we reduce $\tilde{F}_{+}(\tilde{z})$ to $\tilde{{\cal F}}(\tilde{z})$
\begin{equation}
\tilde{{\cal F}}(\tilde{z})=\left(\begin{array}{cc}
-iy_{-}{\cdot\sigma}&2\phi_{b}\\
2\bar{\phi}{}^{a}_{}&\delta^{a}_{~b} 
\end{array}\right)
\end{equation}
With  eq.(\ref{xtbt}) $\tilde{{\cal F}}(\tilde{z})$ is related to
${\cal F}(z)$ as
\begin{equation}
{\cal F}(z)=
\left(\begin{array}{cc}
\bar{\epsilon}^{-1} &0\\
0&-\zeta
\end{array}\right)\tilde{{\cal F}}(-\tilde{z})^{t}
\left(\begin{array}{cc}
\epsilon^{-1} &0\\
0&-\bar{\zeta}
\end{array}\right)
\end{equation}

\subsection{Superinversion in $\Real^{4|4N}$}
In the four-dimensional superspace, $\Real^{4|4N}$, superinversion may
be defined by the reduction of  
$i_{5}$, given by  eq.(\ref{iB})  with $R$ as in
eq.(\ref{Rzeta}). This definition gives  
$z^{{\cal N}}\rightarrow z^{\prime
{\cal N}}=(x^{\prime\mu},\theta^{\prime a\alpha},
\bt^{\prime\da}_{a})$ where
\begin{equation}
\begin{array}{ccc}
x^{\prime\mu}_{\pm}=-\displaystyle{\frac{x_{\mp}^{\mu}}{x_{\mp}^{2}}}
~~~~&~~~~
\theta^{\prime}{}^{at}=\displaystyle{
-i\frac{1}{x_{-}^{2}}\,\epsilon^{-1}x_{-}{\cdot\sigma}\bar{\theta}_{b}\,
\zeta^{ba}}
~~~~&~~~~
\bar{\theta}^{\prime}_{a}=i\displaystyle{\frac{1}{x_{+}^{2}}\,
\bar{\epsilon}^{-1}x_{+}{\cdot\sigma^{t}}\theta^{bt}\,\bar{\zeta}_{ba}}
\end{array}
\end{equation} 
This may be  rewritten as 
\begin{equation}
\begin{array}{ll}
\theta^{\prime}{}^{a}=i\displaystyle{\frac{1}{x_{-}^{2}}
\,\tilde{\bar{\theta}}{}^{a}x_{-}{\cdot\tilde{\sigma}}}~~~~&~~~~
\tilde{\bar{\theta}}{}^{a}
=-\zeta^{ba}\bar{\theta}_{b}^{t}\bar{\epsilon}\\
{}&{}\\
\bar{\theta}^{\prime}_{a}=-i\displaystyle{\frac{1}{x_{+}^{2}}
\,x_{+}{\cdot\tilde{\sigma}}\tilde{\theta}_{a}}~~~~&~~~~
\tilde{\theta}_{a}=\epsilon\theta^{b}{}^{t}\bar{\zeta}_{ba}
\end{array}
\label{convinv}
\end{equation}
which is just the conventional definition of superinversion in
four-dimensions.\newline
In six-dimensional superspace the transformation $i_{5}$ belongs to the
connected superconformal group and restricting to four-dimensions we
may write from eqs.(\ref{GiB},\,\ref{KiB}) 
\begin{equation}
\begin{array}{cc}
G(i_{5})^{-1}
=G_{+}(i_{5})^{-1}+G_{-}(i_{5})^{-1}~~~~~&~~~~~
K(z;i_{5})=K_{+}(z;i_{5})+K_{-}(z;i_{5})
\end{array}
\end{equation}
where
\begin{equation}
\begin{array}{c}
G_{+}(i_{5})^{-1}
=\left(\begin{array}{ccc}
0&\epsilon^{-1}\,p_{-}&0\\
\bar{\epsilon}\,p_{+}&0&0\\
0&0&\bar{\zeta}_{ab}\,\tau_{-}
\end{array}\right)\\
{}\\
K_{+}(z;i_{5})
=\left(\begin{array}{cc}
i(\bar{\epsilon}x_{+}{\cdot\tilde{\sigma}})^{-1}\,\tau_{-}\,&
-2i(x_{-}{\cdot\tilde{\sigma}})^{-1}\bar{\theta}_{c}
\zeta^{cb}\,p_{-}\\
0&v^{a}_{~c}(z)\zeta^{cb}\,\tau_{+}
\end{array}\right)\\
{}\\
v^{a}{}_{b}(z)=\delta^{a}_{~b}
+4i\displaystyle{\frac{1}{x_{-}^{2}}}\theta^{a}x_{-}{\cdot\sigma}
\bar{\theta}_{b}~~~~~~~~~
\det v(z)=\displaystyle{\frac{x_{-}^{2}}{x_{+}^{2}}}
\end{array}
\end{equation}
and  $G_{-}(i_{5})^{-1},K_{-}(z;i_{5})$ are given by
\begin{equation}
\begin{array}{cc}
G_{-}(i_{5})^{-1}=C^{-1}G_{+}(i_{5})^{t}C~~~~&~~~~
K_{-}(z;i_{5})^{t}=\left(\begin{array}{cc}
\tilde{\gamma}{}^{0}&0\\
0&{\cal E}
\end{array}\right)K_{+}(z;i_{5})^{\dagger}\left(\begin{array}{cc}
\gamma^{0}&0\\
0&\bar{{\cal E}}
\end{array}\right)
\end{array}
\end{equation}
Then  $z\stackrel{i_{5}}{\longrightarrow}z^{\prime}$ is  
given by 
\begin{equation}
G_{+}(i_{5})^{-1}Z_{+}(z)
K_{+}(z;i_{5})=Z_{-}(z^{\prime})
\end{equation}
which can be reduced to
\begin{equation}
{\cal G}(i)^{-1}{\cal Z}(z){\cal K}(z;i)
=\left(\begin{array}{cc}
1&0\\
ix^{\prime}_{-}{\cdot\tilde{\sigma}}^{t}&-2\theta^{\prime a}{}^{t}\\
2\bar{\theta}_{b}^{\prime t}& \delta_{b}^{~a}
\end{array}\right)=
\bar{{\cal Z}}(z^{\prime})^{t}
\label{tZp1}
\end{equation}
where
\begin{equation}
\begin{array}{cc}
{\cal G}(i)^{-1}=\left(\begin{array}{ccc}
\bar{\epsilon}&0&0\\
0&\epsilon^{-1}&0\\
0&0&-\bar{\zeta}_{ab}
\end{array}\right)~~~&~~~
{\cal K}(z;i)=\left(\begin{array}{cc}
i(\bar{\epsilon}x_{+}{\cdot\tilde{\sigma}})^{-1}
&2i(x_{-}{\cdot\tilde{\sigma}})^{-1}\bar{\theta}_{c}\zeta^{cb}\\
0&-v^{a}_{~c}(z)\zeta^{cb}
\end{array}\right)
\end{array}
\end{equation}
Similarly we have
\begin{equation}
\bar{{\cal K}}(z;i)\bar{{\cal Z}}(z){\cal
G}(i)={\cal Z}(z^{\prime})^{t}
\label{tZp2}
\end{equation}
where
\begin{equation}
\begin{array}{c}
\bar{{\cal K}}(z;i)=\left(\begin{array}{cc}
-i(x_{-}{\cdot\tilde{\sigma}}\epsilon)^{-1}&0\\
2i\bar{\zeta}_{ac}\theta^{c}(
x_{+}{\cdot\tilde{\sigma}})^{-1}&-\bar{\zeta}_{ac}v^{-1c}{}_{b}(z)
\end{array}\right)\\
{}\\
v^{-1a}{}_{b}(z)=v^{\dagger}{}^{a}{}_{b}(z)=v^{a}{}_{b}(-z)=\delta^{a}_{~b}
-4i\displaystyle{\frac{1}{x_{+}^{2}}}\theta^{a}x_{+}{\cdot\sigma}
\bar{\theta}_{b}
\end{array}
\end{equation}

\subsection{Functions of Two-points in $\Real^{4|4N}$}
For two points, 
the supersymmetric interval $z^{{\cal A}}_{12}$ defined
in eq.(\ref{susyinterval+}) can be reduced $z^{{\cal
A}}_{12}\rightarrow z_{12}^{{\cal N}}=
(x^{\mu}_{12},\theta^{a\alpha}_{12}, 
\bar{\theta}^{\dot{\alpha}}_{12a})=-z_{21}^{{\cal N}}$ 
\begin{equation}
x_{12}^{\mu}=x_{1}^{\mu}-x_{2}^{\mu}+
i\theta^{a}_{1}\sigma^{\mu}\bar{\theta}_{2a}-
i\theta^{a}_{2}\sigma^{\mu}\bar{\theta}_{1a}
\end{equation}
Now we have from eqs.(\ref{Z+},\,\ref{F+-2})
\begin{equation}
\bar{Z}_{+}(z_{2})Z_{+}(z_{1})=F_{+}(z_{12})=
\left(\begin{array}{cc}
ix_{\bar{2}1}{\cdot\tilde{\sigma}}\,\tau_{+}&-2\bar{\theta}_{21b}\,p_{+}\\
-2\theta^{a}_{21}\,\tau_{+}&\delta^{a}_{~b}\,p_{+} 
\end{array}\right)
\label{F+12}
\end{equation}
where
\begin{equation}
x^{\mu}_{\bar{2}1}=x^{\mu}_{2-}-x^{\mu}_{1+}-2i
\theta^{a}_{1}\sigma^{\mu}\bar{\theta}_{2a}
=x_{21}^{\mu}+i\theta^{a}_{21}\sigma^{\mu}\bar{\theta}_{21a}
=(x_{21}^{\mu})_{-}=-(x_{12}^{\mu})_{+}
\end{equation}
Similarly we write
\begin{equation}
x^{\mu}_{\bar{1}2}=x^{\mu}_{1-}-x^{\mu}_{2+}-2i
\theta^{a}_{2}\sigma^{\mu}\bar{\theta}_{1a}
=x_{12}^{\mu}+i\theta^{a}_{12}\sigma^{\mu}\bar{\theta}_{12a}
=(x_{12}^{\mu})_{-}=-(x_{21}^{\mu})_{+}
\end{equation}
In terms of eqs.(\ref{calZ},\,\ref{calF}) (\ref{F+12}) is reduced to 
\begin{equation}
\bar{{\cal Z}}(z_{2}){\cal Z}(z_{1})={\cal F}(z_{12})=
\left(\begin{array}{cc}
ix_{\bar{2}1}{\cdot\tilde{\sigma}}&-2\bar{\theta}_{21b}\\
-2\theta^{a}_{21}&\delta^{a}_{~b}
\end{array}\right)
\label{czcz}
\end{equation}
From eq.(\ref{sdetf})
\begin{equation}
\mbox{sdet}\:{\cal F}(z_{12})=-x^{2}_{\bar{1}2}
\label{sdetczcz}
\end{equation}
Infinitesimally ${\cal F}(z_{12})$ transforms, from eq.(\ref{calZtr}),
as 
\begin{equation}
\delta{\cal F}(z_{12})=\bar{{\cal H}}(z_{2}){\cal F}(z_{12})
-{\cal F}(z_{12}){\cal H}(z_{1})
\end{equation}
From eqs.(\ref{tZp1},\,\ref{tZp2}) under superinversion
\begin{equation}
\bar{{\cal K}}(z_{2};i){\cal F}(z_{12}) 
{\cal K}(z_{1};i)={\cal F}(z^{\prime}_{12})^{t}
\end{equation}
which gives using eq.(\ref{czcz})
\begin{equation}
\begin{array}{cc}
(x_{2-}{\cdot\tilde{\sigma}})^{-1}{}x_{\bar{2}1}{\cdot\tilde{\sigma}}
(x_{1+}{\cdot\tilde{\sigma}})^{-1}=-x_{\bar{1}2}^{\prime}{\cdot\sigma}
~~~~&~~~~~\displaystyle{x_{\bar{1}2}^{\prime
2}=\frac{x_{\bar{2}1}^{2}}{x_{2-}^{2}\,x_{1+}^{2}}}
\end{array}
\end{equation}
By considering ${\cal F}(z_{12})^{-1}$ we may also show 
\begin{equation}
\bar{\zeta}v^{-1}(z_{2})v(z_{21})v(z_{1})\zeta=v(z^{\prime}_{21})^{t}
\end{equation}

\subsection{Functions of Three-points in $\Real^{4|4N}$}
For three points, just as in the six-dimensional case, we may define
$\widetilde{\Z}_{1}{}^{{\cal N}}\in\Real^{4|4N}$ which transforms 
homogeneously at $z_{1}\in\Real^{4|4N}$.  With eq.(\ref{tildeFZ1}) 
\begin{equation}
\tilde{{\cal F}}(\widetilde{\Z}_{1})=\left(\begin{array}{cc}
-iY_{1-}{\cdot\sigma}&2\Phi_{1b}\\
2\bar{\Phi}^{a}_{1}&\delta^{a}_{~b} 
\end{array}\right)
\end{equation}
where
\begin{equation}
\begin{array}{l}
Y_{1-}{\cdot\sigma}
=(x_{\bar{3}1}{\cdot\tilde{\sigma}})^{-1}{}
x_{\bar{3}2}{\cdot\tilde{\sigma}}(x_{\bar{1}2}{\cdot\tilde{\sigma}})^{-1}\\
{}\\
\Phi_{1a}=i(x_{\bar{3}1}{\cdot\tilde{\sigma}})^{-1}\bt_{31a}
-i(x_{\bar{2}1}{\cdot\tilde{\sigma}})^{-1}\bt_{21a}\\
{}\\
\bar{\Phi}^{a}_{1}=i\theta^{a}_{12}(x_{\bar{1}2}{\cdot\tilde{\sigma}})^{-1}
-i\theta^{a}_{13}(x_{\bar{1}3}{\cdot\tilde{\sigma}})^{-1}
\end{array}
\end{equation}
Using
\begin{equation}
x_{\bar{1}3}{\cdot\tilde{\sigma}}
+x_{\bar{2}1}{\cdot\tilde{\sigma}}-4i\bar{\theta}_{21a}\theta^{a}_{13}
=x_{\bar{2}3}{\cdot\tilde{\sigma}}
\end{equation}
one can show the consistency condition
\begin{equation}
Y_{1+}{\cdot\sigma}=Y_{1-}{\cdot\sigma}-4i\Phi_{1a}\bar{\Phi}{}^{a}_{1}=
-(x_{\bar{2}1}{\cdot\sigma})^{-1}x_{\bar{2}3}{\cdot\sigma}
(x_{\bar{1}3}{\cdot\sigma})^{-1}
=(Y_{1-}{\cdot\sigma})^{\dagger}
\end{equation}
which is necessary to define
$\Z_{1}{}^{{\cal
N}}=(X^{\mu},\Theta^{a},\bar{\Theta}_{a})\in\Real^{4|4N}$ with   
$\Theta^{a},\bar{\Theta}_{a}$ defined according to eq.(\ref{xtbt}).\newline
From eq.(\ref{inftildeF}), $\tilde{{\cal F}}(\widetilde{\Z}_{1})$ 
transforms 
infinitesimally as
\begin{equation}
\begin{array}{ll}
\delta \tilde{{\cal F}}(\widetilde{\Z}_{1})=&\left(\begin{array}{cc}   
\tilde{\hat{w}}(z_{1})-\textstyle{\frac{1}{2}}\hat{\lambda}(z_{1})+i
\textstyle{\frac{1}{2}}\hat{\psi}(z_{1})&0\\
0&\hat{t}(z_{1})+i\textstyle{\frac{2}{N}}\hat{\psi}(z_{1})
\end{array}\right)
\tilde{{\cal F}}(\widetilde{\Z}_{1})\\
{}&{}\\
{}&~~-\tilde{{\cal F}}(\widetilde{\Z}_{1})
\left(\begin{array}{cc}   
\hat{w}(z_{1})+\textstyle{\frac{1}{2}}\hat{\lambda}(z_{1})+i
\textstyle{\frac{1}{2}}\hat{\psi}(z_{1})&0\\
0&\hat{t}(z_{1})+i\textstyle{\frac{2}{N}}\hat{\psi}(z_{1})
\end{array}\right)
\end{array}
\end{equation}

\section{Superconformal Invariance of Correlation Functions}
In this section we assume that there exist quasi-primary
 superfields,
$\Psi^{I}(z)$  which   under the superconformal transformation,\,$g$,
 where  $z\stackrel{g}{\longrightarrow}z^{\prime}$,  transform as
\begin{equation}
\begin{array}{cc}
\Psi^{I}\stackrel{g}{\longrightarrow}\Psi^{\prime}{}^{I}~~~&~~~ 
\Psi^{\prime}{}^{I}(z^{\prime})
=\Psi^{J}(z)D^{~I}_{J}(z;g)
\end{array}
\label{primary}
\end{equation}
$D(z;g)$ obeys the group property
so that under the 
successive superconformal transformations,~$g^{\prime\prime}: 
z\stackrel{g}{\longrightarrow}
z^{\prime}\stackrel{g^{\prime}}{\longrightarrow}z^{\prime\prime}$, it 
satisfies 
\begin{equation}
D(z;g)D(z^{\prime};g^{\prime})=D(z;g^{\prime\prime})
\label{Drep}
\end{equation}
and hence also 
\begin{equation}
D(z;g)^{-1}=D(z^{\prime};g^{-1})
\end{equation}
We choose here $D(z;g)$ to be a representation of 
$\mbox{SO}(1,5)\times\mbox{Sp}(N)\times
\mbox{D}$, which is a subgroup of the
stability group at $z=0$,   
where $\mbox{D}$ is the one dimensional group of dilations, and so with
$\Psi^{I}\equiv \Psi^{\rho r}$ it may be factorized 
\begin{equation}
D_{J}^{~I}(z;g)=D_{\rho}^{~\sigma}(\hat{L}(z;g))D_{r}^{~s}(U(z;g))
\Omega(z;g)^{-\eta}
\label{DLU}
\end{equation}
where 
$D_{\rho}^{~\sigma}(\hat{L}),\,D_{r}^{~s}(U)$ are  representations of 
$\mbox{G}_{L},\,\mbox{Sp}(N)$ respectively and since $\Omega(z;g)$
separately satisfies eq.(\ref{Drep}), $\Omega(z;g)^{-\eta}$ forms a
one dimensional representation of $\mbox{D}$ with $\eta$, the scale
dimension  of $\Psi^{\rho r}$. \newline
Infinitesimally 
\begin{equation}
\delta\Psi^{\rho r}(z)=-({\cal L}+\eta\hat{\lambda}(z))\Psi^{\rho r}(z)
-\Psi^{\sigma r}(z)\textstyle{\frac{1}{2}}(s_{AB})_{\sigma}^{~\rho}
\hat{\omega}^{AB}(z)-\Psi^{\rho
s}(z)\textstyle{\frac{1}{2}}(t_{i}^{~j})_{s}^{~r}\hat{T}_{j}^{~i}(z)
\end{equation}
where $s_{AB},\,t_{i}^{~j}$ are matrix generators of
$\mbox{SO}(1,5),\,\mbox{Sp}(N)$ satisfying
\begin{equation}
\begin{array}{c}
[s_{AB},s_{CD}]=-\eta_{AC}s_{BD}+\eta_{AD}s_{BC}
+\eta_{BC}s_{AD}-\eta_{BD}s_{AC}\\
{}\\
{[}{t_{i}^{~j}},{}{t_{k}^{~l}}{]}=
-\delta^{~j}_{k}t_{i}^{~l}+\delta_{i}^{~l}t_{k}^{~j}
-{\cal E}^{jl}t_{ki}+\bar{\cal E}_{ik}t^{jl}
\end{array}
\label{stcom}
\end{equation}
where $t_{ki}=t_{ik}=t_{k}^{~m}\bar{\cal E}_{mi},\,
t^{jl}=t^{lj}=t_{m}{}^{l}{\cal E}^{mj}$.  Thus
\begin{equation}
[\textstyle{\frac{1}{2}}t_{i}^{~j}T_{1j}{}^{i},
\textstyle{\frac{1}{2}}t_{k}^{~l}T_{2l}{}^{k}]=
\textstyle{\frac{1}{2}}t_{i}^{~j}[T_{1},T_{2}]_{j}^{~i}
\end{equation}
From eqs.(\ref{defH},\,\ref{Kcom}) using eq.(\ref{stcom}) we have
\begin{equation}
\delta_{3}\Psi^{\rho r}=[\delta_{2},\delta_{1}]\Psi^{\rho r}
\end{equation}
Superconformal invariance for a general $n$-point  function requires
\begin{equation}
\langle
\Psi_{1}^{\prime I_{1}}(z_{1})\Psi_{2}^{\prime I_{2}}(z_{2}) 
\cdots\Psi_{n}^{\prime I_{n}}(z_{n})\rangle
=\langle \Psi_{1}^{I_{1}}(z_{1})\Psi_{2}^{I_{2}} (z_{2})
\cdots \Psi_{n}^{I_{n}}(z_{n})\rangle
\label{Green}
\end{equation}

\subsection{Two-point Correlation Functions}
The  solution for 
the two-point  function of the quasi-primary superfields,~$\Psi^{\rho
r}$,  has the general  form  
\begin{equation}
\displaystyle{
\langle \Psi^{\rho r}(z_{1})\Psi^{\sigma s}(z_{2})\rangle = 
C_{\Psi}\frac{I^{\rho\sigma}(\hat{{\cal X}}_{12})I^{rs}(V(z_{12}))}{
(\det {\cal X}_{12})^{\frac{1}{2}\eta}}}
\label{2gen}
\end{equation}
where we define 
\begin{equation}
\displaystyle{\hat{{\cal X}}_{12}=
\frac{{\cal X}_{12}}{(\det {\cal X}_{12})^{\frac{1}{4}}}}
\end{equation}
and $I^{\rho\sigma}(\hat{{\cal X}}_{12}),\,I^{rs}(V(z_{12}))$ are
tensors transforming covariantly according to the    
appropriate representations of $\mbox{G}_{L},\,\mbox{Sp}(N)$ which  are
formed by
decomposition of tensor products of   $\hat{{\cal X}}_{12},\,
V(z_{12})$, where  $V(z_{12})$ is given by
eq.(\ref{Vzz}). \newline
Under superconformal transformations 
$I^{\rho\sigma}(\hat{{\cal X}}_{12}),\,I^{rs}(V(z_{12}))$ satisfy from 
eqs.(\ref{green2},\,\ref{green3})
\begin{subeqnarray}
\label{IIpro}
&D(\hat{L}(z_{1};g))^{t}I(\hat{{\cal X}}_{12})D(\hat{L}(z_{2};g))
=I(\hat{{\cal X}}{}^{\prime}_{12})\\
{}\nonumber\\
&D(U(z_{1};g))^{t}I(V(z_{12}))D(U(z_{2};g))
=I(V(z^{\prime}_{12}))
\end{subeqnarray}
As  examples, we consider  
the  spinorial fields,  $\phi^{\alpha}(z),\bar{\phi}_{\alpha}(z)$ 
which transform as 
\begin{subeqnarray}
&\phi^{\prime\alpha}(z^{\prime})=\Omega(z;g)^{-\eta}\phi^{\beta}(z)
\hat{L}_{\beta}^{~\alpha}(z;g)\\
{}\nonumber\\
&\bar{\phi}^{\prime}_{\alpha}(z^{\prime})=\Omega(z;g)^{-\eta}
\hat{L}_{\alpha}^{~\beta}(z^{\prime};g^{-1})\bar{\phi}_{\beta}(z)
\end{subeqnarray}
so that 
$s_{AB}\rightarrow\textstyle{\frac{1}{2}}\gamma_{[A}\tilde{\gamma}_{B]}
$.\newline  
The two-point functions are
\begin{subequations}
\begin{equation}
\begin{array}{cc}
\displaystyle{
\langle \phi^{\alpha}(z_{1})\phi^{\beta}(z_{2})\rangle =
C_{\phi}
\frac{I^{\alpha\beta}(\hat{{\cal X}}_{12})}{
(\det {\cal X}_{12})^{\frac{1}{2}
\eta}}}
~~~~&~~~~
I^{\alpha\beta}(\hat{{\cal X}}_{12})=\hat{{\cal X}}_{12}^{\alpha\beta}
\end{array}
\end{equation}
{}\newline
\begin{equation}
\begin{array}{cc}
\displaystyle{
\langle \bar{\phi}_{\alpha}(z_{1})\bar{\phi}_{\beta}(z_{2})\rangle =
C_{\bar{\phi}}
\frac{\bar{I}_{\alpha\beta}(\hat{{\cal X}}_{12})}{
(\det {\cal X}_{12})^{\frac{1}{2}\eta}}}~~~~&~~~~
\bar{I}_{\alpha\beta}(\hat{{\cal X}}_{12})
=(\hat{{\cal X}}_{21}^{-1})_{\alpha\beta}
\end{array}
\end{equation}
\end{subequations}
$I^{\alpha\beta}(\hat{{\cal X}}_{12}),
\bar{I}_{\alpha\beta}(\hat{{\cal X}}_{12})$
satisfy
\begin{equation}
I^{\alpha\beta}(\hat{{\cal X}}_{12})
\bar{I}_{\gamma\beta}(\hat{{\cal X}}_{12})
=\delta^{\alpha}_{~\gamma}
\end{equation}
For a  vector field, $V^{A}(z)$,  where the representation
of $\mbox{G}_{L}$  is given by
$\hat{R}_{A}^{~B}(z;g)$, we have
\begin{equation}
\begin{array}{cc}
\displaystyle{
\langle V^{A}(z_{1})V^{B}(z_{2})\rangle =
C_{V}\frac{I^{AB}(\hat{{\cal X}}_{12})}{
(\det {\cal X}_{12})^{\frac{1}{2}\eta}}}~~~~&~~~~
I^{AB}(\hat{{\cal X}}_{12})=\textstyle{\frac{1}{4}}\mbox{tr}
(\gamma^{A}\hat{{\cal X}}_{12}\gamma^{B}\hat{{\cal X}}_{21})
\end{array}
\end{equation}
From eqs.(\ref{geg2},\,\ref{gAgA}) and using  
$\hat{{\cal X}}_{12}^{t}
=\hat{{\cal X}}_{21},\,\det \hat{{\cal X}}_{12}=1$ we have
\begin{equation}
(\hat{{\cal X}}_{12}^{-1}\tilde{\gamma}_{A}
\hat{{\cal X}}_{21}^{-1})_{\alpha\beta}
=\textstyle{\frac{1}{2}}
\epsilon_{\alpha\beta\gamma\delta}(\hat{{\cal X}}_{21}
\gamma_{A}\hat{{\cal X}}_{12})^{\gamma\delta}
\end{equation}
which implies with eq.(\ref{geg1}) 
\begin{equation}
I_{AB}(\hat{{\cal X}}_{12})
=\textstyle{\frac{1}{4}}\mbox{tr}
(\gamma_{A}\hat{{\cal X}}_{12}\gamma_{B}\hat{{\cal X}}_{21})=
\textstyle{\frac{1}{4}}\mbox{tr}
(\tilde{\gamma}_{A}
\hat{{\cal X}}_{21}^{-1}\tilde{\gamma}_{B}\hat{{\cal X}}_{12}^{-1})
\label{IIAB}
\end{equation}
Hence $I^{AB}(\hat{{\cal X}}_{12})$ satisfies
\begin{equation}
I^{AB}(\hat{{\cal X}}_{12})I_{CB}(\hat{{\cal X}}_{12})
=\delta^{A}_{~C}
\end{equation}
Note that $I(\hat{{\cal X}}_{12})\propto R(z_{12};i)$, 
where $R(z;i)$ is given by eq.(\ref{Rsuperinversion}).  \newline
For a  scalar field, $S(z)$ and 
a symplectic field,  $\psi^{i}(z)$,  which transforms as
\begin{equation}
\psi^{\prime}{}^{i}(z^{\prime})
=\Omega(z;g)^{-\eta}\psi^{j}(z)U_{j}^{~i}(z;g)
\end{equation}
the two-point functions are
\begin{equation}
\displaystyle{
\langle S(z_{1})S(z_{2})\rangle = 
C_{S}\frac{1}{(\det {\cal X}_{12})^{\frac{1}{2}\eta}}}
\end{equation}
{}\newline
\begin{equation}
\begin{array}{cc}
\displaystyle{
\langle \psi^{i}(z_{1})\psi^{j}(z_{2})\rangle = 
C_{\psi}
\frac{V^{ij}(z_{12})}{
(\det {\cal X}_{12})^{\frac{1}{2}\eta}}}~~~~&~~~~
V^{ij}(z_{12})=V_{k}^{~j}(z_{12}){\cal E}^{ki}
\end{array}
\end{equation}
The uniqueness of the two-point function can be shown using the method
in \cite{paper1}.

\subsection{Three-point Correlation Functions}
The  solution for 
the three-point  function of the quasi-primary
superfields,~$\Psi^{\rho r}$, 
has the general  form~\cite{paper1}  
\begin{equation}
\displaystyle{
\langle \Psi_{1}^{\rho r}(z_{1})\Psi_{2}^{\sigma s}(z_{2})
\Psi_{3}^{\tau t}(z_{3})\rangle = 
\frac{I^{\sigma\sigma^{\prime}}(\hat{{\cal X}}_{21})
I^{\tau\tau^{\prime}}(\hat{{\cal X}}_{31})I^{ss^{\prime}}(V(z_{21}))
I^{tt^{\prime}}(V(z_{31}))H^{\rho
r}{}_{\sigma^{\prime}s^{\prime}\tau^{\prime} t^{\prime}}(\widetilde{\Z}_{1})
}{
(\det {\cal X}_{31})^{\frac{1}{2}\eta_{3}}
(\det {\cal X}_{12})^{\frac{1}{2}\eta_{2}}}}
\label{3gen}
\end{equation}
where  $\widetilde{\Z}_{1}{}^{\tilde{{\cal A}}}
=(\Y_{1}^{A},\Phi_{1}^{i})\in\Real^{6|8N}_{-}$ is given by 
eqs.(\ref{W1},\,\ref{Ypro}). \newline
Superconformal invariance~(\ref{Green}) is  
equivalent, from eqs.(\ref{green4},\,\ref{expX},\,\ref{IIpro}), to   
\begin{subeqnarray}
\label{equiv}
\begin{array}{c}
H^{\rho^{\prime}r}{}_{\sigma s\tau  t}(\widetilde{\Z})
D_{\rho^{\prime}}{}^{\rho}(\hat{L})
=D_{\sigma}{}^{\sigma^{\prime}}(\hat{L})
D_{\tau}{}^{\tau^{\prime}}(\hat{L})
H^{\rho
r}{}_{\sigma^{\prime}s\tau^{\prime}t}(\widetilde{\Z}{}^{\prime})\\
{}\\
\widetilde{\Z}{}^{\prime\tilde{{\cal A}}}
=( Y^{B}\hat{R}_{B}{}^{A}(\hat{L}),\,\hat{L}^{-1}\Phi^{i})
\end{array}\\
{}\nonumber\\
{}\nonumber\\
\begin{array}{c}
H^{\rho r^{\prime}}{}_{\sigma s\tau  t}(\widetilde{\Z})
D_{r^{\prime}}{}^{r}(U)=
D_{s}{}^{s^{\prime}}(U)
D_{t}{}^{t^{\prime}}(U)H^{\rho
r}{}_{\sigma s^{\prime}\tau 
t^{\prime}}(\widetilde{\Z}{}^{\prime\prime})\\
{}\\
\widetilde{\Z}{}^{\prime\prime\tilde{{\cal A}}}=(
Y^{A},\,\Phi^{j}U_{j}^{~i})
\end{array}\\
{}\nonumber\\
{}\nonumber\\
\begin{array}{c}
H^{\rho r}{}_{\sigma s\tau  t}(\widetilde{\Z})
=\lambda^{\eta_{2}+\eta_{3}-\eta_{1}}{}
H^{\rho r}{}_{\sigma s\tau
t}(\widetilde{\Z}{}^{\prime\prime\prime})~~~~~~~~~~~\\
{}\\
\widetilde{\Z}{}^{\prime\prime\prime\tilde{{\cal A}}}
=( \lambda Y^{A},\,\lambda^{\frac{1}{2}}\Phi^{i})~~~~~~~~~
\end{array}
\end{subeqnarray}
Note that $\hat{L}\in\mbox{G}_{L},\,U\in\mbox{Sp}(N),\,\lambda\in\Real$
and $\hat{R}_{B}{}^{A}(\hat{L})$ is given
by eq.(\ref{hatR}).  \newline
In general there may be  a finite number, $n$, 
linearly independent solutions
of eq.(\ref{equiv}) 
so that the three-point function depends on $n$ 
parameters.   
\newline
\begin{center}
\large{\textbf{Acknowledgements}}
\end{center}
I am  deeply indebted  to Hugh Osborn for his careful guidance 
throughout  the time from the inception of this research to its  
writing up  in this paper.  \newline
This work was partly supported by Cambridge Overseas Trust.


\newpage
\appendix
\begin{center}
\Large{\textbf{Appendix}}
\end{center}
\section{Notations \& Useful Equations}
Using eqs.(\ref{gaga},\,\ref{contract}) one can derive  
\begin{subeqnarray}
\label{tildetilde}
(\gamma^{A})_{\alpha\beta}(\gamma_{A})_{\gamma\delta}=
2c\,\epsilon_{\alpha\beta\gamma\delta}\label{gAgA}\\
{}\nonumber\\
(\tilde{\gamma}^{A})^{\alpha\beta}(\tilde{\gamma}_{A})^{\gamma\delta}=
2\frac{1}{c}\,\epsilon^{\alpha\beta\gamma\delta}
\end{subeqnarray}
where $\epsilon_{1234}=\epsilon^{1234}=1$ and $c$ is a constant, equal
to the Pfaffian of $\gamma^{0}$, which can be without loss of 
generality taken to be $-1$ to get eqs.(\ref{geg1},\,\ref{geg2}).\newline
Due to  the identities\footnote{We put $\epsilon^{012345}=1$ and
$[\,]$ means  anti-symmetrizing indices with ``strength one''.} 
\begin{equation}
\begin{array}{cc}
\gamma^{[A}\tilde{\gamma}^{B}\gamma^{C]}=-\textstyle{\frac{1}{6}}
\epsilon^{ABC}{}_{DEF}\,\gamma^{D}\tilde{\gamma}^{E}\gamma^{F}~~~&~~~
\tilde{\gamma}^{[A}\gamma^{B}\tilde{\gamma}^{C]}=\textstyle{\frac{1}{6}}
\epsilon^{ABC}{}_{DEF}\,\tilde{\gamma}^{D}\gamma^{E}\tilde{\gamma}^{F}
\end{array}
\end{equation}
there are only $10$ independent $\gamma^{[A}\tilde{\gamma}^{B}\gamma^{C]}$ and 
$\tilde{\gamma}^{[A}\gamma^{B}\tilde{\gamma}^{C]}$  separately and 
 both of them form   bases 
of $4 \times 4$ symmetric matrices  with the completeness relation 
\begin{equation}
(\gamma^{[A}\tilde{\gamma}^{B}\gamma^{C]})_{\alpha\beta}
(\tilde{\gamma}_{[A}\gamma_{B}\tilde{\gamma}_{C]})^{\gamma\delta}=
-24(\delta^{~\gamma}_{\alpha}\delta^{~\delta}_{\beta}
+\delta^{~\delta}_{\alpha}\delta^{~\gamma}_{\beta}) 
\end{equation}
The coefficient on the right hand side may be determined by 
\begin{equation}
\mbox{tr}(\gamma^{[A}\tilde{\gamma}^{B}\gamma^{C]}
\tilde{\gamma}_{[D}\gamma_{E}\tilde{\gamma}_{F]})=4\epsilon^{ABC}{}_{DEF}
-24\delta^{[A}_{~D}\delta^{B}_{~E}\delta^{C]}_{~F}
\end{equation}
A particular convenient choice for   $\gamma^{A},\tilde{\gamma}^{A}$   is
\begin{equation}
\begin{array}{cll}
\gamma^{\mu}=\left(\begin{array}{cc}
                 0&-\sigma^{\mu}{}^{t}\\
                  \sigma^{\mu} &0 \end{array}\right)~~~&~~~
\gamma^{4}=\left(\begin{array}{cc}
                           \bar{\epsilon} &0 \\
                           0&\epsilon \end{array}\right)~~~&~~~
\gamma^{5}=i\left(\begin{array}{cc}
                           \bar{\epsilon}&0\\
                           0&-\epsilon \end{array}\right)\\
{}&{}&{}\\
\tilde{\gamma}^{\mu}=\left(\begin{array}{cc}
                 0&\tilde{\sigma}^{\mu}\\
                  -\tilde{\sigma}^{\mu}{}^{t} &0 \end{array}\right)~~~&~~~
\tilde{\gamma}^{4}=-\left(\begin{array}{cc}
                           \bar{\epsilon}^{-1} &0 \\
                           0&\epsilon^{-1} \end{array}\right)
~~~&~~~\tilde{\gamma}^{5}=i\left(\begin{array}{cc}
                           \bar{\epsilon}^{-1}&0\\
                           0&-\epsilon^{-1} \end{array}\right)
\end{array}
\label{ourgamma}
\end{equation}
where
$\epsilon_{\alpha\beta},\,\bar{\epsilon}_{\dot{\alpha}\dot{\beta}}$ are
the $2\times 2$ anti-symmetric matrices,  $
\epsilon_{12}=\bar{\epsilon}_{12}=1$ with  inverses,  \newline
$(\epsilon^{-1})^{\alpha\beta},(\bar{\epsilon}^{-1})^{\dot{\alpha}\dot{\beta}}$
and $(\sigma^{\mu})_{\alpha\dot{\alpha}},
(\tilde{\sigma}^{\mu})^{\dot{\alpha}\alpha},\,\mu=0,1,2,3$  are  
$2\times 2$  matrices satisfying  
\begin{equation}
\begin{array}{ll}
\sigma^{\mu}\tilde{\sigma}^{\nu}+
\sigma^{\nu}\tilde{\sigma}^{\mu}=2\eta^{\mu\nu}~~~&~~~
\eta^{\mu\nu}=\mbox{diag}(+1,-1,-1,-1)\\
{}&{}\\
\sigma^{\mu}_{\alpha\dot{\alpha}}\tilde{\sigma}_{\mu}^{\dot{\beta}\beta}
=2\delta_{\alpha}^{~\beta}\delta_{\dot{\alpha}}^{~\dot{\beta}}~~~~&~~~~
\sigma^{\mu}_{\alpha\dot{\alpha}}\sigma_{\mu}{}_{\beta\dot{\beta}}
=2\epsilon_{\alpha\beta}\bar{\epsilon}_{\dot{\alpha}\dot{\beta}}\\
{}&{}\\
\epsilon\tilde{\sigma}^{\mu}{}^{t}\bar{\epsilon}=-\sigma^{\mu}~~~~&~~~~
\bar{\epsilon}^{-1}\sigma^{\mu}{}^{t}\epsilon^{-1}=-\tilde{\sigma}^{\mu}
\end{array}
\end{equation}
We define the conjugate of $D_{i\alpha},\tilde{D}_{i}^{\alpha}$ to be 
\begin{equation}
\begin{array}{cc}
\displaystyle{
\bar{D}^{i}_{\alpha}=-\frac{\partial~}{\partial\bar{\theta}^{\alpha}_{i}}+
i(\gamma^{A}\theta^{i})_{\alpha}\frac{\partial~}{\partial
x^{A}}}~~~&~~~
\displaystyle{
\bar{\tilde{D}}{}^{i\alpha}=
-\frac{\partial~}{\partial\bar{\tilde{\theta}}_{i\alpha}}+
i(\tilde{\gamma}^{A}\tilde{\theta}^{i})^{\alpha}
\frac{\partial~}{\partial x^{A}}}
\end{array}
\end{equation}
so that they  satisfy the pseudo-Majorana condition 
\begin{equation}
\begin{array}{cc}
D_{i\alpha}=\bar{D}^{j}_{\alpha}\bar{\cal E}_{ji}~~~&~~~
\tilde{D}_{i}^{\alpha}=\bar{\tilde{D}}{}^{j\alpha}\bar{\cal E}_{ji}
\end{array}
\end{equation}

\section{Derivation of 
Eqs.(\ref{btheta},\,\ref{Wtheta1},\,\ref{Wtheta2},\,\ref{atheta1},\,\ref{atheta2})}
We recall eqs.(\ref{master},\,\ref{tildeh})
\begin{eqnarray}
&D_{i\alpha}\tilde{{\rm h}}^{\beta\gamma}=\textstyle{\frac{1}{3}}(
\delta_{\alpha}^{~\beta}D_{i\delta}\tilde{{\rm h}}^{\delta\gamma}-
\delta_{\alpha}^{~\gamma}D_{i\delta}\tilde{{\rm h}}^{\delta\beta})\\
&{}\nonumber\\
&\tilde{{\rm h}}(x,\theta)=\tx_{-}{\rm b}(\theta)\tx_{+}+W(\theta)\tx_{+}
+\tx_{-}W^{t}(\theta)+\tilde{{\rm A}}(\theta)
\end{eqnarray}
Writing  $\Psi_{\alpha\beta\gamma}=D_{i\alpha}{\rm
b}_{\beta\gamma}(\theta),~\Psi_{\alpha}
=\tx^{\beta\gamma}\Psi_{\beta\gamma\alpha}$, the $x^{2}$ terms read
\begin{equation}
3x^{2}\Psi_{\alpha\beta\gamma}=\x_{\beta\alpha}\Psi_{\gamma}-
\x_{\gamma\alpha}\Psi_{\beta}
\end{equation}
and so
\begin{equation}
3x^{2}(\Psi_{\alpha\beta\gamma}+\Psi_{\beta\alpha\gamma}+
\Psi_{\gamma\beta\alpha})=2\x_{\beta\gamma}\Psi_{\alpha}
\end{equation}
Contracting with $\tx^{\gamma\beta}$ gives
\begin{equation}
2\Psi_{\alpha}=3\tx^{\gamma\beta}\Psi_{\alpha\beta\gamma}
\end{equation}
which implies $\tilde{\gamma}_{A}^{\beta\gamma}(2D_{i\beta}{\rm
b}_{\gamma\alpha}+3D_{i\alpha}{\rm b}_{\beta\gamma})=0$, and by
eq.(\ref{contract}) we get
\begin{equation}
3D_{i\alpha}{{\rm b}}_{\beta\gamma}+D_{i\beta}{{\rm
b}}_{\gamma\alpha}+D_{i\gamma}{{\rm b}}_{\alpha\beta}=0
\end{equation}
which gives $D_{i\alpha}{{\rm b}}_{\beta\gamma}=-\frac{3}{2}
D_{i[\alpha}{{\rm b}}_{\beta\gamma]}=0$. Thus, ${{\rm b}}(\theta)$ is
independent of $\theta$.\newline
Now the terms linear in $x$  read
\begin{equation}
3D_{i\alpha}(W(\theta)\tilde{\gamma}^{A})^{[\beta\gamma ]}=
\delta_{\alpha}^{~\beta}D_{i\delta}
(W(\theta)\tilde{\gamma}^{A})^{[\delta\gamma]}-
\delta_{\alpha}^{~\gamma}D_{i\delta}
(W(\theta)\tilde{\gamma}^{A})^{[\delta\beta]}
\end{equation}
Contracting with $\gamma_{A\delta\epsilon}$ gives
\begin{equation}
3D_{i\alpha}W^{[\gamma}_{[\delta}\delta^{\beta ]}_{~\epsilon ]}=
D_{i\eta}W^{\eta}_{~[\delta}\delta^{[\beta}_{~\epsilon
]}\delta^{\gamma ]}_{~\alpha}+D_{i[\delta}W^{[\beta}_{~\epsilon
]}\delta^{\gamma ]}_{~\alpha}
\end{equation}
Contracting with $\delta^{\delta}_{~\gamma}$ again, we get
\begin{equation}
2(D_{i\alpha}W^{\beta}_{~\gamma}-\delta_{\alpha}^{~\beta}D_{i\gamma}
W^{\delta}_{~\delta})=D_{i\delta}W^{\delta}_{~\{\alpha}
\delta^{\beta}_{~\gamma\}}-D_{i\{\alpha}W^{\beta}_{~\gamma\}}
-3D_{i\{\alpha}\delta^{\beta}_{~\gamma\}}W^{\delta}_{~\delta}
\label{DW}
\end{equation}
The right hand side is symmetric over $\alpha\leftrightarrow\gamma$ and so 
\begin{equation}
D_{i\alpha}W^{\beta}_{~\gamma}-\delta_{\alpha}^{~\beta}D_{i\gamma}
W^{\delta}_{~\delta}=D_{i\gamma}W^{\beta}_{~\alpha}
-\delta_{\gamma}^{~\beta}D_{i\alpha}W^{\delta}_{~\delta}
\end{equation}
Contracting with $\delta^{\alpha}_{~\beta}$ gives
$D_{i\alpha}W^{\alpha}_{~\beta}=4D_{i\beta}W^{\beta}_{~\beta}$.
Substituting this back into eq.(\ref{DW}) gives
\begin{equation}
D_{i\alpha}W^{\beta}_{~\gamma}-
\delta_{\alpha}^{~\beta}D_{i\gamma}W^{\delta}_{~\delta}=-
5(D_{i\gamma}W^{\beta}_{~\alpha}-
\delta_{\gamma}^{~\beta}D_{i\alpha}W^{\delta}_{~\delta})
\end{equation}
which implies eq.(\ref{Wtheta1}) and
\begin{equation}
D_{j\delta}D_{i\alpha}W^{\beta}_{~\gamma}=-
\delta_{\alpha}^{~\beta}D_{i\gamma}
D_{j\delta}W^{\epsilon}_{~\epsilon}=-
D_{i\gamma}D_{j\alpha}W^{\beta}_{~\delta}
=\delta^{\beta}_{~\gamma}D_{j\alpha}D_{i\delta}W^{\epsilon}_{~\epsilon}
\end{equation}
Contracting with $\delta_{\beta}^{~\gamma}$ gives
\begin{equation}
D_{j\beta}D_{i\alpha}W^{\gamma}_{~\gamma}
=4D_{j\alpha}D_{i\beta}W^{\gamma}_{~\gamma}
\end{equation}
which implies $D_{j\beta}D_{i\alpha}W^{\gamma}_{~\gamma}=0$ and so  
eq.(\ref{Wtheta2}).\newline
Acting $D_{j\delta}$ on eq.(\ref{remain}) reads
\begin{equation}
\textstyle{\frac{9}{2}}
D_{j\delta}D_{i\alpha}\tilde{{\rm A}}^{\beta\gamma}=D_{i\delta}D_{j\epsilon}
\tilde{{\rm A}}^{\epsilon[\beta}\delta_{\alpha}^{~\gamma ]}+
\delta_{\delta}^{~[\beta}\delta_{\alpha}^{~\gamma ]}D_{i\epsilon}
D_{j\eta}\tilde{{\rm A}}^{\epsilon\eta}
\label{DDa}
\end{equation}
Contracting with $\delta^{\alpha}_{~\beta}$ gives
\begin{equation}
3D_{j\delta}D_{i\alpha}\tilde{{\rm A}}^{\alpha\gamma}+D_{i\delta}D_{j\alpha}
\tilde{{\rm A}}^{\alpha\gamma}=-\delta^{~\gamma}_{\delta}D_{i\epsilon}
D_{j\eta}\tilde{{\rm A}}^{\epsilon\eta}
\end{equation}
The right hand side of which is symmetric over $i\leftrightarrow j$,
and so 
\begin{equation}
D_{i\alpha}D_{j\beta}\tilde{{\rm A}}^{\beta\gamma}
=D_{j\alpha}D_{i\beta}\tilde{{\rm A}}^{\beta\gamma}
=
-\textstyle{\frac{1}{4}}\delta_{\alpha}^{~\gamma}D_{i\epsilon}
D_{j\eta}\tilde{{\rm A}}^{\epsilon\eta}
\end{equation}
Substituting this back into eq.(\ref{DDa}) reads
eq.(\ref{atheta1}). Eq.(\ref{atheta1}) implies
\begin{equation}
D_{i\alpha}D_{j\beta}D_{k\gamma}\tilde{{\rm A}}^{\beta\gamma}=
D_{j\beta}D_{k\gamma}D_{i\alpha}\tilde{{\rm A}}^{\beta\gamma}= 
-\textstyle{\frac{1}{4}}D_{j\alpha}D_{k\beta}D_{i\gamma}\tilde{{\rm A}}^{\beta\gamma}
\end{equation}
and so
$D_{i\alpha}D_{j\beta}D_{k\gamma}\tilde{{\rm A}}^{\beta\gamma}=0$. 
Therefore, we get  eq.(\ref{atheta2}). 
\section{$(N,0)$ Superconformal Algebra}
We write the superconformal generators in general as 
\begin{equation}
\chi=a^{A}P_{A}+\bar{\varepsilon}_{i}Q^{i}+\lambda D
+\textstyle{\frac{1}{2}}\omega^{AB}M_{AB}+b^{A}K_{A}+
\bar{\rho}_{i}S^{i}+\textstyle{\frac{1}{2}}T_{i}^{~j}A_{j}^{~i}
\end{equation}
where the $\mbox{Sp}(N)$ generators, $A_{i}^{~j}$, satisfy
$A^{\dagger}=-A,~A^{t}{\cal E} +{\cal E} A=0$. \newline
The $(N,0)$ superconformal algebra can now be obtained by imposing
\begin{equation}
[\chi_{1},\chi_{2}]=-i\chi_{3}
\end{equation}
where $\chi_{i}$ is defined by substituting 
$(a^{A},\varepsilon^{i},\lambda,\omega^{AB},b^{A},\rho^{i},T_{i}^{~j})$
by the corresponding coefficient appearing in eq.(\ref{MMcom}).  
From this expression, we can read off the following $(N,0)$ superconformal algebra. 
\begin{itemize}
\item Poincar\'{e} algebra
\begin{equation}
\begin{array}{cc}
[P_{A},P_{B}]=0 & [M_{AB},P_{C}]=i(\eta_{AC}P_{B}-\eta_{BC}P_{A})\\
{}&{}\\
\multicolumn{2}{c}{[M_{AB},M_{CD}]=
i(\eta_{AC}M_{BD}-\eta_{AD}M_{BC}-\eta_{BC}M_{AD}+\eta_{BD}M_{AC})}
\end{array}
\label{poincare}
\end{equation}
\item Supersymmetry algebra
\begin{equation}
\begin{array}{cc}
[P_{A},Q^{i}]=0~~~&~~~
\{Q^{i},Q^{j}\}=2{\cal E}^{ij}
\gamma^{A}P_{A} \\
{}&{}\\
\multicolumn{2}{c}{
[M_{AB},Q^{i}]=i\textstyle{\frac{1}{2}}\gamma_{[A}\tilde{\gamma}_{B]}Q^{i}}
\end{array}
\end{equation}
\item Special superconformal algebra
\begin{equation}
\begin{array}{ll}
[K_{A},K_{B}]=0~~~ &~~~
[M_{AB},K_{C}]=i(\eta_{AC}K_{B}-\eta_{BC}K_{A})\\
{}&{}\\
\displaystyle{[K_{A},S^{i}]=0} ~~~&~~~ 
\{S^{i},S^{j}\}=2{\cal E}^{ij}\tilde{\gamma}^{A}K_{A}\\
{}&{}\\
\multicolumn{2}{c}{
[M_{AB},S^{i}]=i\textstyle{\frac{1}{2}}\tilde{\gamma}_{[A}\gamma_{B]}S^{i}}
\end{array}
\end{equation}
\item Cross terms between $(P,Q)$ and $(K,S)$
\begin{equation}
\begin{array}{c}
[P_{A},K_{B}]=2i(M_{AB}+\eta_{AB}D)\\
{}\\
\displaystyle{[P_{A},S^{i}]=\tilde{\gamma}_{A}Q^{i}}~~~~~~~~~
[K_{A},Q^{i}]=\gamma_{A}S^{i}\\
{}\\
\{Q^{i}_{\alpha},S^{j\beta}\}=i{\cal E}^{ij}
(2\delta_{\alpha}^{~\beta}D
+(\gamma^{[A}\tilde{\gamma}^{B]})_{\alpha}^{~\beta}
M_{AB})
-4i\delta_{\alpha}^{~\beta}A^{ij}
\end{array}
\end{equation}
\item Dilations
\begin{equation}
\begin{array}{cc}
[D,P_{A}]=-iP_{A}~~~&~~~[D,K_{A}]=iK_{A}\\
{}&{}\\
{[D,Q^{i}]=-i\textstyle{\frac{1}{2}}Q^{i}}~~~&~~~
[D,S^{i}]=i\textstyle{\frac{1}{2}}S^{i}\\
{}&{}\\
\multicolumn{2}{c}{[D,D]=[D,M_{AB}]=[D,A_{i}^{~j}]=0}
\end{array}
\end{equation}
\item R-symmetry, $\mbox{Sp}(N)$
\begin{equation}
\begin{array}{c}
[A_{i}^{~j},A_{k}^{~l}]=i
(\delta^{~j}_{k}A_{i}^{~l}-\delta_{i}^{~l}A_{k}^{~j}
+{\cal E}^{jl}A_{ki}-\bar{\cal E}_{ik}A^{jl})\\
{}\\
{[A_{i}^{~j},Q^{k}]=i(
{\cal E}^{jk}\bar{Q}_{i}^{t}-\delta_{i}^{~k}Q^{j})}\\
{}\\
{[A_{i}^{~j},S^{k}]=i
({\cal E}^{jk}\bar{S}_{i}^{t}-\delta_{i}^{~k}S^{j})}\\
{}\\
\displaystyle{[A_{i}^{~j},P_{A}]=[A_{i}^{~j},K_{A}]=[A_{i}^{~j},M_{AB}]=0}
\end{array}
\label{SpN}
\end{equation}
where $A_{ij}=A_{ji}=A_{i}^{~k}\bar{{\cal E}}_{kj},\,
A^{ij}=A^{ji}=A_{k}{}^{j}{\cal E}^{ki}$.
\end{itemize}

\section{Realization of $\mbox{O}(2,6)$ structure in $M$}
We exhibit explicitly the relation of the six-dimensional conformal
group to $\mbox{O}(2,6)$ by introducing eight-dimensional gamma
matrices with $R=0,1,\cdots,7$
\begin{equation}
\left(\begin{array}{cc}
      0&\Sigma^{R}\\
   \tilde{\Sigma}^{R}&0
\end{array}\right)
\label{8gamma}
\end{equation}
$\Sigma^{R},\tilde{\Sigma}^{R}$ satisfy
\begin{equation}
\Sigma^{R}\tilde{\Sigma}^{S}+
\Sigma^{S}\tilde{\Sigma}^{R}=
2G^{RS}
\end{equation}
where $G^{RS}
=\mbox{diag}(+1,-1,-1,\cdots,-1,+1)$. In particular, here we
choose $\Sigma^{R},\tilde{\Sigma}^{R}$ as
\begin{equation}
\begin{array}{cll}
\Sigma^{A}=\left(\begin{array}{cc}
                 \tilde{\gamma}^{A} &0\\
                     0&\gamma^{A}
                  \end{array}\right)~~~~&~~~
\Sigma^{6}=\left(\begin{array}{cc}
                     0&i\\
                     -i&0
                  \end{array}\right)~~~~&~~~
\Sigma^{7}=\left(\begin{array}{cc}
                     0&i\\
                      i&0
                  \end{array}\right)\\
{}&{}&{}\\
\tilde{\Sigma}^{A}=\left(\begin{array}{cc}
                      \gamma^{A} &0\\
                     0&\tilde{\gamma}^{A}
                  \end{array}\right)~~~~&~~~
\tilde{\Sigma}^{6}=-\Sigma^{6} ~~~~&~~~
\tilde{\Sigma}^{7}=-\Sigma^{7}
\end{array}
\end{equation}
$\Sigma_{R},\tilde{\Sigma}_{R}$ satisfy
\begin{equation}
\begin{array}{c}
\left(\begin{array}{cc}
        0&\tilde{\gamma}^{0}\\
           \gamma^{0}&0
        \end{array}\right)\Sigma_{R}\left(\begin{array}{cc}
        0&\gamma^{0}\\
       \tilde{\gamma}^{0}&0
        \end{array}\right)=\tilde{\Sigma}_{R}{}^{\dagger}=\Sigma^{R}\\
{}\\
\left(\begin{array}{cc}
        0&1\\
        1&0
        \end{array}\right)\Sigma_{R}\left(\begin{array}{cc}
        0&1\\
         1&0
        \end{array}\right)=-\tilde{\Sigma}_{R}{}^{t}
\end{array}
\label{bc}
\end{equation}
For the matrix,~$M$, given in eq.(\ref{Mform}), we may now express 
the $8\times 8$ part in terms of $\Sigma^{[R}\tilde{\Sigma}^{S]}$
\begin{equation}
\left(\begin{array}{cc}
\omega+\textstyle{\frac{1}{2}}\lambda &-i\ta\\
-i{\rm b} & \tilde{\omega}-\textstyle{\frac{1}{2}}\lambda
\end{array}\right)
=\textstyle{\frac{1}{4}}\omega_{RS}
\Sigma^{[R}\tilde{\Sigma}^{S]}
\end{equation}
where $\omega_{67},\,\omega_{A6},\,\omega_{A7}$ are given by
\begin{equation}
\begin{array}{ccc}
\omega_{67}=\lambda~~~&~~~~\omega_{A6}=a_{A}-b_{A}~~~&~~~~
\omega_{A7}=a_{A}+b_{A}
\end{array}
\end{equation}
$\Sigma^{RS}\equiv
\textstyle{\frac{1}{2}}\Sigma^{[R}\tilde{\Sigma}^{S]}$ 
generates the Lie algebra of $\mbox{O}(2,6)$
\begin{equation}
[\Sigma^{RS},\Sigma^{TU}]=
-G^{RT}\Sigma^{SU}+G^{RU}\Sigma^{ST}
+G^{ST}\Sigma^{RU}-G^{SU}\Sigma^{RT}
\end{equation}
Eqs.(\ref{Mdagger},\,\ref{Mt}), the conditions on $M$, are 
satisfied by eq.(\ref{bc}).\newline
The result on
superinversion~(\ref{resulti}) corresponds to the reflection of the
sixth axis.

\section{Superconformally Covariant Operators}
In general acting on a quasi-primary superfield, $\Psi^{\rho r}(z)$, 
with the spinor derivative, $D_{i\alpha}$, does not lead to a
quasi-primary field.  
For a superfield, $\Psi^{\rho r}$, from
eqs.(\ref{comDL},\,\ref{Ddelta})  we have
\begin{equation}
\begin{array}{ll}
D_{i\alpha}\delta\Psi^{\rho r}=&
-({\cal L}+(\eta+\textstyle{\frac{1}{2}})\hat{\lambda})
D_{i\alpha}\Psi^{\rho r}
+\tilde{\hat{\omega}}_{\alpha}{}^{\beta}D_{i\beta}\Psi^{\rho r}
-D_{i\alpha}\Psi^{\sigma r}
\textstyle{\frac{1}{2}}(s_{AB}\hat{\omega}^{AB})_{\sigma}^{~\rho}\\
{}&{}\\
{}&{}+\hat{T}_{i}^{~j}D_{j\alpha}\Psi^{\rho r}
-D_{i\alpha}\Psi^{\rho
s}\textstyle{\frac{1}{2}}(t_{j}^{~k}\hat{T}_{k}^{~j})_{s}^{~r}+
2\bar{\hat{\rho}}_{j\beta}(\Psi Y^{j\beta}_{i\alpha})^{\rho r}
\end{array}
\label{infDPSI}
\end{equation}
We may connect the generator of $\mbox{G}_{L}$ to $\mbox{SO}(1,5)$ by
\begin{equation}
s_{\alpha}{}^{\beta}\equiv -\textstyle{\frac{1}{2}}s_{AB}
(\gamma^{[A}\tilde{\gamma}{}^{B]})_{\alpha}{}^{\beta}
\end{equation}
where
\begin{equation}
[s_{\alpha}{}^{\beta},s_{\gamma}{}^{\delta}]=2\delta_{\alpha}^{~\delta}
s_{\gamma}{}^{\beta}-2\delta_{\gamma}^{~\beta}s_{\alpha}{}^{\delta}
\end{equation}
so that  
$s_{\alpha}{}^{\beta}\tilde{\hat{\omega}}_{\beta}{}^{\alpha}
=s_{AB}\hat{\omega}^{AB}$
and then 
\begin{equation}
Y^{j\beta}_{i\alpha}=\eta\delta_{i}^{~j}\delta_{\alpha}^{~\beta}
-\delta_{i}^{~j}s_{\alpha}{}^{\beta}+2t_{i}^{~j}\delta_{\alpha}^{~\beta}
\end{equation}
To ensure that $D_{i\alpha}\Psi^{\rho r}$ is quasi-primary it is
necessary that the terms proportional to $\hat{\rho}$ vanish and this can be
achieved by restricting $D_{i\alpha}\Psi^{\rho r}$ to an irreducible
representation of $\mbox{G}_{L},\mbox{Sp}(N)$ and choosing a
particular value of $\eta$ so that $\Psi Y=0$. The change of the scale dimension,
$\eta\rightarrow\eta+\frac{1}{2}$,  in eq.(\ref{infDPSI}) is   also
apparent  from eq.(\ref{Dtrans}) 
\begin{equation}
D_{i\alpha}=\Omega(z;g)^{1/2}\hat{L}_{\alpha}^{~\beta}(z;g)U_{i}^{~j}(z;g)
D^{\prime}_{j\beta}
\end{equation}
As an illustration we consider tensorial
fields, $\Psi^{\alpha_{1}\cdots\alpha_{l}}_{
\beta_{1}\cdots\beta_{m}i_{1}\cdots i_{n}}$, which transform as
\begin{equation}
\begin{array}{ll}
\delta\Psi^{\alpha_{1}\cdots\alpha_{l}}_{
\beta_{1}\cdots\beta_{m}i_{1}\cdots i_{n}}=&\displaystyle{
-({\cal L}+\eta\hat{\lambda})\Psi^{\alpha_{1}\cdots\alpha_{l}}_{
\beta_{1}\cdots\beta_{m}i_{1}\cdots i_{n}}}\\
{}&{}\\
{}&\,\displaystyle{
-\sum^{l}_{p=1}\Psi^{\alpha_{1}\cdots\gamma\cdots\alpha_{l}}_{
\beta_{1}\cdots\beta_{m}i_{1}\cdots i_{n}}\tilde{\hat{\omega}}_{\gamma}
{}^{\alpha_{p}}+\sum^{m}_{q=1}\tilde{\hat{\omega}}_{\beta_{q}}{}^{\gamma}
\Psi^{\alpha_{1}\cdots\alpha_{l}}_{
\beta_{1}\cdots\gamma\cdots\beta_{m}i_{1}\cdots i_{n}}}\\
{}&{}\\
{}&\,\displaystyle{+\sum^{n}_{r=1}\hat{T}_{i_{r}}{}^{j}
\Psi^{\alpha_{1}\cdots\alpha_{l}}_{
\beta_{1}\cdots\beta_{m}i_{1}\cdots j\cdots
i_{n}}}
\end{array}
\end{equation}
In this case we have
\begin{equation}
\begin{array}{ll}
(\Psi Y^{j\beta}_{i\alpha})^{\alpha_{1}\cdots\alpha_{l}}_{
\beta_{1}\cdots\beta_{m}i_{1}\cdots i_{n}}
=&2\delta_{i}^{~j}\left(\displaystyle{-\sum_{p=1}^{l}}
\,\delta_{\alpha}^{~\alpha_{p}}
\Psi^{\alpha_{1}\cdots\beta\cdots\alpha_{l}}_{
\beta_{1}\cdots\beta_{m}i_{1}\cdots i_{n}}
+\displaystyle{\sum_{q=1}^{m}}\,\delta_{\beta_{q}}^{~\beta}
\Psi^{\alpha_{1}\cdots\alpha_{l}}_{
\beta_{1}\cdots\alpha\cdots\beta_{m}i_{1}\cdots i_{n}}\right)\\
{}&{}\\
{}&{}+2\delta_{\alpha}^{~\beta}\left(
\displaystyle{\sum_{r=1}^{n}}\,(
\bar{{\cal E}}_{ii_{r}}{\cal E}^{jk}
\Psi^{\alpha_{1}\cdots\alpha_{l}}_{
\beta_{1}\cdots\beta_{m}i_{1}\cdots k\cdots i_{n}}-\delta_{i_{r}}^{~j}
\Psi^{\alpha_{1}\cdots\alpha_{l}}_{
\beta_{1}\cdots\beta_{m}i_{1}\cdots i\cdots i_{n}})\right)\\
{}&{}\\
{}&{}+(\eta+\textstyle{\frac{1}{2}}l-\textstyle{\frac{1}{2}}m)
\delta_{i}^{~j}\delta_{\alpha}^{~\beta}
\Psi^{\alpha_{1}\cdots\alpha_{l}}_{
\beta_{1}\cdots\beta_{m}i_{1}\cdots i_{n}}
\end{array}
\label{rhoY}
\end{equation}
In particular, 
eq.(\ref{rhoY}) shows that the following are quasi-primary
\begin{subeqnarray}
D_{(i(\alpha}\Psi_{
\beta_{1}\cdots\beta_{m})i_{1}\cdots i_{n})}
~~~~~~~~~~~~~~~~~~~~\mbox{if}~~~
\eta=2n-\textstyle{\frac{3}{2}}m\label{DPsi}~~~~~\\
{}\nonumber\\
D_{(i[\alpha}\Psi_{
\beta_{1}\cdots\beta_{m}]i_{1}\cdots i_{n})}
~~~~~~~~~~~~~~~~~~~~\mbox{if}~~~
\eta=2n+\textstyle{\frac{5}{2}}m~~~~~\label{m3}\\
{}\nonumber\\
D_{(i|\alpha |}\Psi^{[\alpha\alpha_{1}\cdots\alpha_{l}]}_{i_{1}\cdots i_{n})}
~~~~~~~~~~~~~~~~~~~~~~~~~\mbox{if}~~~
\eta=\textstyle{\frac{15}{2}}-\textstyle{\frac{5}{2}}l+2n\label{l3}\\
{}\nonumber\\
D_{(i|\alpha |}\Psi^{(\alpha\alpha_{1}\cdots\alpha_{l})}_{i_{1}\cdots i_{n})}
~~~~~~~~~~~~~~~~~~~~~~~~~~~~~~~~~~~\mbox{if}~~~
\eta=\textstyle{\frac{15}{2}}+\textstyle{\frac{3}{2}}l+2n\\
{}\nonumber\\
D_{(i|\alpha|}\Psi^{\beta}_{i_{1}\cdots i_{n})}-\textstyle{\frac{1}{4}}
\delta_{\alpha}^{~\beta}D_{(i|\gamma|}\Psi^{\gamma}_{i_{1}\cdots i_{n})}~~~~~~~~~~~~~\mbox{if}~~~\eta=2n-\textstyle{\frac{1}{2}}~~~~~~~
\end{subeqnarray}
where $(\,),\,[\,]$ denote the usual
symmetrization,\,anti-symmetrization respectively and obviously 
eqs.(\ref{m3},\,\ref{l3}) are nontrivial if $m,l\leq 3$. \newline
Now we consider the case where  more than one spinor derivative,
$D_{i\alpha}$, act on a quasi-primary superfield. In this case, acting
on a general tensor field, there are inhomogeneous terms proportional
to $\bar{\hat{\rho}}_{j\beta}$, as in eq.(\ref{infDPSI}), but also 
\begin{equation}
D_{i\alpha}{\bar{\hat{\rho}}}_{j\beta}=\textstyle{\frac{1}{2}}
i{\rm b}_{\alpha\beta}\bar{{\cal
E}}_{ij}
\end{equation}
The latter terms can be eliminated by symmetrizing all the symplectic
indices, $i,j,\cdots$, while the $\bar{\hat{\rho}}$ terms  may vanish 
by  a suitable choice of the scale dimension, $\eta$. \newline
Hence,   the following are quasi-primary
\begin{subeqnarray}
D_{(i_{1}[\alpha_{1}}\cdots D_{i_{k}\alpha_{k}}
\Psi_{\beta_{1}\cdots\beta_{m}]j_{1}\cdots j_{n})}
~~~~~~~~~~~~~~~~~~~~\mbox{if}~~~
\eta=\textstyle{\frac{5}{2}}m+2n+2k-2~~~~~\\
{}\nonumber\\
D_{(i_{1}|\alpha_{1}|}\cdots D_{i_{k}|\alpha_{k}|}
\Psi^{[\alpha_{1}\cdots\alpha_{k}\beta_{1}\cdots\beta_{l}]}_{j_{1}\cdots
j_{n})}~~~~~~~~~~~~~~~~~~~\mbox{if}~~~
\eta=8+2n-\textstyle{\frac{5}{2}}l-\textstyle{\frac{1}{2}}k~~~~~~
\end{subeqnarray}
Note that
\begin{equation}
D_{(i(\alpha}D_{j)\beta)}=0
\end{equation}


\bibliographystyle{unsrt}
\bibliography{reference}

\end{document}